\documentclass[nofootinbib,aps,11pt]{revtex4-1}
\usepackage{graphicx}
\usepackage{subfigure} 
\usepackage{hyperref}
\usepackage{cancel}
\usepackage{amssymb}
\usepackage{textcomp}
\usepackage{amsmath}
\usepackage{bm}
\usepackage{times}
\usepackage{epsfig}
\usepackage{color}

\newcommand{\eq}{{\rm eq}}
\newcommand{\CP}{{\rm CP}}
\newcommand{\GeV}{{\rm GeV}}
\newcommand{\DM}{{\rm DM}}
\newcommand{\SM}{{\rm SM}}
\begin{document}
\title{\Large Common Origin of Dark Matter and Leptogenesis in $U(1)_{B-L}$}
\bigskip
\author{Ang Liu$^1$}
\author{Feng-Lan Shao$^1$}
\email{shaofl@mail.sdu.edu.cn}
\author{Zhi-Long Han$^2$}
\email{sps\_hanzl@ujn.edu.cn}
\author{Yi Jin$^2$}
\author{Honglei Li$^2$}
\email{sps\_lihl@ujn.edu.cn}
\affiliation{$^1$School of Physics and Physical Engineering, Qufu Normal University, Qufu, Shandong 273165, China\\
	$^2$School of Physics and Technology, University of Jinan, Jinan, Shandong 250022, China}
\date{\today}
\begin{abstract} 
In this paper, we investigate the common parameter space of dark matter and leptogenesis in the $U(1)_{B-L}$ symmetry. This model involves a complex scalar $\phi$, sterile neutrinos $N$, and Majorana dark matter $\chi$, where only dark matter $\chi$ is charged under the $Z_2$ symmetry. Masses of $N$ and $\chi$ are generated via the Yukawa interactions to $\phi$ after breaking of the $U(1)_{B-L}$ symmetry. TeV scale sterile neutrinos $N$ are responsible for the generation of baryon asymmetry through the resonance leptogenesis mechanism. The new particles in the $U(1)_{B-L}$ have a significant impact on the dilution of $N$, thus on leptogenesis. Meanwhile, the annihilation processes of dark matter $\chi$ are almost identical to that of $N$, which indicates that both leptogenesis and dark matter are closely related to  satisfying the observed results simultaneously. Under various theoretical and experimental constraints, the viable common parameter space of dark matter and leptogenesis is obtained for both global and local $U(1)_{B-L}$ symmetry.

\end{abstract}
\maketitle

\section{Introduction}

The standard model (SM) is one of the most successful theories, which provides excellent interpretations for particle physics. However, some sharp questions cannot be explained by SM. For instance, observations of neutrino oscillations \cite{Super-Kamiokande:1998kpq, SNO:2002tuh} indicate that neutrinos have tiny masses under the constraint from cosmology $\sum m_\nu<0.12$ eV \cite{Planck:2018vyg}.  Another issue is that the amount of CP violation in SM is too small to account for the observed baryon asymmetry \cite{Fong:2012buy}. Meanwhile, various cosmological and astrophysical observations require the existence of particle dark matter (DM) \cite{Bertone:2004pz,Cirelli:2024ssz}. These provide concrete evidence for new physics beyond the SM. One appealing way is seeking the common origin of these new physics \cite{Asaka:2005pn,Ma:2006fn,Falkowski:2011xh,Falkowski:2017uya,Hugle:2018qbw,Chianese:2019epo,Liu:2020mxj}.

The famous type-I seesaw mechanism \cite{Minkowski:1977sc,Mohapatra:1979ia} can naturally explain tiny neutrino masses by introducing sterile neutrinos $N$. Meanwhile, as described by the leptogenesis mechanism \cite{Fukugita:1986hr, Davidson:2008bu, Blanchet:2012bk}, the lepton asymmetry could be created by the out-of-equilibrium decays of $N$ with $\CP$-asymmetry parameter in the early universe, afterwards which can be converted to the baryon asymmetry through the $(B+L)$-violating sphaleron transitions \cite{Manton:1983nd, Klinkhamer:1984di, Kuzmin:1985mm, Khlebnikov:1988sr, Harvey:1990qw}.  But the accompanying question is the mass scale of $N$, namely the seesaw scale, which can be widely distributed between eV and $10^{14}$ GeV from \cite{Mohapatra:2005wg, Drewes:2013gca}. Typically, the standard thermal leptogenesis with hierarchical sterile neutrinos requires  $m_N\gtrsim10^{9}$ GeV \cite{Davidson:2002qv}, which is not testable at current or even future colliders. One viable pathway for TeV scale leptogenesis is the resonant mechanism with nearly degenerate sterile neutrinos \cite{Flanz:1996fb, Pilaftsis:1997dr, Pilaftsis:1997jf, Pilaftsis:2003gt, Dev:2017wwc}, where the TeV scale sterile neutrino could lead to a rich phenomenon \cite{Deppisch:2015qwa, Cai:2017mow, deGouvea:2013zba, Alekhin:2015byh}.

Another drawback of canonical type-I seesaw is the lack of DM candidates. A keV scale sterile neutrino could serve as a decaying DM \cite{Dodelson:1993je,Datta:2021elq}, but it is tightly constrained by observations \cite{Ng:2019gch}. A stable sterile neutrino DM is possible with discrete symmetry \cite{Ma:2006km}, which usually suffers tight constraints from lepton flavor violation \cite{Kubo:2006yx}. Instead of being DM, the sterile neutrino is an ideal messenger between the dark sector and SM \cite{Pospelov:2007mp,Gonzalez-Macias:2016vxy,Escudero:2016ksa,Batell:2017cmf,Blennow:2019fhy,Ballett:2019pyw,Biswas:2021kio,Coy:2021sse,Borah:2021pet,Fu:2021uoo,Das:2022oyx,Coito:2022kif,Biswas:2022vkq,Liu:2023kil}, which can lead to naturally suppressed DM-nucleon scattering cross section \cite{Escudero:2016ksa} and observable indirect detection signature \cite{Batell:2017rol}. Currently, the most popular DM candidate is a weakly interacting massive particle (WIMP) generated through the freeze-out mechanism. Therefore, many experiments are dedicated to searching for evidence of their existence \cite{XENON:2018voc,XENON:2023sxq, PandaX-II:2017hlx, PandaX-4T:2021bab, Boveia:2018yeb,Adam:2021rrw, AMS:2013fma,Fermi-LAT:2015att}.

In seeking for the common origin of neutrino mass, baryon asymmetry and DM, a meaningful thing to consider is  what symmetry follows the new physics.  One natural option is the $B-L$ symmetry \cite{Marshak:1979fm, Mohapatra:1980qe}. The phenomenology of WIMP DM  and leptogenesis in the framework of $U(1)_{B-L}$ extended type-I seesaw are separately analyzed in Ref. \cite{Okada:2010wd,Escudero:2016tzx,Okada:2018ktp} and Ref. \cite{Iso:2010mv,Heeck:2016oda,Dev:2017xry}, respectively. Since both DM relic density and baryon asymmetry via leptogenesis are produced around the TeV scale, they can be closely correlated by the sterile neutrinos and other new particles in $U(1)_{B-L}$. In this paper, we further explore the common parameter space for DM and leptogenesis in the type-I seesaw with $U(1)_{B-L}$ symmetry.

The $U(1)_{B-L}$ extended type-I seesaw introduces sterile neutrinos $N$ to generate tiny neutrino masses. TeV scale $m_N$ is required to produce the baryon asymmetry through the resonant leptogenesis. A dark fermion $\chi$, which is odd under the exact $Z_2$ symmetry, is regarded as the DM candidate. In the minimal $B-L$ model, the DM $\chi$ is actually one of the sterile neutrinos \cite{Okada:2010wd}. A complex scalar $\phi$  breaks the $B-L$ symmetry, whilst brings masses to $N$ and $\chi$. In this framework, the annihilation processes of $N$ and $\chi$ mainly affect the final baryon asymmetry and relic density, which are directly related via the conversion process $NN\leftrightarrow\chi\chi$. We will demonstrate the common parameter space for DM and leptogenesis, which can simultaneously satisfy the corresponding observation values under various constraints.

The structure of this paper is organized as follows.  In Sec. \ref{SEC:FW}, we briefly introduce the framework of the model and certain constraints. The calculation of lepton asymmetry and relic density of DM as well as the analysis of some phenomenological constraints  are revealed in Sec. \ref{SEC:GUS} for the global $U(1)_{B-L}$ scenario, and in Sec. \ref{SEC:LUS} for the local one. Finally, the conclusion is presented in Sec. \ref{SEC:CL}.

\section{The Framework}\label{SEC:FW}
\begin{table}
	\begin{center}\Large
		\begin{tabular}{|c|c c c |c c c c | c c |} 
			\hline
			&$q_L$ & $u_R$ & $d_R$ &$L_L$  & $e_R$ &~$N$~ & ~$\chi$~ & ~$H$~ & ~$\phi$~ \\ \hline
			$SU(2)_L$ &2 &1 & 1&2 &1 &1 & 1 & 2 & 1 \\ \hline
			$U(1)_Y$ &$+\frac{1}{6}$ &$+\frac{2}{3}$& $-\frac{1}{3}$&$-\frac{1}{2}$~ &-1 & 0 & 0 & $\frac{1}{2}$ & $0$ \\ \hline
			$U(1)_{B-L}$&$+\frac{1}{3}$&$+\frac{1}{3}$ &$+\frac{1}{3}$ & -1 &-1& -1 & -1 & 0 & $2$~ \\\hline
			$Z_2$ &$+$ &$+$& $+$& $+$ & $+$ & $+$ & $-$ & $+$& $+$ \\
			\hline
		\end{tabular}
	\end{center}
	\caption{Relevant particle contents and corresponding charge assignments. 
		\label{Tab:Particle}}
\end{table} 

 This model extends complex scalar $\phi$, sterile neutrinos $N$, and DM $\chi$ beyond the SM. The particle contents and the corresponding charge assignments are listed in Table \ref{Tab:Particle}.  Regarding the descriptions of Refs. \cite{Garcia-Cely:2013wda,  Garcia-Cely:2013nin, Escudero:2016tzx}, the scalar $\phi$ will acquire a non-zero vacuum expectation value (VEV), which spontaneously breaks the $U(1)_{B-L}$ symmetry. Ultimately, only $\chi$ is odd under the $Z_2$ symmetry, which is naturally  identified as the stable DM candidate. 
 
 The scalar potential is 
 \begin{eqnarray} 
 	V & = & \mu_{H}^2 H^\dag H
 	+\mu_\phi^{2} \phi^\dag \phi - \lambda_H (H^\dag H)^2 - \lambda_\phi (\phi^\dag \phi)^2 - \lambda_{H\phi} (H^\dag H)(\phi^\dag \phi),
 \end{eqnarray}
 where $H$ is the standard Higgs doublet. $\mu_{H}$ and $\mu_{\phi}$ satisfy the relationship $\mu_{H}^2=\lambda_H v_H^2+\frac{1}{2}\lambda_{H\phi}v_\phi^2$  and $\mu_{\phi}^2=\lambda_\phi v_\phi^2+\frac{1}{2}\lambda_{H\phi}v_H^2$ under the condition of the minimization of the scalar potential \cite{Escudero:2016tzx}.
 
The relevant Yukawa interaction can be written as \cite{Escudero:2016tzx}
 \begin{eqnarray}\label{Eqn:fl}
 	\mathcal{L}&\supset& -\frac{\lambda_N}{\sqrt{2}}\phi\bar{N} P_R N- \frac{\lambda_\chi}{\sqrt{2}}\phi\bar{\chi}P_R\chi 
 	-y_\nu \bar{L}_L H P_R N+h.c..
 \end{eqnarray}
 The tiny neutrino mass is generated via the third term in  Eq.~\eqref{Eqn:fl}, and can be expressed as
 \begin{equation}
 	m_\nu = - \frac{v_H^2}{2} y_\nu~ m_{N}^{-1} y_\nu^T.
 \end{equation}
Here, $v_H = 246~\GeV$ is the VEV of SM Higgs doublet $H$.  Following the Casas-Ibarra parametrization, the Yukawa matrix $y_\nu$ can be expressed as \cite{Casas:2001sr}
\begin{equation}
 y_\nu=\frac{\sqrt{2}}{v_H}\sqrt{\hat m_N} R \sqrt{\hat m_\nu} U_{\rm PMNS},
\end{equation}
where $\hat{m}_N$ and $\hat{m}_\nu$ are the diagonalized mass matrix for sterile and light neutrinos. $U_{\rm PMNS}$ is the neutrino mixing matrix. It is possible to achieve relatively large components of $y_\nu$ by adjusting the complex angle of the orthogonal matrix $R$ \cite{Barman:2022scg}.  But our focus is not this, the parameter related to leptogenesis is the effective neutrino mass $\tilde{m}\equiv v_H^2(y_\nu^{\dagger} y_\nu)/m_N$, which determines the decay width of $N\to HL$.

The normal thermal leptogenesis requires $m_N\gtrsim10^{9}$ GeV to generate a sufficient amount of the baryon asymmetry \cite{Davidson:2002qv}, which is far beyond the typical WIMP scale. However, the resonant leptogenesis can solve this conflict by TeV scale $m_N$ \cite{Flanz:1996fb, Pilaftsis:1997dr, Pilaftsis:1997jf, Pilaftsis:2003gt, Dev:2017wwc}.  The CP-asymmetry parameter $\varepsilon_{\rm CP}$ can be enhanced by the self-energy corrections for nearly degenerate sterile neutrinos \cite{Iso:2010mv}.  Furthermore, the condition $\Delta m_{N21}^2=m_{N2}^2-m_{N1}^2\simeq m_{N2}\Gamma_{N1}\ll m_{N2}^2$ could result in  order unity of $\varepsilon_{\rm CP}$. In the subsequent calculations, $\varepsilon_{\rm CP}$ will be used as an input parameter.
  
\subsection{The Global $U(1)_{B-L}$ Scenario}

In the global $U(1)_{B-L}$ scenario, the complex scalar $\phi$ can be defined as $\phi=(v_\phi + \tilde{\rho} +i\eta)/\sqrt{2}$ with the CP-even real scalar $\tilde{\rho}$ and CP-odd massless real scalar $\eta$ named Majoron, which is the Goldstone boson of the spontaneous breaking of the global $U(1)_{B-L}$ symmetry \cite{Chikashige:1980ui}. After the spontaneous symmetry breaking, there are two CP-even scalars $h$ and $\rho$, whose masses are
\begin{eqnarray} 
	m_h^2=2\lambda_H v_H^2 \cos^2\theta + 2\lambda_{\phi}v_\phi^2\sin^2\theta - \lambda_{H\phi} v_H v_\phi \sin2\theta, \\
	m_\rho^2=2\lambda_H v_H^2 \sin^2\theta + 2\lambda_{\phi}v_\phi^2\cos^2\theta + \lambda_{H\phi} v_H v_\phi \sin2\theta. 
\end{eqnarray}
Here $h$ is identical to the 125 GeV SM Higgs boson. $\theta$ is the mixing angle between two real CP-even scalars $h$ and $\rho$. Then quartic couplings can  be expressed as functions of masses and mixing angle,
\begin{eqnarray} 
	\lambda_H&=&\frac{m_h^2\cos^2\theta+m_\rho^2\sin^2\theta}{2v_H^2}, \\
	\lambda_\phi&=&\frac{m_h^2\sin^2\theta+m_\rho^2\cos^2\theta}{2v_\phi^2},\\
	\lambda_{H\phi}&=&\frac{(m_\rho^2-m_h^2)\sin2\theta}{2v_H v_\phi},
\end{eqnarray}
where the stability of the scalar potential require $\lambda_{H\phi}^2<4\lambda_{H}\lambda_{\phi}$. This condition can be  easily fulfilled as long as $m_{\rho,h}^2>0$.
The chiral fermions will obtain  Majorana masses after $U(1)_{B-L}$ symmetry breaking. In the global $U(1)_{B-L}$  symmetry scenario, the Lagrangian of Eq.~\eqref{Eqn:fl} becomes 
\begin{eqnarray}
	\mathcal{L}_{\rm global}&=& -\frac{m_N}{2}\bar{N}N-\frac{\lambda_{N}}{2}(-h\sin\theta+\rho\cos\theta)\bar{N}N-\frac{\lambda_{N}}{2}i\eta\bar{N}\gamma^5N \nonumber \\
	&&-\frac{m_\chi}{2}\bar{\chi}\chi-\frac{\lambda_{\chi}}{2}(-h\sin\theta+\rho\cos\theta)\bar{\chi}\chi-\frac{\lambda_{\chi}}{2}i\eta\bar{\chi}\gamma^5\chi
	\nonumber \\
	&&- (y_\nu \bar{L}_L H P_R N+h.c.),
\end{eqnarray}
where $m_N=\lambda_{N}v_\phi$ and $m_\chi=\lambda_{\chi}v_\phi$. The free parameters are $\left\{m_\rho, m_\eta, m_N, m_\chi, \tilde{m}, v_\phi, \theta,  \varepsilon_{\rm CP} \right\}$ in the global scenario, then all required couplings $\lambda_\phi, \lambda_{H\phi}, \lambda_N, \lambda_{\chi}$ can be expressed as functions of these parameters. Since different couplings have dissimilar powers in the renormalization group equation \cite{Dev:2015vjd}, we use the general perturbation limits as $\lambda_{N,\chi}<\sqrt{4\pi}$ and $\lambda_{\phi}<4\pi$ \cite{Dev:2017xry}.

\subsection{The Local $U(1)_{B-L}$ Scenario}

Different from previous studies \cite{Garcia-Cely:2013wda, Garcia-Cely:2013nin, Escudero:2016tzx}, we also consider the local scenario.  In this case, the massless Goldstone mode is eaten by the gauge boson $Z^\prime$, which acquires a mass $m_{Z^\prime}=2g_{B-L}v_\phi$. Here $g_{B-L}$ is the gauge coupling associated with the $U(1)_{B-L}$ gauge group. The relevant interactions can be rewritten as
\begin{eqnarray}
	\mathcal{L}_{\rm local}&=& -\frac{m_N}{2}\bar{N}N-\frac{\lambda_{N}}{2}(-h\sin\theta+\rho\cos\theta)\bar{N}N-
	\frac{g_N^A}{2} Z^\prime_\mu\bar{N}\gamma^\mu\gamma^5N \nonumber \\
	&-&\frac{m_\chi}{2}\bar{\chi}\chi-\frac{\lambda_{\chi}}{2}(-h\sin\theta+\rho\cos\theta)\bar{\chi}\chi-\frac{g_\chi^A}{2}  Z^\prime_\mu\bar{\chi}\gamma^\mu\gamma^5\chi 
	\nonumber \\
	&-& g_f^V Z^\prime_\mu\bar{f}\gamma^\mu f -(y_\nu \bar{L}_L H P_R N+h.c.),
\end{eqnarray}
where $g_i^{A,V}=g_{B-L}Q_i$ with $Q_i$ the the $U(1)_{B-L}$ charge of particle $i$. $f$ denotes the SM fermions. We need to clarify here that the vectorial coupling $g_{N,\chi}^V$ vanishes for Majorana particles $N$ and $\chi$, as well as the axial coupling $g_{f}^A$ with the disappearing $U(1)_{B-L}$ charge of SM Higgs doublet $H$ \cite{Kahlhoefer:2015bea}. In the local scenario, the free input parameters are $\left\{m_\rho, m_{Z^\prime}, m_N, m_\chi, \tilde{m}, v_\phi, \theta,  \varepsilon_{\rm CP}\right\}$. Slightly different from the global scenario, $m_{\rho, N, \chi}\leq \sqrt{4\pi} v_\phi$ needs to be satisfied to ensure the perturbative couplings, i.e., $\lambda_{N,\chi}<\sqrt{4\pi}$ and $\lambda_{\phi}<2\pi$ \cite{Kahlhoefer:2015bea, Duerr:2016tmh}.

\section{The Global $U(1)_{B-L}$ Scenario}\label{SEC:GUS}

\subsection{Relic Density and Leptogenesis}\label{SEC:GRD}  
\begin{figure}
	\begin{center}
		\subfigure[]{
			\centering	
			\includegraphics[width=0.45\linewidth]{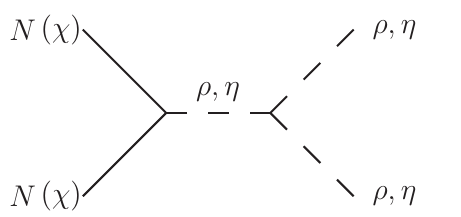}		
		}
		\subfigure[]{
			\centering	
			\includegraphics[width=0.45\linewidth]{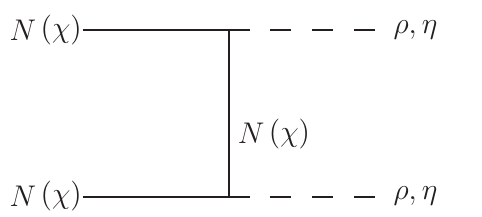}		
		}
		\subfigure[]{
			\centering	
			\includegraphics[width=0.45\linewidth]{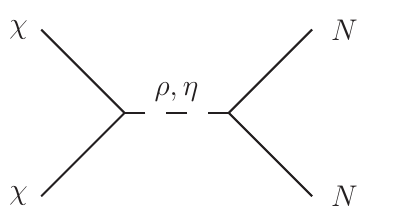}		
		}
		\subfigure[]{
			\centering	
			\includegraphics[width=0.45\linewidth]{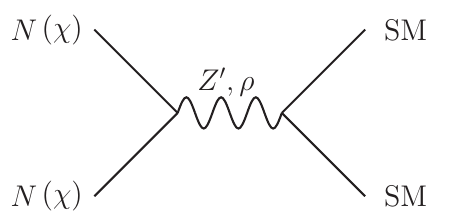}		
		}
	\end{center}
	\caption{The dominant Feynman diagrams for $N$ and $\chi$ annihilation involving new particles in $U(1)_{B-L}$. Among them, subfigure (a)-(c) are universal for the global and the local scenarios, while the latter needs to replace $\eta$ with $Z^\prime$. Meanwhile, mixing can lead to processes such as $NN(\chi\chi)\to \rho ,h\to {\rm SM SM}$ in both scenarios,  yet the $\rho$ mediated one is dominant for TeV scale $N,\chi$ as shown in panel (d). The $Z^\prime$ mediated channel in panel (d) only appears in the local scenario.  }
	\label{FIG:FD}
\end{figure}

In the following calculation, we assume that the new scalars $\rho$ and $\eta$ invariably stay in thermal equilibrium by coupling to $H$. Additionally, the CP-odd scalar $\eta$ with $m_\eta=0$ is employed throughout this work for simplicity. The relevant Feynman diagrams are shown in Fig.~\ref{FIG:FD}. Additionally, almost all contributions of processes $NN(\chi\chi)\to\rho\to \SM\SM$ with large $m_{N,\chi}$ come from the boson final states of $V(V=W,Z,h)$. Therefore, we use $NN(\chi\chi)\to VV$ to represent this process below. Following Refs.~\cite{Plumacher:1996kc,Iso:2010mv,Dev:2017xry}, we can write the relevant Boltzmann equations as
\begin{eqnarray}\label{Eqn:GBE}
	\frac{dY_N}{dz} &= & -\frac{z}{sH(m_N)}\left(\frac{Y_N}{Y_N^\eq}-1\right)(\gamma_{N\to HL}+2\gamma_{NL\to qt}+4\gamma_{Nt\to qL})\nonumber \\
	&-&\frac{z}{sH(m_N)} \left(\left(\frac{Y_N}{Y_N^\eq}\right)^2-1\right)(2\gamma_{NN\to\rho\rho,\eta\eta}+2\gamma_{NN\to\rho\eta,h\eta}+2\gamma_{NN\to VV})\nonumber \\
	&+&\frac{z}{sH(m_N)}\left(\left(\frac{Y_\chi}{Y_\chi^\eq}\right)^2-\left(\frac{Y_N}{Y_N^\eq}\right)^2\right)2\gamma_{\chi\chi\to NN},\\
\frac{dY_\chi}{dz} &= & -\frac{z}{sH(m_N)} \left(\left(\frac{Y_\chi}{Y_\chi^\eq}\right)^2-1\right)(2\gamma_{\chi\chi\to\rho\rho,\eta\eta}+2\gamma_{\chi\chi\to\rho\eta,h\eta}+2\gamma_{\chi\chi\to VV})\nonumber \\
	&-&\frac{z}{sH(m_N)}\left(\left(\frac{Y_\chi}{Y_\chi^\eq}\right)^2-\left(\frac{Y_N}{Y_N^\eq}\right)^2\right)2\gamma_{\chi\chi\to NN},\\
\frac{dY_{B-L}}{dz} &= & \frac{z}{sH(m_N)} \left(\varepsilon_{\CP}\left(\frac{Y_N}{Y_N^\eq}-1\right)-\frac{Y_{B-L}}{2Y_L^{\eq}}\right)\gamma_{N\to HL}\nonumber \\
	&-&\frac{z}{sH(m_N)}\frac{Y_{B-L}}{Y_L^{\eq}}\left(\frac{Y_N}{Y_N^{\eq}}\gamma_{NL\to qt}+2\gamma_{Nt\to qL}\right),
\end{eqnarray}
where $z=m_N/T$, the entropy density $s=2\pi^2 g_s T^3/(45z^3)$. The
Hubble expansion rate at temperature $T=m_N$ is $H(m_N)=\sqrt{4\pi^3g_*/45} m_N^2/m_{pl}$ with the Planck mass $m_{pl}=1.22\times10^{19}~\GeV$. Here we take both the number of relativistic degrees of freedom for the entropy density $g_s$ and energy density $g_\star$ as 106.75.

In the above Boltzmann equations, the contributions of the annihilation processes mediated by $\rho$ to the final SM particles are taken into account with a small mixing angle $\theta\simeq0.05$, which is favored by DM direct detection.  Moreover, the lepton asymmetry could be converted to the baryon asymmetry through the sphaleron processes, i.e., $Y_{B}=\frac{28}{79}Y_{B-L}$. The DM relic density can be written as $\Omega_{\chi} h^2=m_{\chi}s_0  Y_{\chi}h^2/\rho_c$ with the entropy density of today $s_0=2891.2~\rm cm^{-3}$ and the critical density of the universe $\rho_c=1.0538 \times 10^{-5} h^2 ~\rm GeV/cm^3$.

 The abundance at thermally equilibrium of $N,\chi,L$ can be expressed as
\begin{eqnarray}
Y_N^{\eq}=\frac{45z^2}{2\pi^4g_s}\mathcal{K}_2(z),~Y_\chi^{\eq}=\frac{45z^2m_\chi^2}{2\pi^4g_sm_N^2}\mathcal{K}_2\left(\frac{m_\chi}{m_N}z\right),~Y_L^{\eq}=\frac{3m_N^3}{2s\pi^2}\zeta(3),
\end{eqnarray}
where $\mathcal{K}_2$ and $\zeta$ are the second modified Bessel function of the second kind and the Riemann zeta function, respectively. The thermalized interaction rates for decay and scatter processes are \cite{Plumacher:1996kc}
\begin{eqnarray}
\gamma_{1\to2+3}&=&n_1^{\eq}\frac{\mathcal{K}_1}{\mathcal{K}_2}\Gamma_{1\to2+3},\\
\gamma_{1+2\to3+4}&=&\frac{T}{64\pi^2}\int_{(m_1+m_2)^2}^{\infty}ds^{\prime} \hat{\sigma}(1+2\to3+4)\sqrt{s^{\prime}}\mathcal{K}_1\left( \frac{\sqrt{s^{\prime}}}{T}\right)
\end{eqnarray}
where $n^{\eq}$ is the number density in thermal equilibrium, $\Gamma$ is the decay width. $T$, $s^{\prime}$ and $\hat{\sigma}$ represent the temperature, the squared center-of-mass energy and the reduced cross section, respectively. The reduced cross sections of $NN\to\rho\rho,\eta\eta,\rho\eta,h\eta,VV$ and $\chi\chi\to\rho\rho,\eta\eta,\rho\eta,h\eta,VV,NN$ can be found in Appendix~\ref{SEC:RCS}.

The decay width of $N\to HL$ can be expressed as
\begin{eqnarray}
	\Gamma_{N\to HL}&=&\frac{m_N^2\tilde{m}}{8\pi v_H^2}.
\end{eqnarray}
Here we fix the effective neutrino mass $\tilde{m}\simeq\sqrt{\Delta m_{\rm atm}^2}\simeq50 ~\rm meV$ throughout this paper, which is in a natural-seesaw regime. In addition, the specific representation of $\gamma_{NL\to qt}$ and $\gamma_{Nt\to qL}$ could be found in Ref.~\cite{Plumacher:1996kc}, and some washout scattering processes with tiny contributions have been overlooked \cite{Borah:2021mri}, e.g., $HL\leftrightarrow \bar{H}\bar{L}$.

\begin{figure} 
	\begin{center}
		\includegraphics[width=0.45\linewidth]{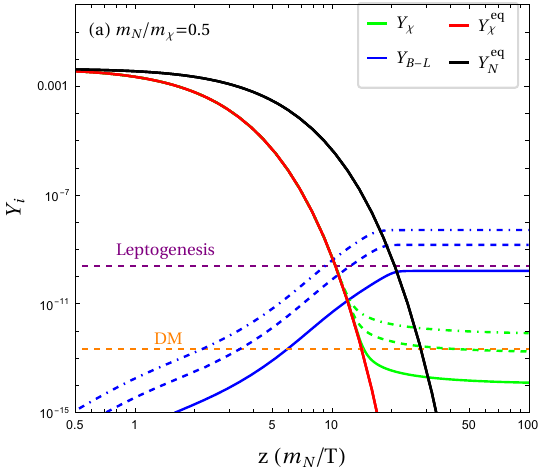}
		\includegraphics[width=0.45\linewidth]{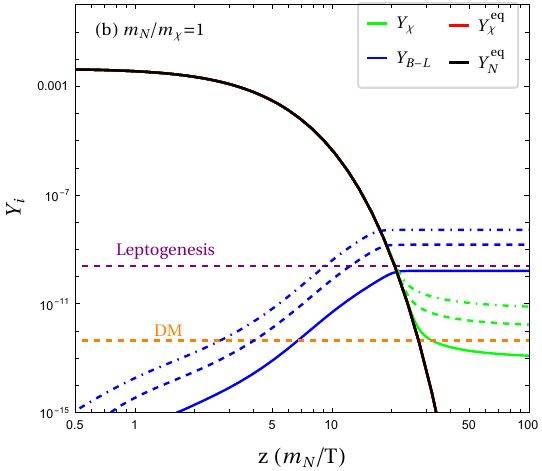}
		\includegraphics[width=0.45\linewidth]{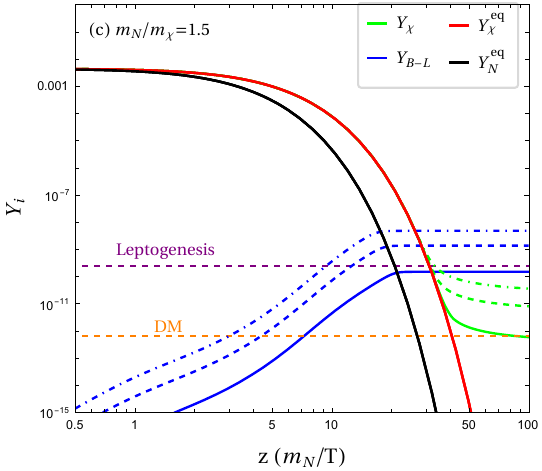}
		\includegraphics[width=0.45\linewidth]{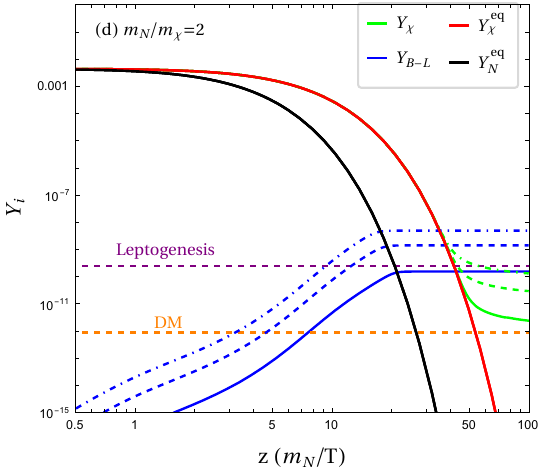}
	\end{center}
	\caption{The evolutions of various abundances $Y_i$ with $m_N=1000 ~\GeV$, $m_\rho=500 ~\GeV$, $\varepsilon_{\CP}=0.1$, and $\theta=0.05$ in the global scenario. Subfigures (a)-(d) correspond to four cases with mass ratios $m_N/m_\chi=0.5, 1, 1.5$, and 2 respectively. The solid black and red lines represent the abundances evolution of $N$ and $\chi$ in thermal equilibrium. The dotted lines in purple and orange are the observations of baryon asymmetry and DM \cite{Planck:2018vyg}, namely $Y_{B}^{\rm obs} \simeq 8.7 \times 10^{-11}$ and $\Omega_{\DM} h^2\simeq0.12$. Solid, dashed, and dot-dashed lines in green and blue represent $Y_{B-L}$ and $Y_\chi$ with $v_{\phi}=1000~\GeV, 2000~\GeV, 3000~\GeV$, respectively.
	}
	\label{FIG:fig2}
\end{figure}

The evolutions of various abundances for some benchmarks are shown in Fig.~\ref{FIG:fig2}. In these four cases, the evolution of $Y_N$ is slightly departure from thermal equilibrium, so $Y_N^{\eq}$ in all panels represent approximately the evolution of $Y_N$. In panel (a) of Fig.~\ref{FIG:fig2}, $Y_{B-L}$ needs $v_{\phi}$ to be slightly greater than 1000 GeV to satisfy the observation of baryon asymmetry. But this result is too small for observational $Y_\chi$, which requires $v_{\phi}$ around 2000 GeV. Both leptogenesis and DM cannot be satisfied simultaneously by changing $v_{\phi}$ alone, because $Y_\chi$ and $Y_{B-L}$ are coherently rising with increasing $v_{\phi}$, which is directly caused by the decreasing of $\lambda_\chi$ and $\lambda_N$.  The simplest way to realize common origin is varying the CP-asymmetry parameter $\varepsilon_{\CP}$, which means that $\varepsilon_{\CP}$ should be less than 0.1 with $v_\phi\sim2000$ GeV in case (a). In the following three subfigures (b)-(d) of Fig.~\ref{FIG:fig2}, $Y_{B-L}$ does not show significant changes with the decrease of $m_\chi$, which means that the transition process $NN\to\chi\chi$ is not the dominant contribution for the dilution of $N$. In contrast, $Y_\chi$ continuously increases as $m_\chi$ decreases for fixed $v_\phi$. Therefore, an alternative pathway to satisfy both leptogenesis and DM is modifying $m_\chi$, e.g., case (c) with $m_N/m_\chi=1.5$ and $v_\phi\sim1000$ GeV. For lighter DM as in case (d), $v_\phi<1000$ GeV is required to satisfy the observed relic density, meanwhile $\varepsilon_{\CP}$ should be larger than 0.1 for leptogenesis.

\begin{figure} 
	\begin{center}
		\includegraphics[width=0.45\linewidth]{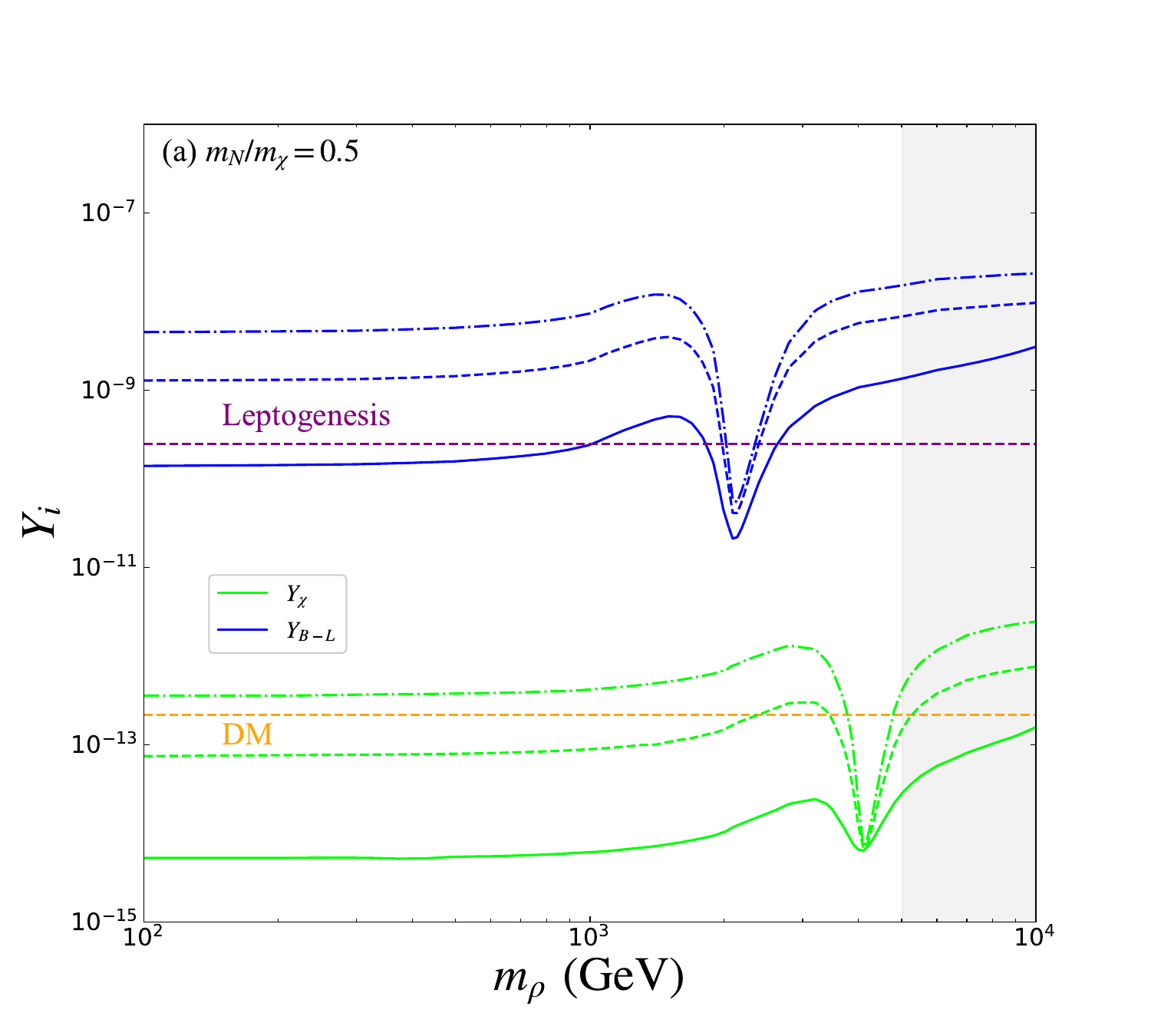}
		\includegraphics[width=0.45\linewidth]{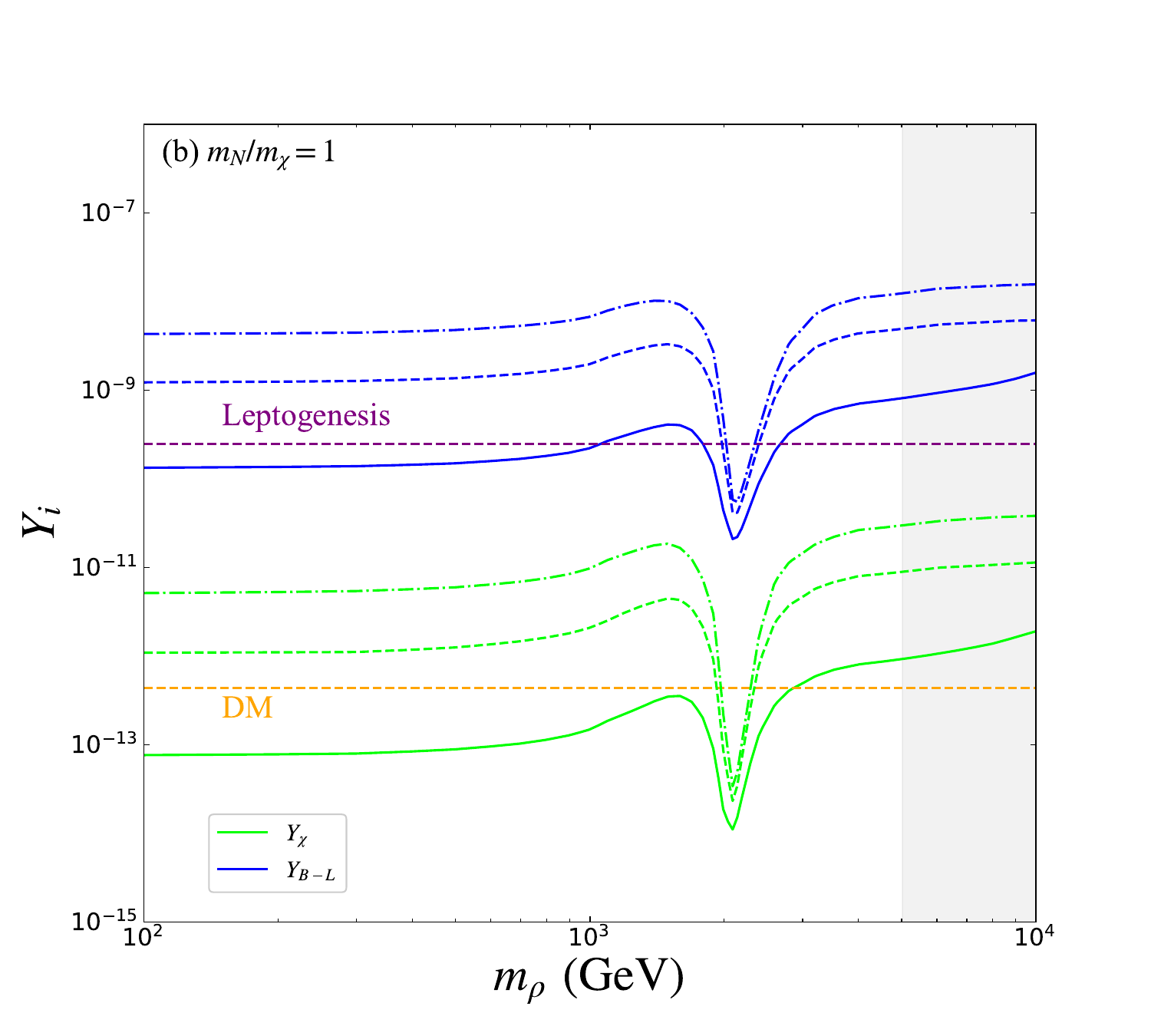}
		\includegraphics[width=0.45\linewidth]{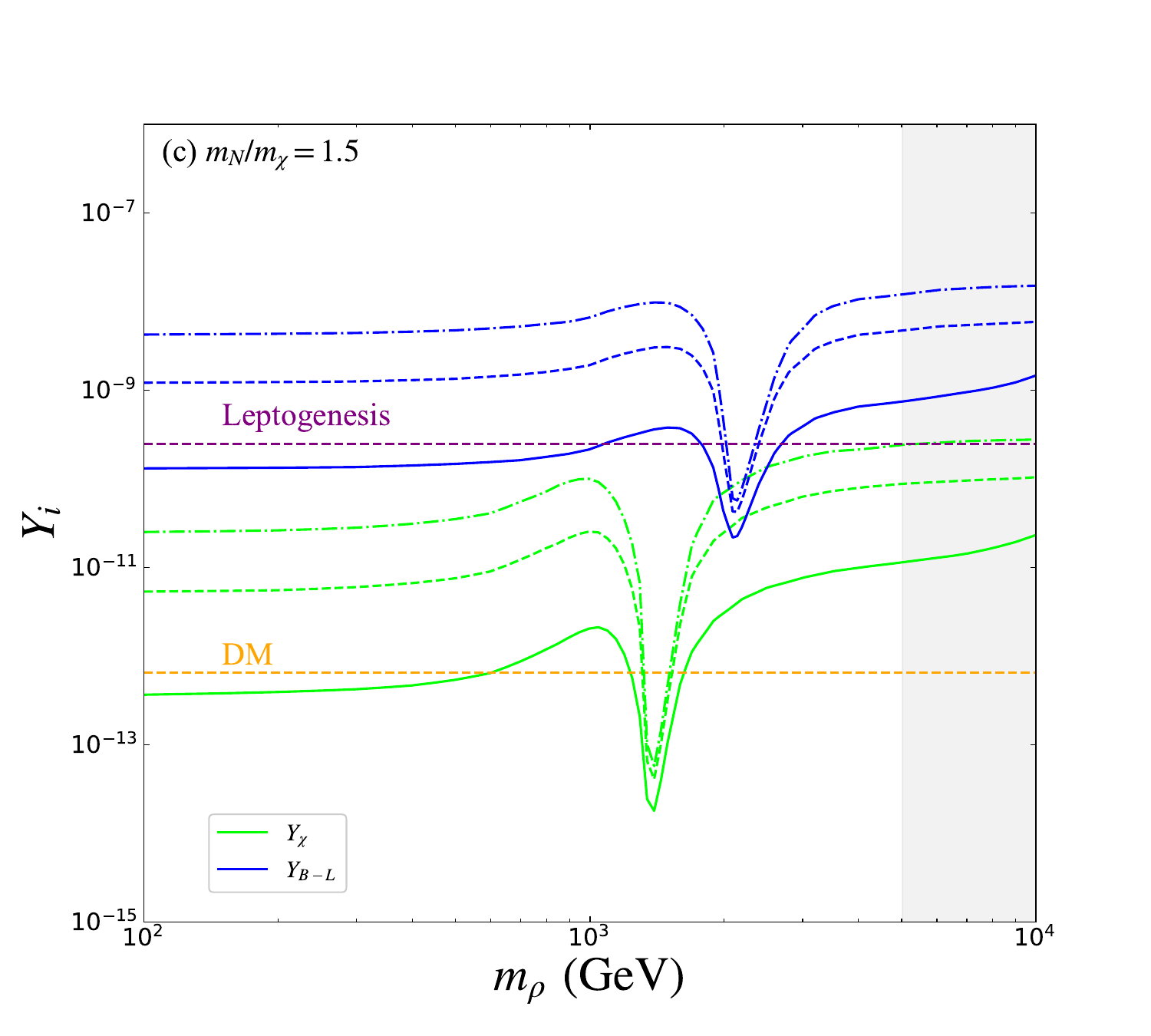}
		\includegraphics[width=0.45\linewidth]{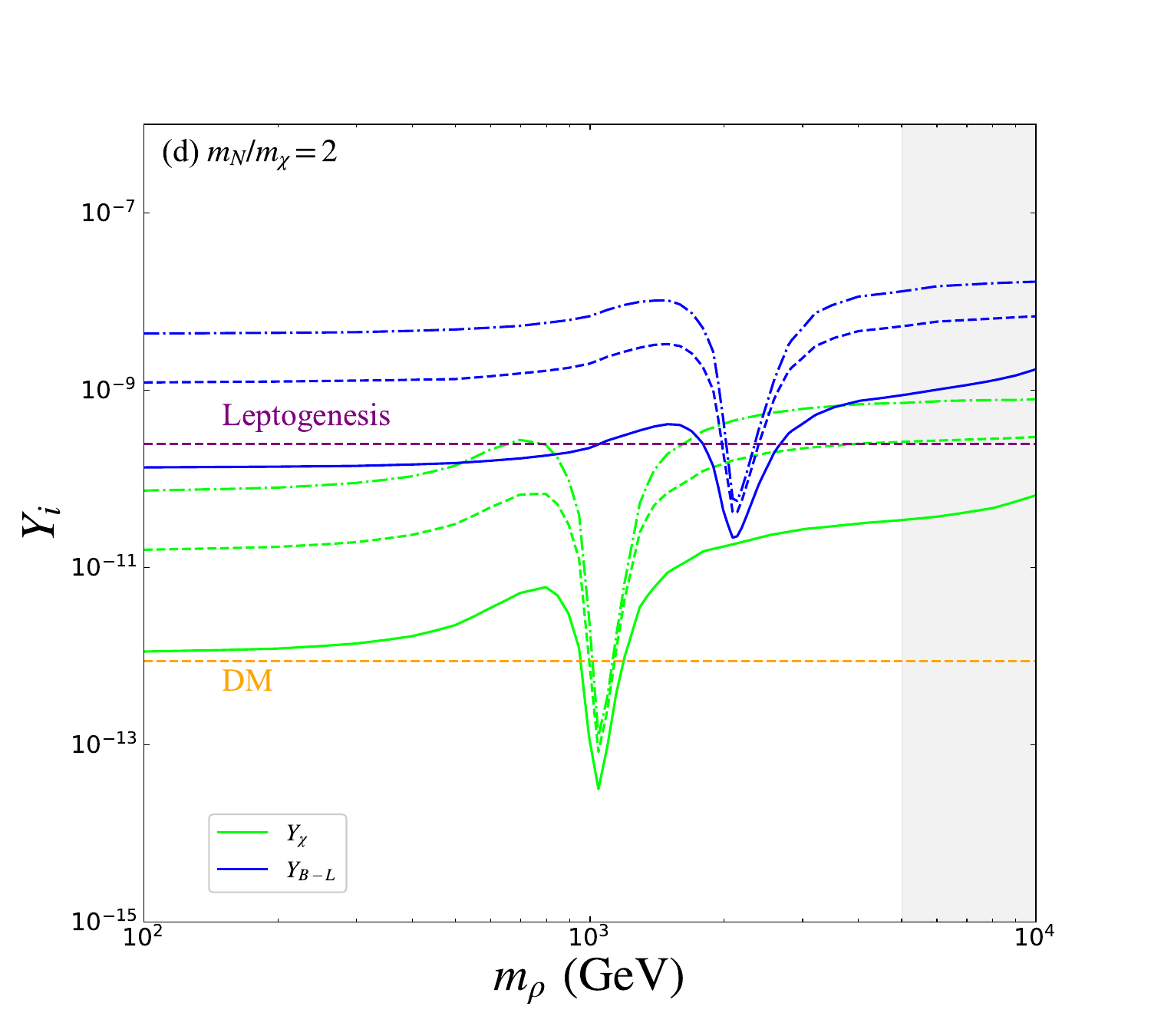}
	\end{center}
	\caption{$Y_i$ as a function of $m_\rho$ at present in the global scenario. $m_N$ is fixed as 1000 GeV,  $\varepsilon_{\CP}=0.1$, and $\theta=0.05$. Subfigures (a)-(d) correspond to four cases with mass ratios $m_N/m_\chi=0.5, 1, 1.5$, and 2 respectively. The purple, orange, blue, and green curves are the same as in Fig.~\ref{FIG:fig2}. The gray area indicates that $\lambda_\phi$ has exceeded the perturbative limit $\lambda_\phi<4\pi$ for $v_{\phi}=1000$~GeV, while the limits for $v_{\phi}=2000$ GeV and 3000 GeV are outside the scope of $m_\rho$.
	}
	\label{FIG:fig3}
\end{figure}

We then illustrate the impact of $m_\rho$ in Fig.~\ref{FIG:fig3}. The correct $Y_{B-L}$ and $Y_\chi$ occur at the resonance region $m_\rho\simeq2m_{N}$ and $m_\rho\simeq2m_{\chi}$ separately for $v_{\phi}=3000$ GeV and $m_N/m_\chi=0.5$, which is clear that leptogenesis and DM can not be satisfied simultaneously in this case. When $v_\phi=2000$ GeV, there are three solutions for correct $Y_\chi$, with two near the resonance region and the third one lower than the resonance region. By slightly modifying $\varepsilon_{\CP}$, it is possible to obtain the correct $Y_{B-L}$ and $Y_\chi$ at the same time. There will be three values of $\rho$ matching the observed $Y_{B-L}$ when $v_\phi=1000$ GeV.  However excessive dilution of $\chi$ caused by the stronger annihilation processes can only successfully explain the observed $Y_\chi$ when $m_\rho\gg m_{N,\chi}$, which is not allowed by perturbation of $\lambda_\phi$. Therefore, we mainly consider $v_\phi>1000$~GeV in the following studies.  In panels (b), (c), and (d), $Y_{B-L}$ can be hardly changed due to the tiny impact of $NN \to \chi\chi$ on the evolution of $N$. Meanwhile, the DM abundance $Y_\chi$ increases as $m_\chi$ decreases, also with the resonance region more concentrated. By comparing these four cases, we prefer the easiest approach to satisfy both leptogenesis and DM simultaneously by varying $\varepsilon_{\CP}$, especially for light $m_\rho$ region. Since both  $Y_{B-L}$ and $Y_\chi$ favor the resonance region,  $m_N\sim m_\chi$ is required to match the observed values.

\subsection{Perturbation and CP-asymmetry} \label{SEC:GC} 

\begin{figure} 
	\begin{center}
		\includegraphics[width=0.45\linewidth]{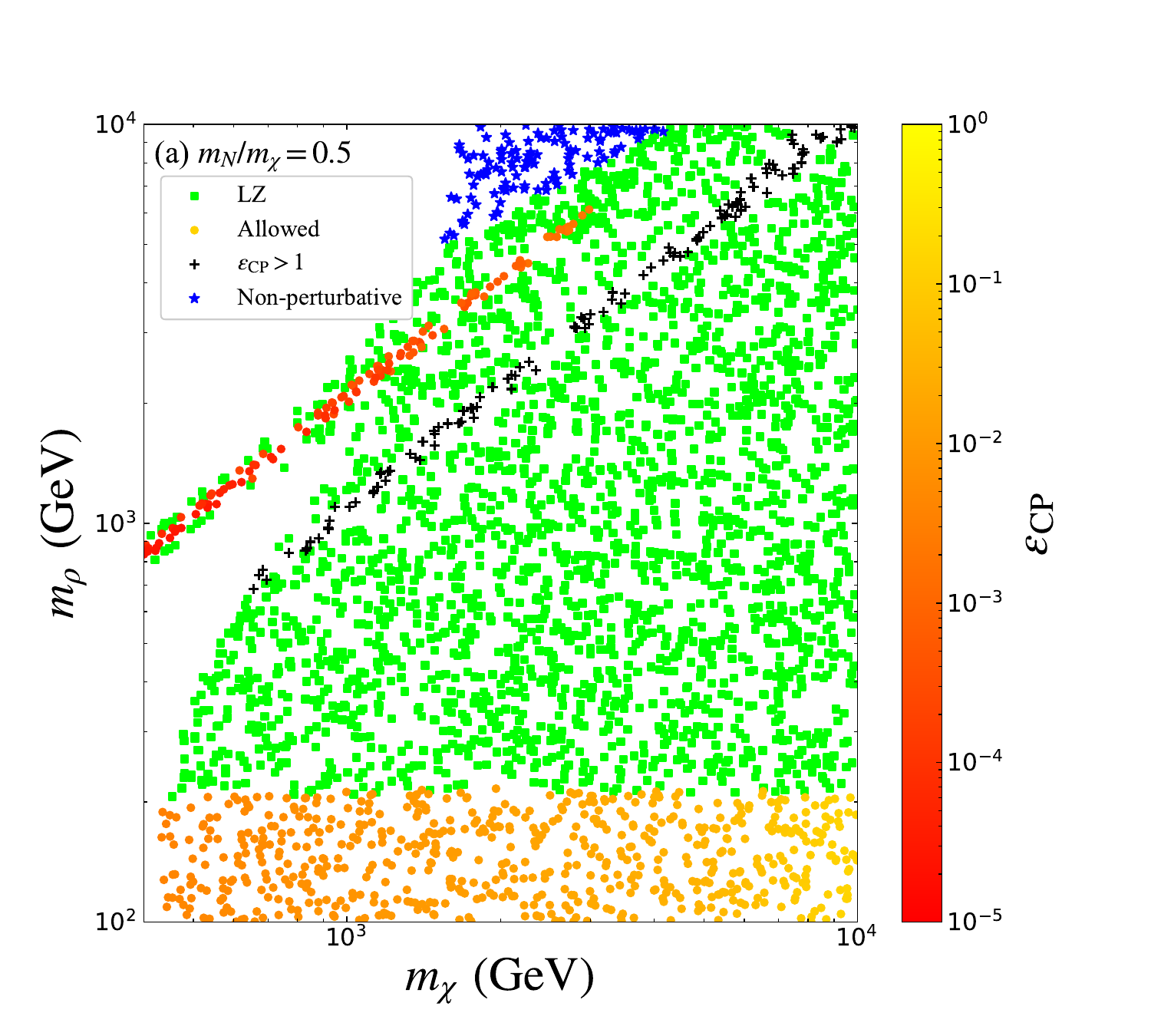}
		\includegraphics[width=0.45\linewidth]{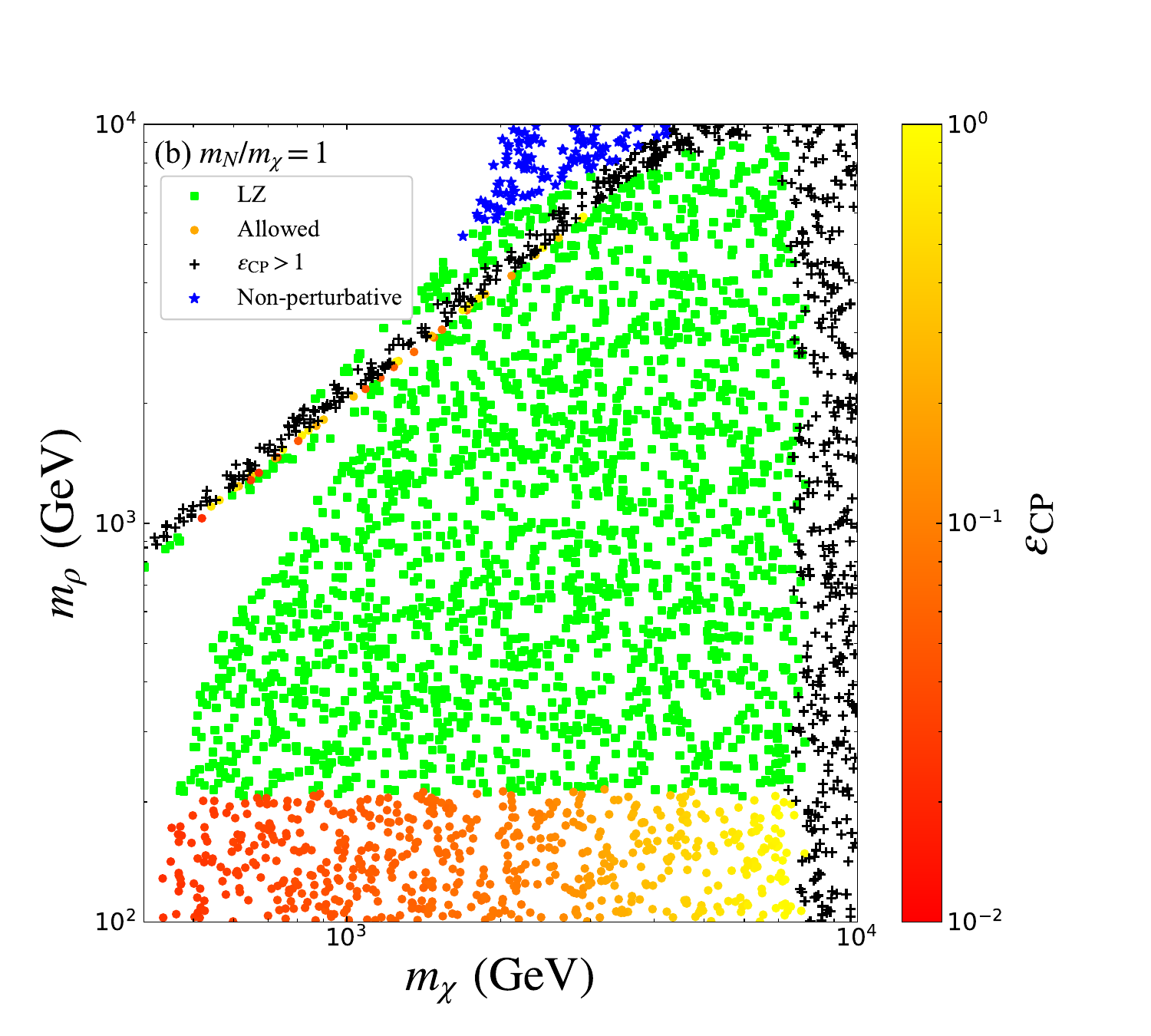}
		\includegraphics[width=0.45\linewidth]{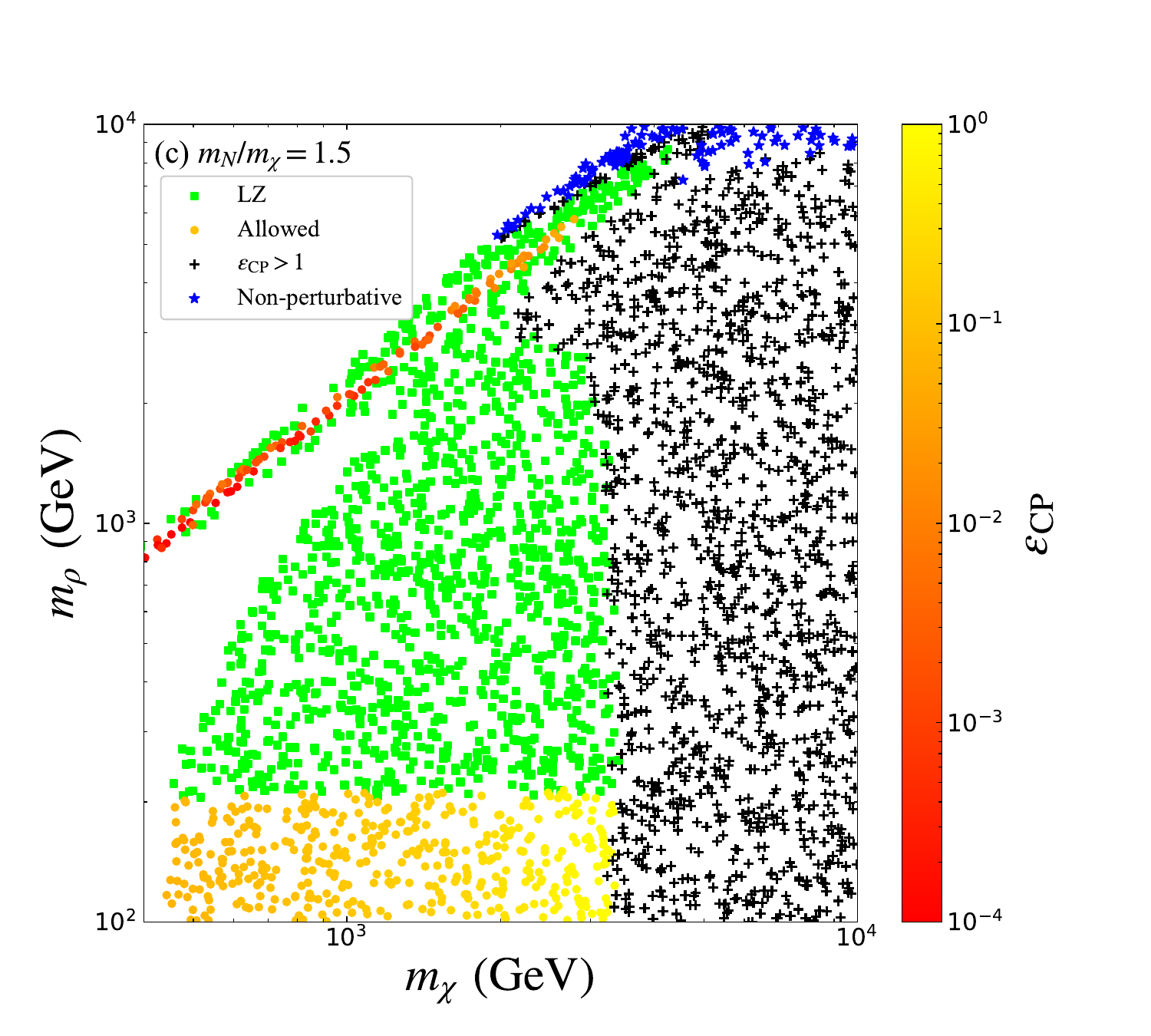}
		\includegraphics[width=0.45\linewidth]{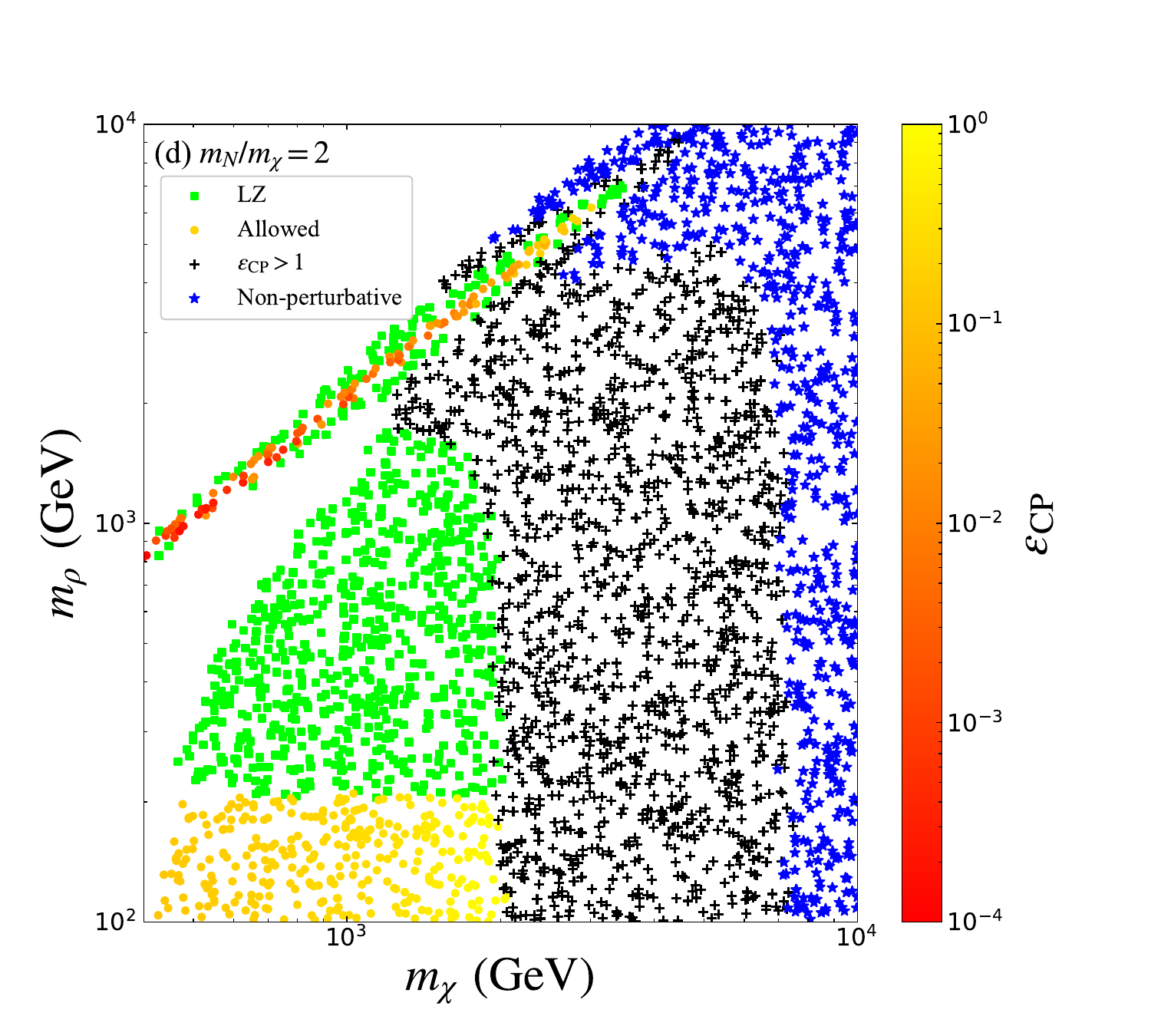}
	\end{center}
	\caption{Perturbation constraints and exceeding $\varepsilon_{\CP}$ on $m_\chi-m_\rho$ parameter space in the global scenario. (a)-(d) correspond to four cases with mass ratios $m_N/m_\chi=0.5, 1, 1.5$, and 2 respectively. The blue stars and black crosses express the regions excluded by perturbation constraints and exceeding $\varepsilon_{\CP}$, respectively. The parameter space composed of lime squares is disallowed by the present direct detection LZ~\cite{LZ:2022lsv} limit. The remaining allowed samples are colored by $\varepsilon_{\CP}$ (yellow to red). }
	\label{FIG:fig4}
\end{figure}

To obtain the allowed parameter space for the common origin of leptogenesis and DM, we then explore the following scope
\begin{eqnarray}
	\begin{aligned}\label{Eq:PS}
		m_{\chi}\in[400,10000]~\GeV, m_\rho\in[100,10000]~\GeV, v_{\phi}\in[1000,10000]~\GeV, \varepsilon_{\CP}\in[10^{-6},1].
	\end{aligned}
\end{eqnarray}
The mixing angle $\theta$ is important for DM direct detection and collider signatures. For simplicity, we fix $\theta=0.05$ for illustration. Similar to the above benchmark scenarios, masses of sterile neutrinos are determined by the ratio $m_N/m_\chi=0.5,1,1.5$, and 2 for illustration, because the too large mass difference between $m_N$ and $m_\chi$ is not favored for the common origin.  Here we do not consider the lighter DM mass $m_\chi<400$~GeV, which may disrupt the decay process $N\to H L$ and even sphaleron transitions for the scenarios as $m_N/m_\chi=0.5$. Meanwhile, $v_\phi<1000$~GeV usually leads to the DM relic density under the experimental value due to too large coupling. During the scan, the abundances of DM and baryon are required within the $3\sigma$ range of the Planck observed result~\cite{Planck:2018vyg}, i.e., $\Omega_{\chi}h^2\in[0.117,0.123],~Y_B\in[8.6,8.8]\times10^{-11}$. 

The first thing to discuss is the perturbation constraints related to the couplings, namely, $\lambda_{N,\chi}<\sqrt{4\pi}$ and $\lambda_{\phi}<4\pi$  \cite{Dev:2017xry}. For certain samples, the dilution effects of $Y_{B-L}$ are so strong that $\varepsilon_{\CP}>1$ is required, which can not be realized even by resonance leptogenesis. The scanning results are shown in Fig.~\ref{FIG:fig4}. As shown in panel (a), the parameter space with $m_\rho\gtrsim200$ GeV except for the DM resonance region $m_\rho\simeq 2m_{\chi}$ with $\theta=0.05$, is excluded by direct detection limit of LZ experiment ~\cite{LZ:2022lsv}. Furthermore, samples in the resonance regions with $m_\rho\gtrsim3000$~GeV will be excluded by LZ due to smaller $v_\phi$. It should be noted that the LZ exclusion region mainly depends on the value of $\theta$, which will satisfy the current LZ limit when $\theta\lesssim0.01$. See more detailed discussion in Sec.~\ref{SEC:DD}. As we fixed $\theta=0.05$ for all four scenarios in Fig.~\ref{FIG:fig4}, the LZ exclusion regions are near the same.

Excessive inhibition of $Y_{B-L}$ results in the need for a larger $\varepsilon_{\CP}$ to meet the observed values, sometimes even greater than 1 in the resonance region of $m_\rho\simeq m_\chi\simeq2m_N$ for scenario $m_N/m_\chi=0.5$. In addition, some parameter space with $m_\rho>2m_\chi$ will be disallowed by perturbation constraints, namely $\lambda_{\phi}\geqslant4\pi$.  Certainly, it is inevitable that the resonance effect of DM will appear around $m_\rho\simeq 2m_\chi\simeq4m_N$, where $\varepsilon_{\CP}$ will exhibit relatively smaller values $\varepsilon_{\CP}\lesssim10^{-3}$.
This is because the correct DM relic density is satisfied with relatively smaller $y_\chi$, thus larger $v_{\phi}$, which leads to a larger value of $Y_{B-L}$ in the non-resonance region for leptogenesis. So the baryon asymmetry can be met by simply reducing $\varepsilon_{\CP}$. In the special scenario with $m_N/m_\chi=1$, the overlapping resonance regions of leptogenesis and DM results in $\varepsilon_{\CP}\sim\mathcal{O}(0.1)$ to meet the observations.

Besides the resonance region, DM and leptogenesis can also be simultaneously satisfied for light scalar as $m_\rho\lesssim200$~GeV. Within this light scalar region, the required $\varepsilon_{\CP}$ increases for larger $m_N$. In the scenario with $m_N/m_\chi=0.5$, $\varepsilon_{\CP}\lesssim1.7\times10^{-1}$ should be satisfied when $m_\chi<10^4$ GeV. Increasing the mass ratio $m_N/m_\chi$ leads to larger required values of $\varepsilon_{\CP}$ for certain $m_\chi$. As shown in panel (b) of Fig.~\ref{FIG:fig4},  $m_\chi\gtrsim7400$~GeV is excluded by $\varepsilon_{\CP}>1$ due to too large $m_N$ in the light scalar region. The upper bound on $m_\chi$ from $\varepsilon_{\CP}>1$ decreases as $m_N/m_\chi$ increases, e.g., $m_\chi\gtrsim3400$ GeV for $m_N/m_\chi=1.5$ and $m_\chi\gtrsim2000$ GeV for $m_N/m_\chi=2$ are excluded. In scenario $m_N/m_\chi=2$, the perturbation constraint on $\lambda_N$ can also exclude heavy mass regions, but it is clearly less stringent than the limit from $\varepsilon_{\CP}>1$.

\subsection{Collider Signature and Direct Detection}\label{SEC:DD}

\begin{figure}
	\begin{center}
		\includegraphics[width=0.45\linewidth]{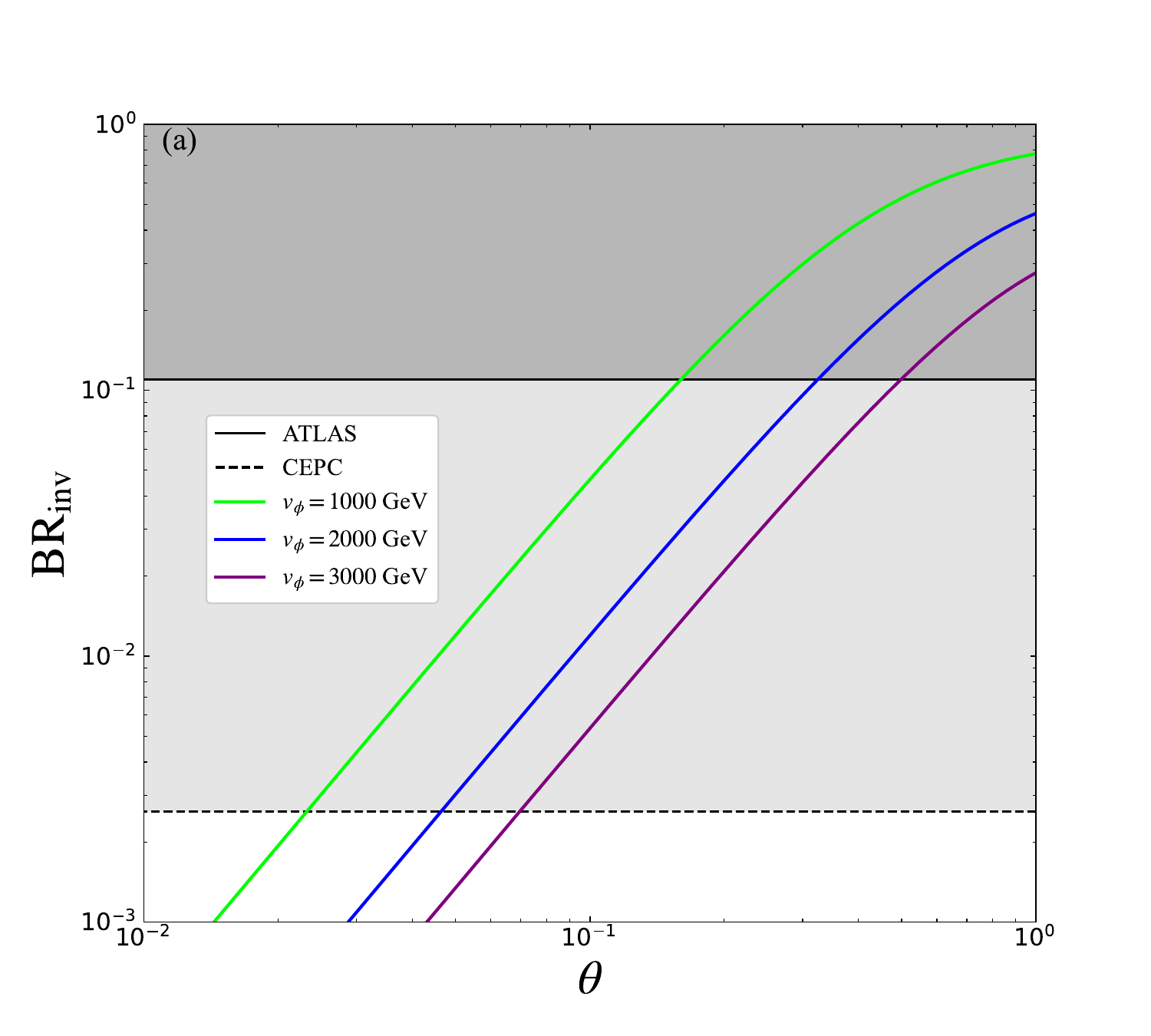}
		\includegraphics[width=0.45\linewidth]{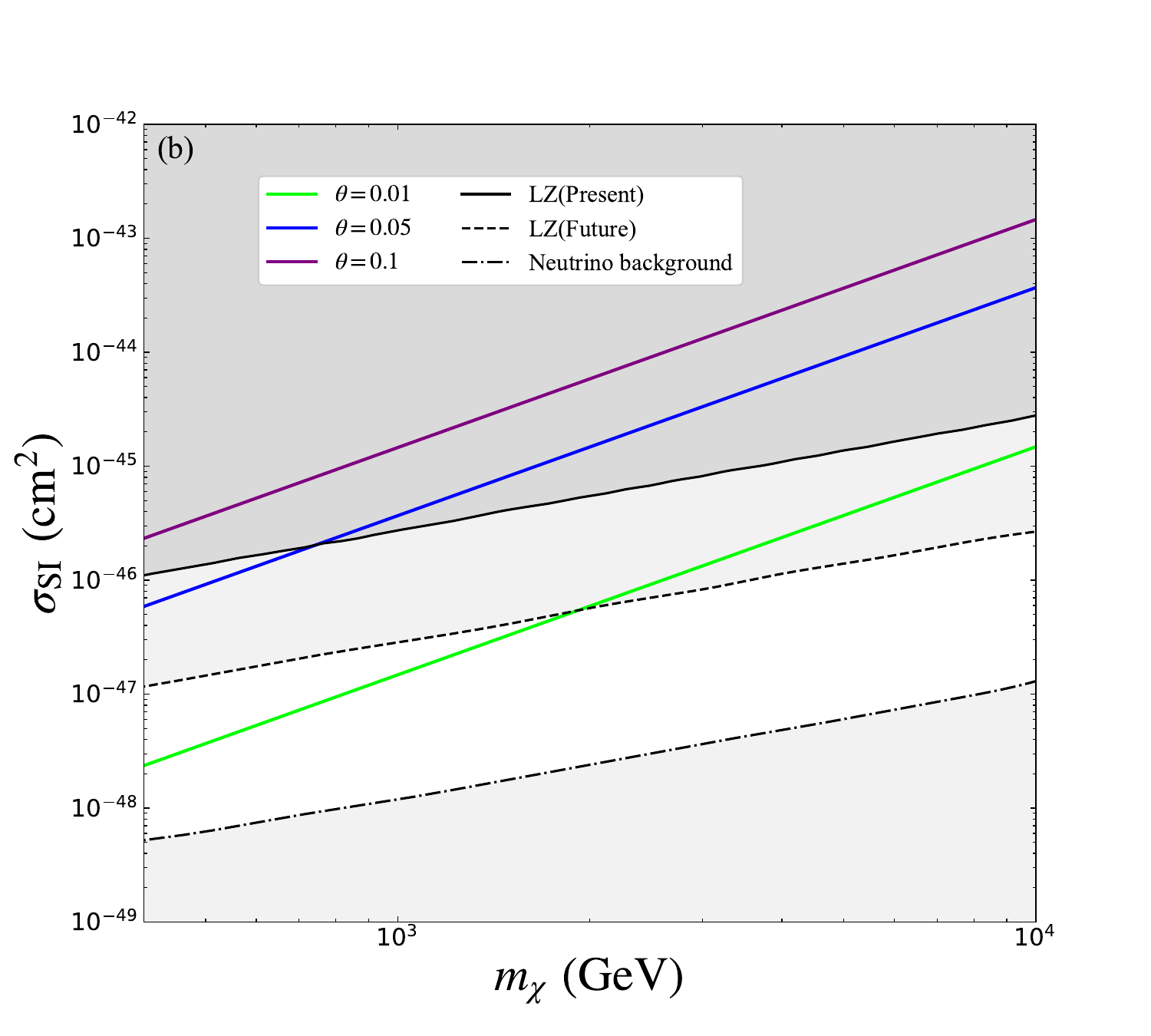}
	\end{center}
	\caption{Predictions for the branching ratio of Higgs invisible decay BR$_\text{inv}$ (a) and spin-independent cross section $\sigma_{\rm SI}$ (b). The dark gray region in panel (a) is excluded by ATLAS search of Higgs invisible decay~\cite{ATLAS:2020kdi}, and the black dashed line represents the predicted result of future CEPC~\cite{Tan:2020ufz}. The lime, blue, and purple curves stand for $v_{\phi}=1000, 2000$, and 3000~GeV, respectively. In subfigure (b), $v_{\phi}$ and $m_\rho$ are fixed as 2000 GeV and 1000~GeV.	The black solid and dashed lines represent the present~\cite{LZ:2022lsv} and future~\cite{LZ:2015kxe} LZ limits. The bottom dot-dashed line is the lower limitation on the discovery of WIMP in direct detection experiments due to the neutrino backgrounds~\cite{Billard:2013qya}. The lime, blue, and purple curves express $\theta=0.1, 0.05$, and 0.01, respectively.
	}
	\label{FIG:fig5}
\end{figure}

In this specific model, the SM Higgs $h$ can decay into $\rho\rho$, $\eta\eta$, $\chi\chi$, $\nu N$, and $NN$ if kinetically allowed. See Ref.~\cite{Escudero:2016tzx} for a detailed discussion. On the other hand, the singlet scalar $\rho$ is also promising at colliders \cite{Dev:2017xry,Robens:2016xkb}, which depends on the mixing angle $\theta$. In this paper, we have fixed $\theta=0.05$, which is allowed by current experimental limits \cite{Robens:2022cun}. 

In the global scenario, one distinct particle is the Majoron $\eta$, which is invisible at colliders. With the parameter space in Eq.~\eqref{Eq:PS}, we focus on the invisible decay $h\to \eta\eta$, since this channel is always allowed.  The corresponding branching ratio has been constrained by the ATLAS experiment as~\cite{ATLAS:2020kdi},
\begin{equation}\label{Eqn:hid}
	\rm BR_{\rm inv}=\frac{\Gamma_{h\to\eta\eta}}{\Gamma_{h\to\eta\eta}+\Gamma_{\rm SM}}<0.11,
\end{equation}
where $\Gamma_{\rm SM}=4$ MeV is the standard Higgs width. The
theoretical Higgs invisible decay width into Majoron is given by~\cite{Garcia-Cely:2013wda,  Garcia-Cely:2013nin, Escudero:2016tzx},
\begin{equation}\label{Eqn:gh}
	\Gamma_{h\to\eta\eta}=\frac{\sin^2\theta~m_h^3}{32\pi v_{\phi}^2}.
\end{equation}

The elastic scattering of DM $\chi$ on nuclei mediated by $\rho$ and SM Higgs $h$ through  mixing are tightly constrained by the DM direct detection experiments~\cite{LZ:2022lsv,LZ:2015kxe}. The spin-independent scattering cross section is calculated as~\cite{Garcia-Cely:2013wda,  Garcia-Cely:2013nin, Escudero:2016tzx}
\begin{equation}\label{Eqn:dd}
	\sigma_{\rm SI}=\frac{C^2(\sin2\theta)^2 m_n^4m_\chi^4}{4\pi v_H^2 v_\phi^2(m_n+m_\chi)^2}\left(\frac{1}{m_h^2}-\frac{1}{m_\rho^2}\right)^2,
\end{equation}
where $C\simeq0.3$ is a nucleon matrix element dependent constant \cite{Belanger:2013oya}, and $m_n$ represents the nucleon mass. 

It can be seen from Eq.~\eqref{Eqn:gh} and Eq.~\eqref{Eqn:dd} that the Higgs invisible  decay width and elastic scattering cross section sensitively rely on the mixing angle $\theta$. 
Theoretical predictions for the branching ratio of Higgs invisible decay and scattering cross section for benchmark cases are shown in Fig.~\ref{FIG:fig5}. In panel (a), the constraint ability of ATLAS on $\theta$ weakens as $v_{\phi}$ increases. Typically, $\theta\lesssim\mathcal{O}(0.1)$ is required to satisfy the experimental limit. In the future, the CEPC could probe  BR$_\text{inv}\gtrsim2.6\times10^{-3}$ \cite{Tan:2020ufz}, which corresponds to $\theta\gtrsim0.02$ for TeV scale $v_\phi$. Meanwhile, the direct detection limit on $\sigma_{\rm SI}$ has already set more stringent constraints than Higgs invisible decay. As shown in panel (b) of Fig.~\ref{FIG:fig5}, $\theta=0.1$ is excluded by the current LZ limit, although it is allowed by BR$_\text{inv}$. In the following studies, we will fix $\theta=0.05$ for illustration. A much smaller $\theta\lesssim0.01$ is easy to satisfy Higgs invisible and direct detection limits but is also hard to be detected at future experiments. 

\begin{figure}
	\begin{center}
		\includegraphics[width=0.45\linewidth]{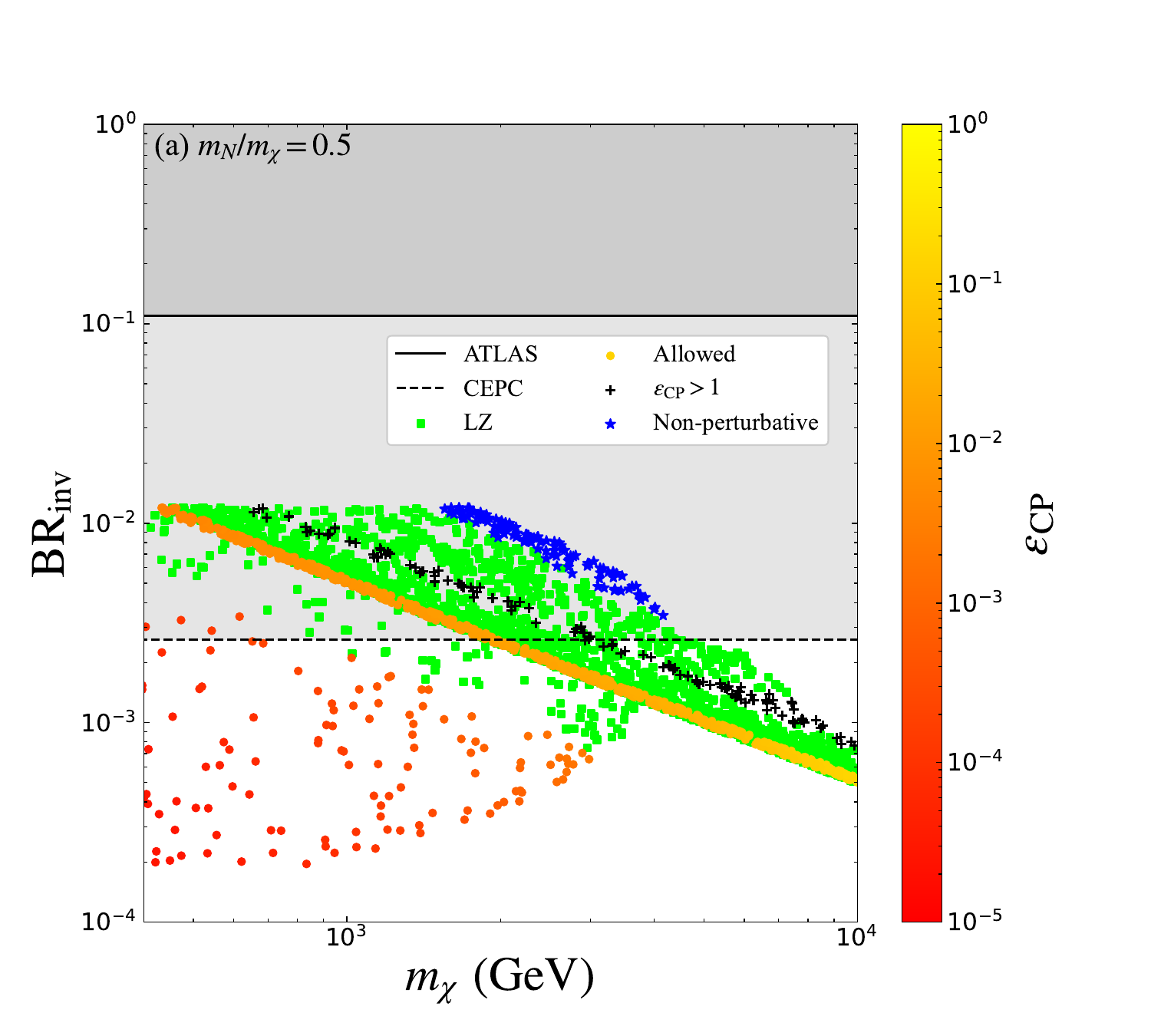}
		\includegraphics[width=0.45\linewidth]{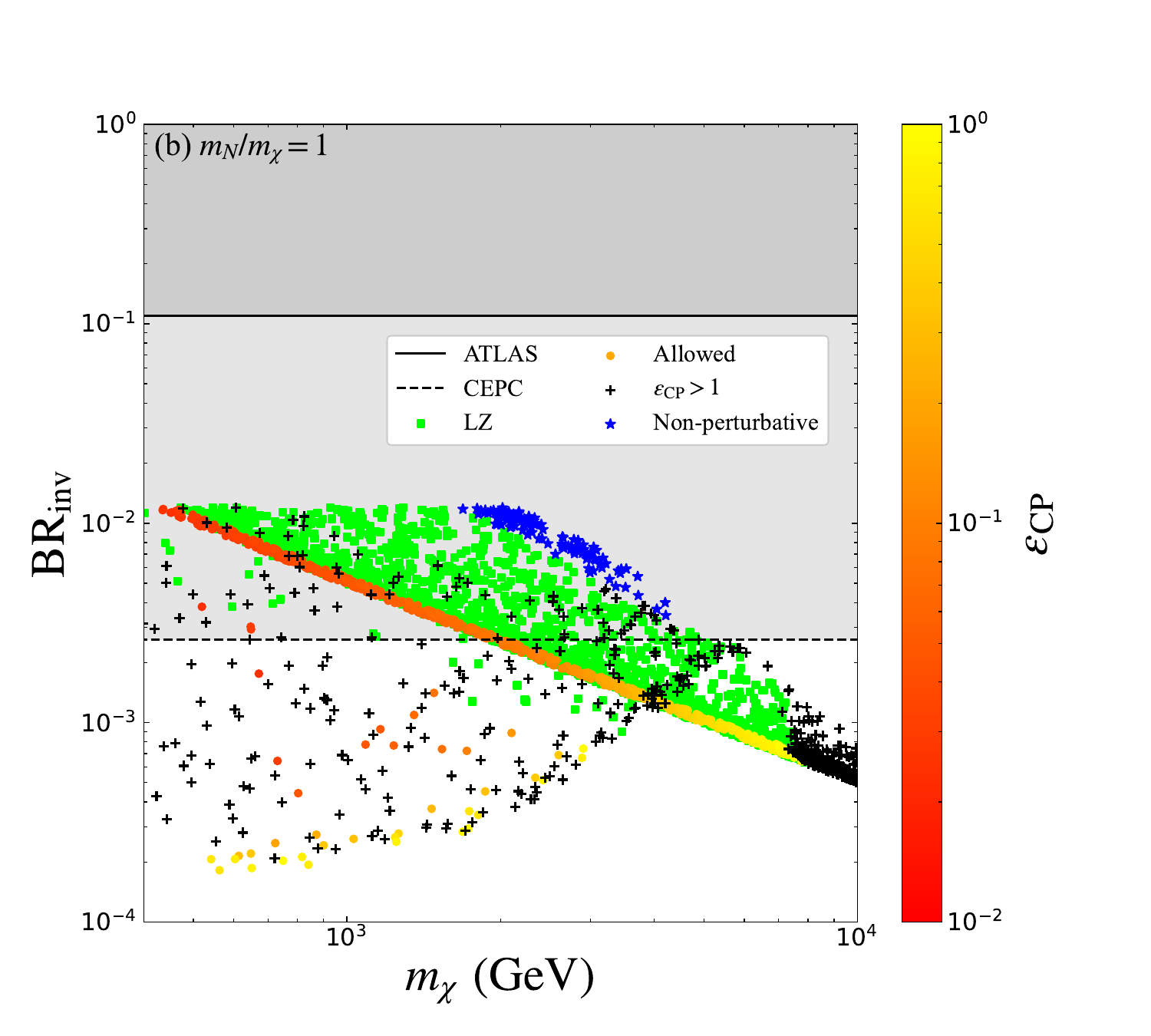}
		\includegraphics[width=0.45\linewidth]{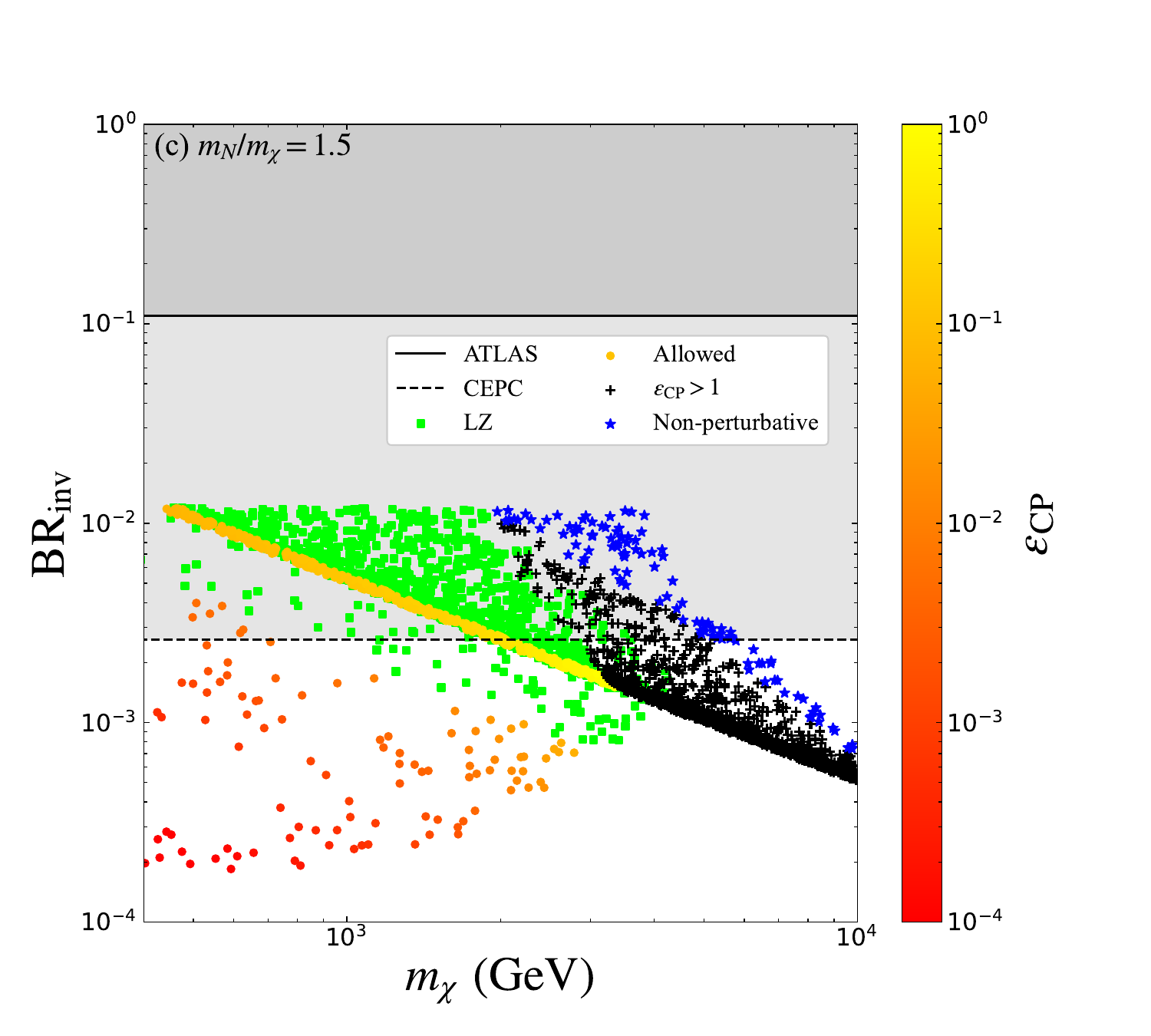}
		\includegraphics[width=0.45\linewidth]{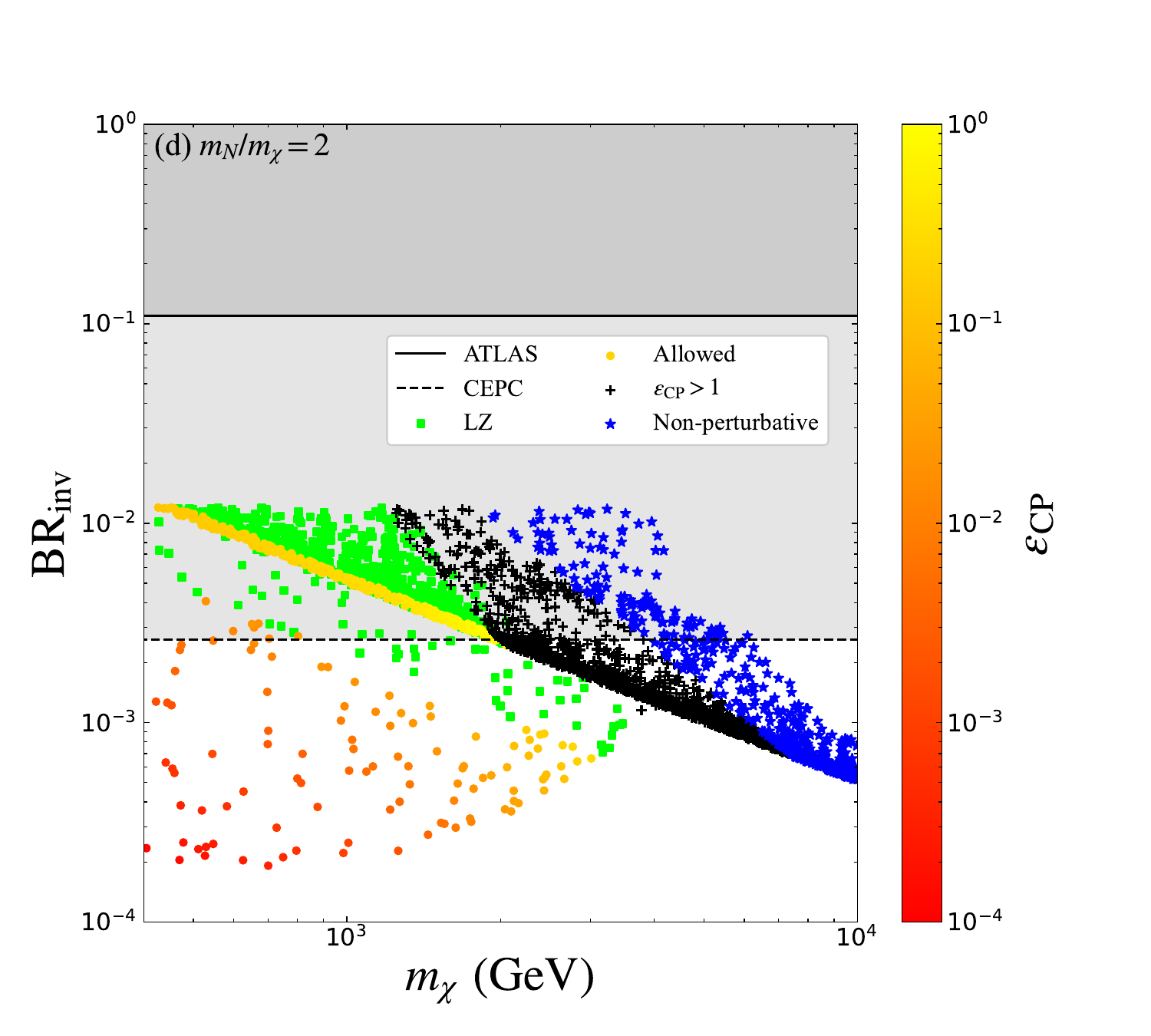}
	\end{center}
	\caption{Branching ratio of Higgs invisible decay in the global scenario. We have fixed $\theta=0.05$ in the calculation. The legends and markers of samples in the four subfigures are consistent with those in Fig.~\ref{FIG:fig4}. The solid and dashed lines correspond to the experimental limits on Higgs invisible decay of current ATLAS ~\cite{ATLAS:2020kdi} and future CEPC~\cite{Tan:2020ufz}.}
	\label{FIG:fig6}
\end{figure}

Fig.~\ref{FIG:fig6} displays the theoretical branching ratio of Higgs invisible decay for the four scenarios with $m_N/m_\chi=0.5,1,1.5,2$ within the parameter space in Eq.~\eqref{Eq:PS}. By fixing $\theta=0.05$, the predicted branching ratio of Higgs invisible decay is less than $10^{-2}$, which is one order of magnitude smaller than current ATLAS limitation.  The future CEPC experiment could probe a large part of the allowed samples. Under the direct detection LZ limit, allowed samples are also separated into two regions. For light scalar $\rho$ in the non-resonance region (mainly yellow samples as shown in Fig.~\ref{FIG:fig4}), i.e., $m_\rho\lesssim200$ GeV, there is a strong correlation between BR$_{\rm inv}$ and DM mass $m_\chi$. Therefore, we may determine the DM mass by precise measurement of BR$_{\rm inv}$ once light $\rho$ is discovered. In the resonance region (mainly red samples as shown in Fig.~\ref{FIG:fig4}), the predicted  BR$_{\rm inv}$ is smaller than those in the non-resonance region for a certain value of $m_\chi$. And only $m_\chi\lesssim1000$ GeV could lead to observable BR$_{\rm inv}$ at CEPC.  Although the theoretical BR$_{\rm inv}$ is in the range of $[2\times10^{-4},10^{-2}]$ for all the four scenarios, allowed samples under current constraints are different, especially for heavy DM in the non-resonance region. For instance, scenario $m_N/m_\chi=2$ predicts BR$_{\rm inv}\in[3\times10^{-3},10^{-2}]$ in the non-resonance region, since $m_\chi\gtrsim2000$ GeV is already excluded by $\varepsilon_{\CP}>1$ and non-perturbative. 

 \begin{figure}
	\begin{center}
		\includegraphics[width=0.45\linewidth]{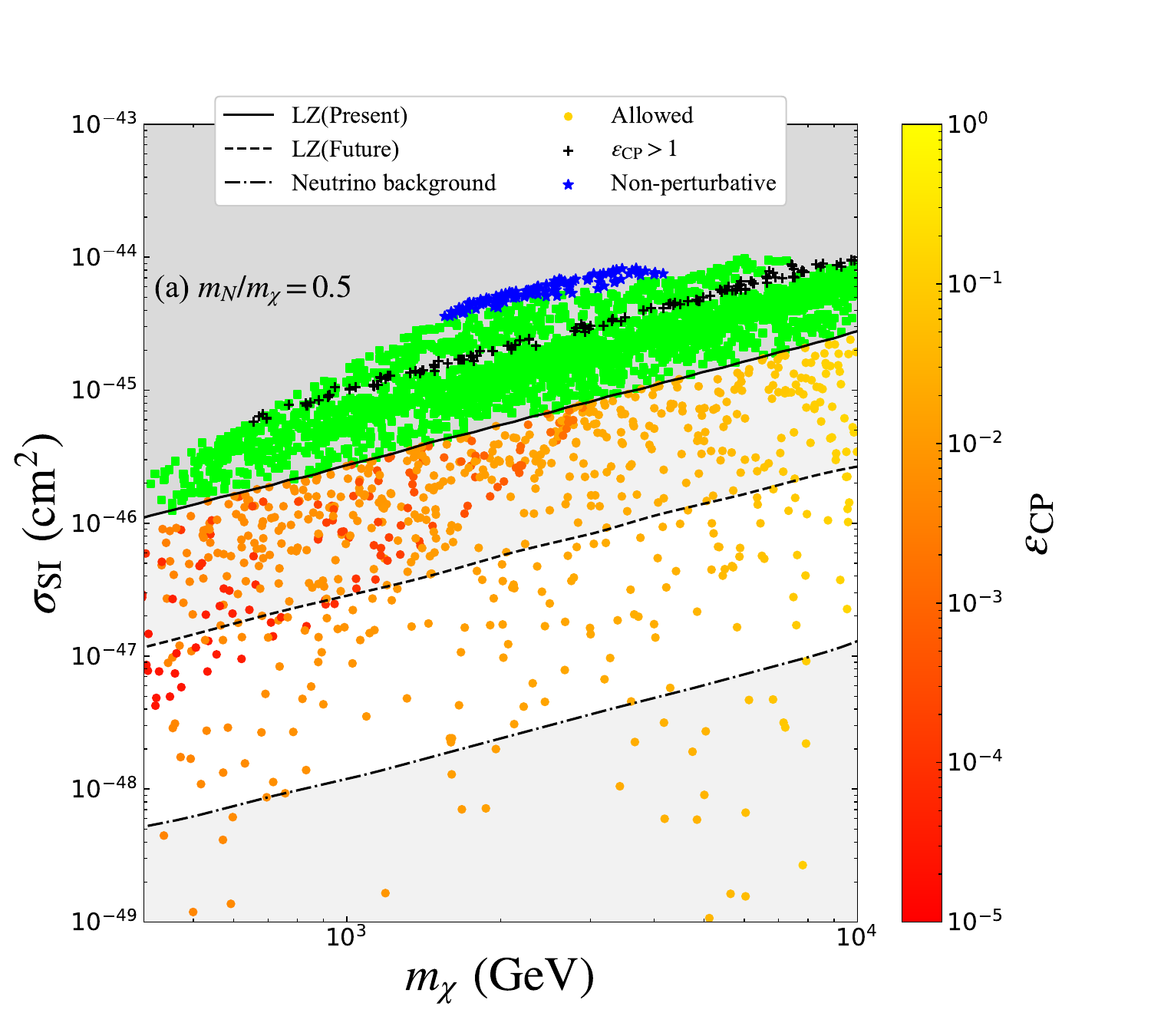}
		\includegraphics[width=0.45\linewidth]{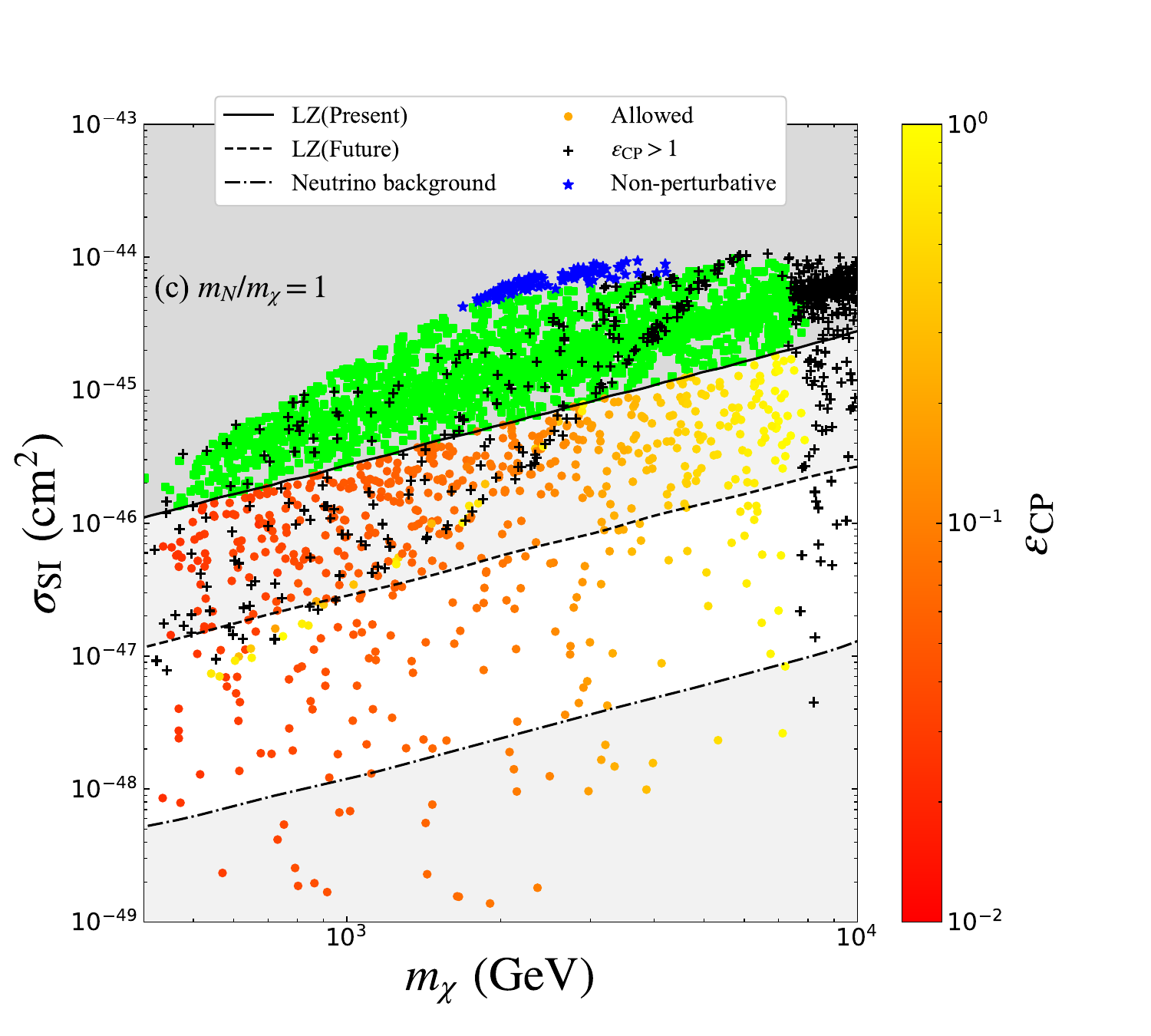}
		\includegraphics[width=0.45\linewidth]{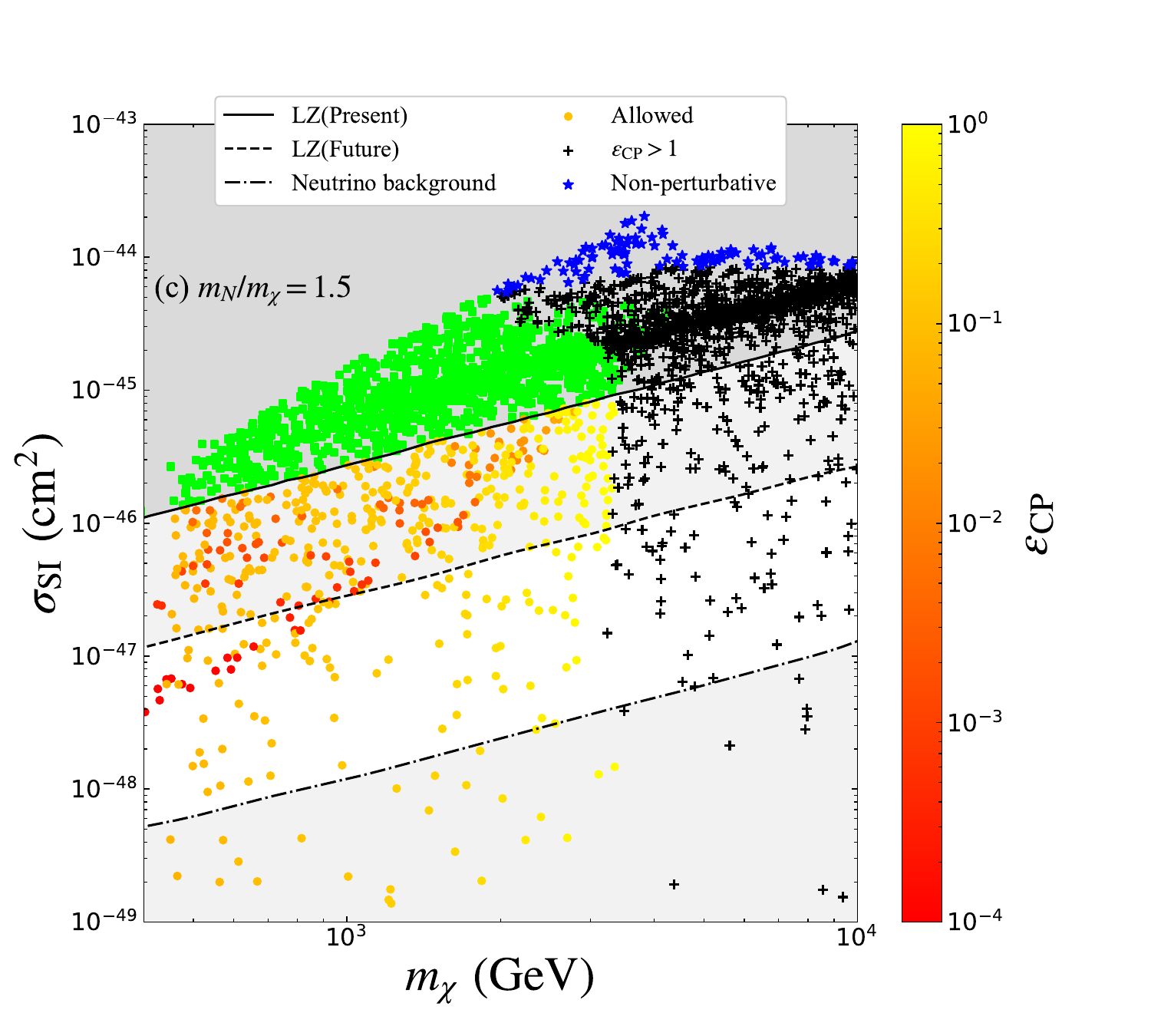}
		\includegraphics[width=0.45\linewidth]{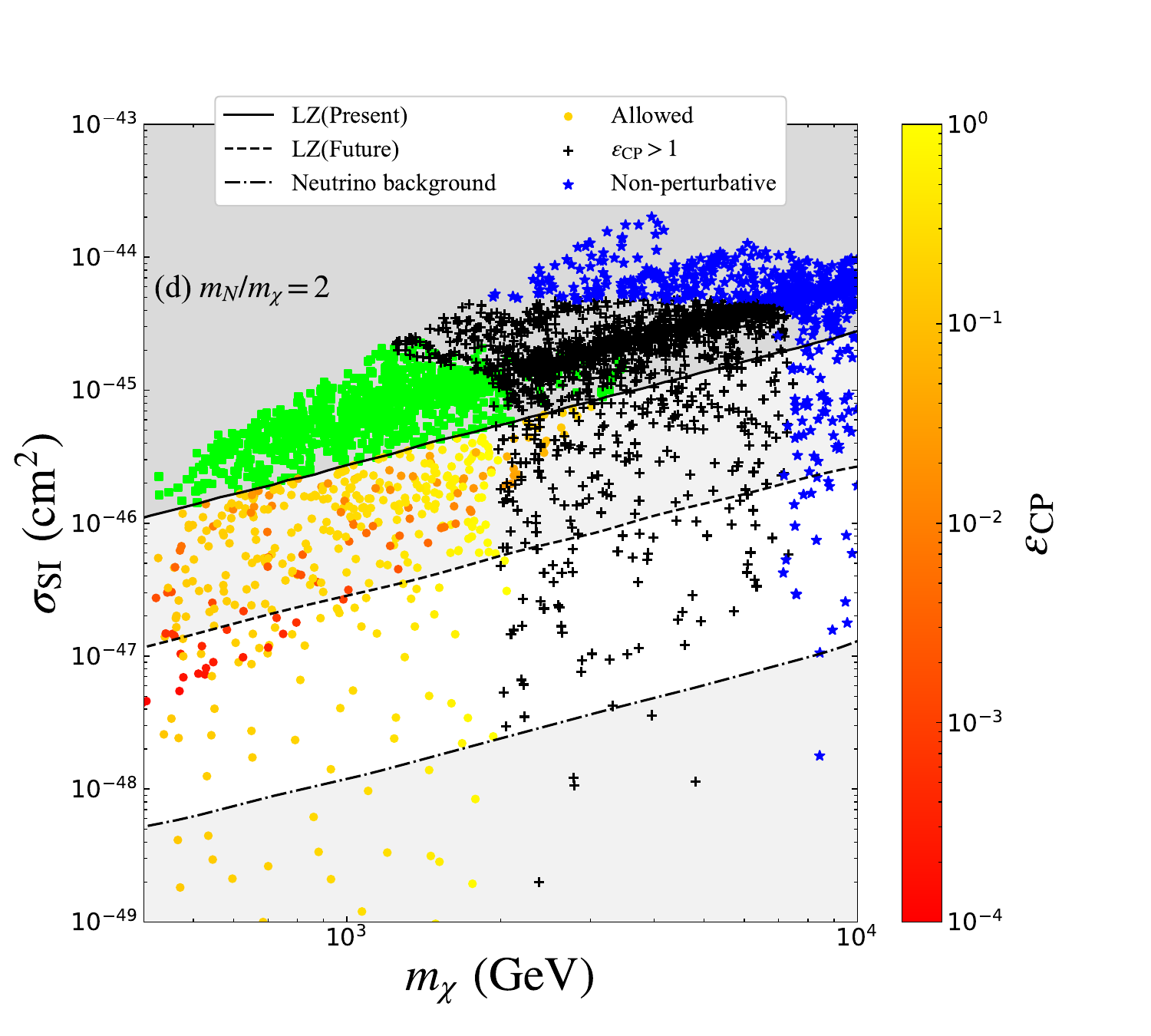}
	\end{center}
	\caption{Spin-independent scattering cross section in the global scenario with $\theta=0.05$. Legends and markers appearing in these subfigures have the same meaning as that in Fig.~\ref{FIG:fig4}. The black solid, dashed and dot-dashed lines represent the present LZ limit~\cite{LZ:2022lsv}, future LZ limit~\cite{LZ:2015kxe}, and neutrino background ~\cite{Billard:2013qya}.
	}
	\label{FIG:fig7}
\end{figure} 

The scanning results of the scattering cross sections are displayed in Fig.~\ref{FIG:fig7}, which are tightly constrained by the LZ limit with $\theta=0.05$. For example, the present LZ limit at most excludes the lime samples with a cross section larger than $10^{-46} \rm {cm}^2$ when $m_\chi \sim 400$ GeV. The future LZ experiment can further push this limit down by one order of magnitude. Therefore, most samples with $\theta=0.05$ are within future experimental reach. With relatively heavy $m_\rho$, we find that the samples in the resonance region (mainly red samples) usually predict detectable direct detection cross section. On the other hand, the samples in the non-resonance region (mainly yellow samples) could lead to suppressed scattering cross section due to cancellation for $m_\rho\sim m_h$ as in Eq.~\eqref{Eqn:dd}.

As shown in Fig.~\ref{FIG:fig5}, $\theta$ plays a crucial role in the elastic scattering cross section. For excessively large $\theta$, such as $\theta=0.1$, we have checked that almost all samples will be excluded by the present LZ limit.  However, the constraints from the direct detection experiment are extremely relaxed when $\theta$ takes a smaller value like 0.01. When $m_N/m_\chi$ increase, as shown in panels (b), (c), and (d), an increasing number of samples that escape the constraints of direct detection experiments  will be excluded by unitarity constraints and exceeding $\varepsilon_{\CP}$, although the exclusion range of LZ has not changed much.

\subsection{Indirect Detection}

\begin{figure} 
 	\begin{center}
 		\includegraphics[width=0.45\linewidth]{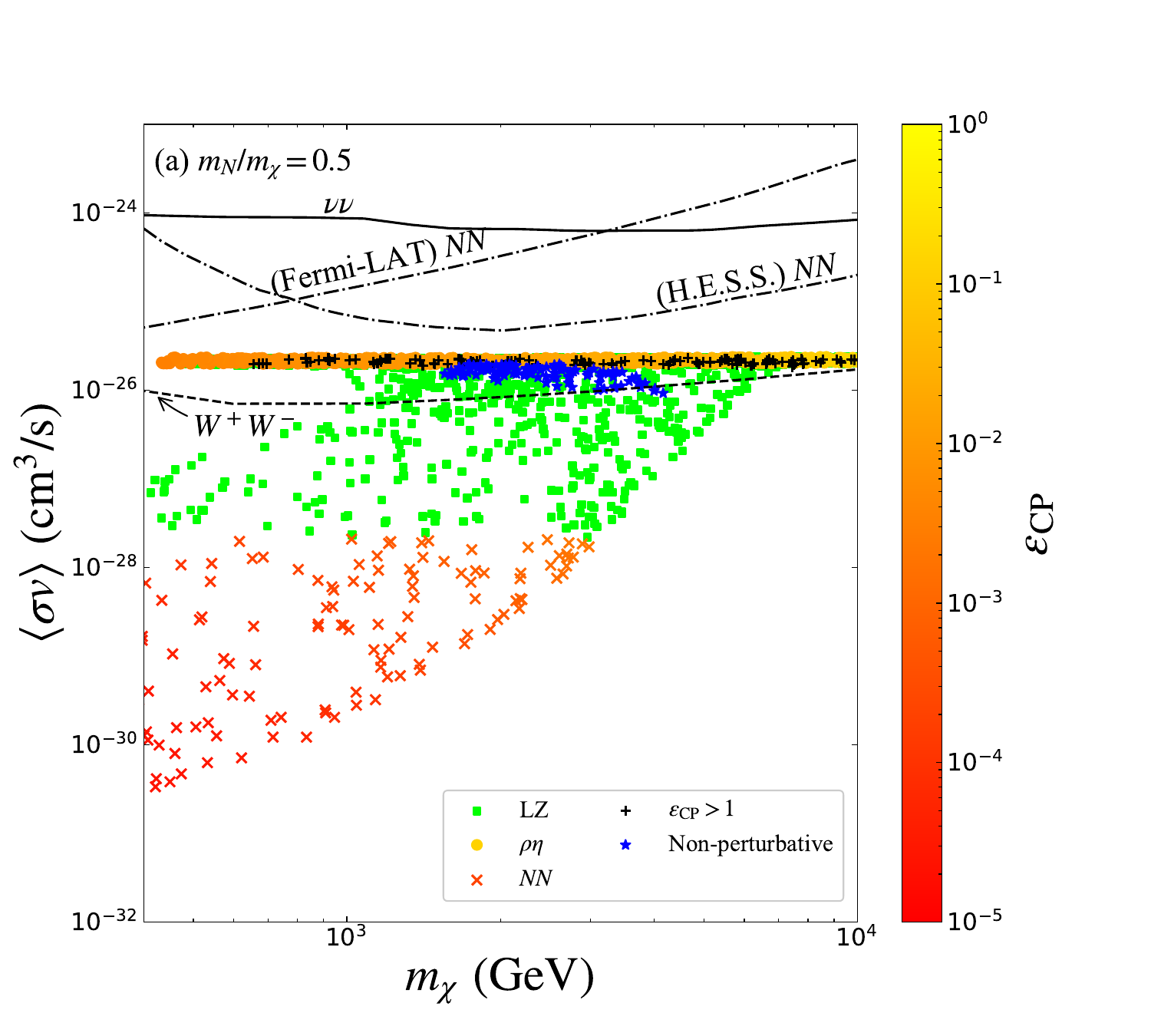}
 		\includegraphics[width=0.45\linewidth]{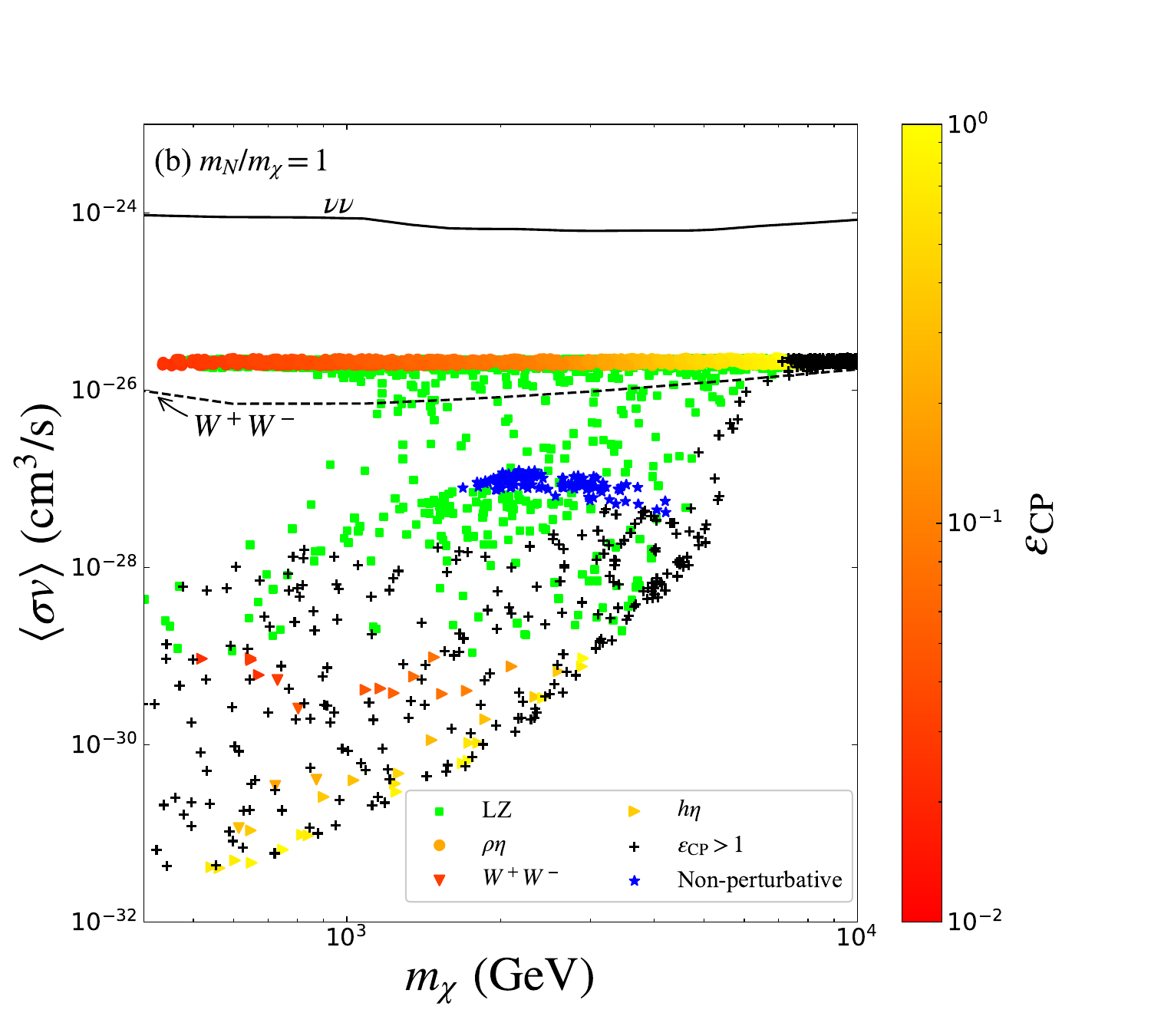}
 		\includegraphics[width=0.45\linewidth]{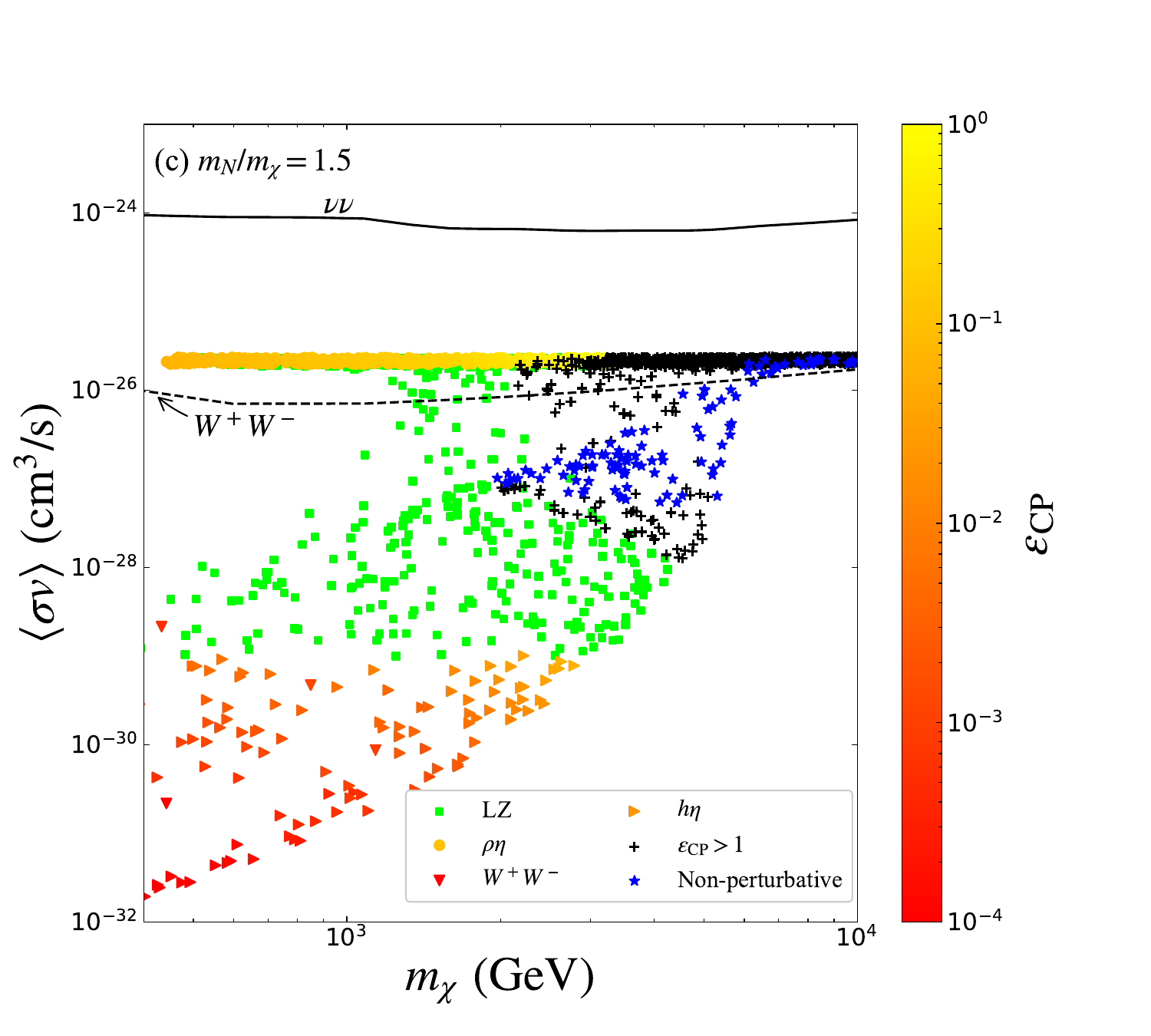}
 		\includegraphics[width=0.45\linewidth]{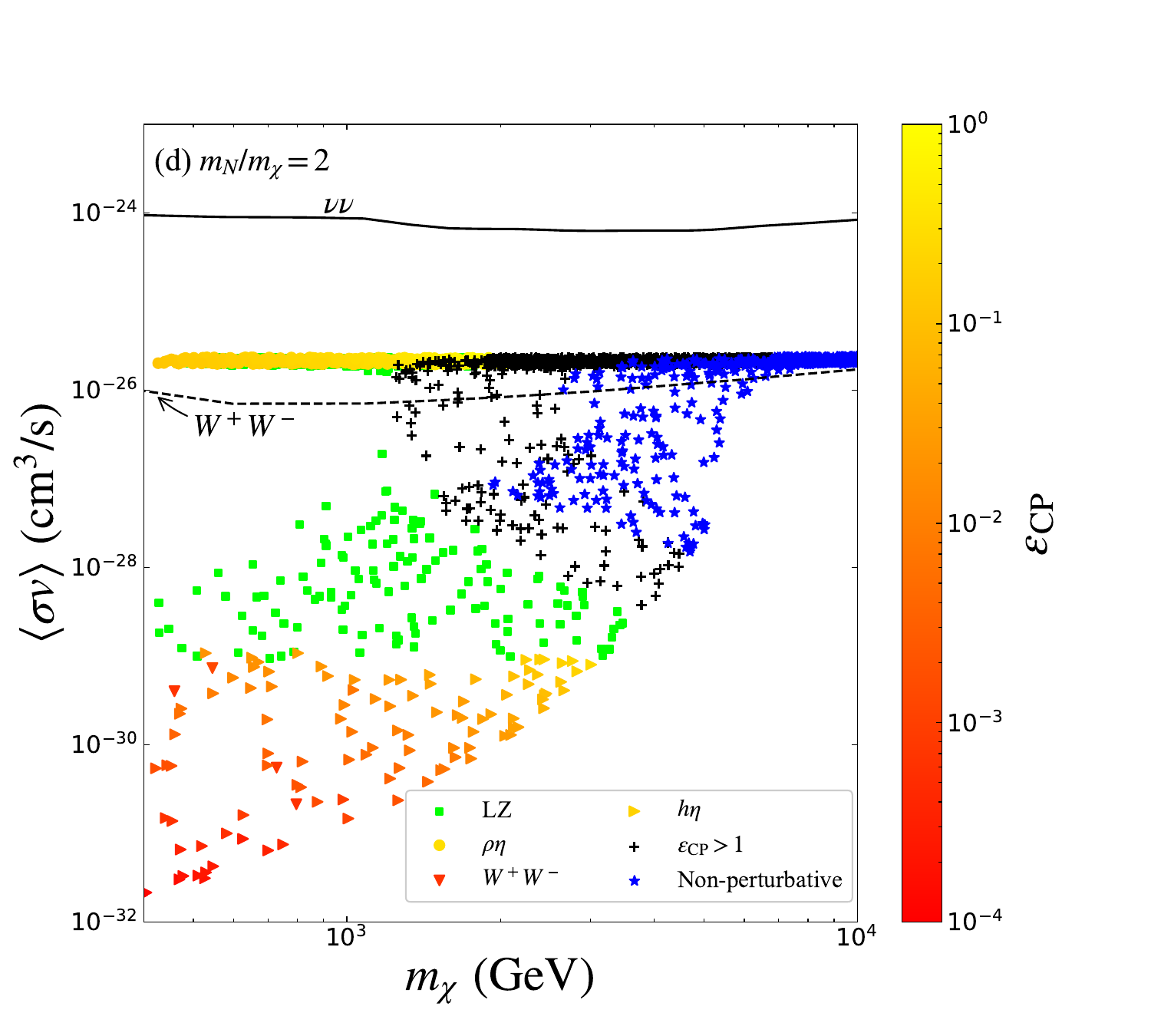}
 	\end{center}
 	\caption{Constraints from the indirect detection experiments in the global scenario. The solid lines stand for the upper limit on $\left<\sigma v\right>$ for pair annihilation of DM into neutrinos by ANTARES~\cite{Gozzini:2023nka}. The dot-dashed lines in panel (a) express the bound on $NN$ final state from Fermi-LAT and H.E.S.S. ~\cite{Campos:2017odj}. The dashed lines represent the projected upper limits of $W^+W^-$ final states from the Cherenkov Telescope Array (CTA)~\cite{CTA:2020qlo, Mangipudi:2021ivm}. The definitions of the blue squares, black crosses, and lime stars are the same as those in Fig.~\ref{FIG:fig4}. The remaining allowed samples are colored by $\varepsilon_{\CP}$ while shaped by different final products. 
 	}
 	\label{FIG:fig8}
 \end{figure}

The annihilation of DM into SM particles provides an alternative pathway to probe DM.  In the global scenario, DM $\chi$ can directly annihilate into the final states of $NN$, $\rho\rho$, $\rho\eta$, $\eta\eta$, and $VV$, which is shown in Fig.~\ref{FIG:FD}. At present, the process $\chi\chi\to \rho\rho,\eta\eta$ , and $VV$ are $p$-wave suppressed by DM relative velocity, thus can not be detected by indirect detection. So we focus on the $s$-wave processes $\chi\chi\to NN,\rho\eta$. Once kinetically allowed, the $\chi\chi\to \rho\eta$ channel is the dominant one \cite{Escudero:2016tzx}, therefore allowed samples in the non-resonance region with light $m_\rho$ are dominant by this channel. On the other hand, the $\chi\chi\to NN$ channel becomes dominant only in the resonance region when $m_\rho \simeq 2m_\chi$. Meanwhile, the $\chi\chi\to NN$ channel is kinetically forbidden in scenarios (b), (c), and (d) with $m_N\geq m_\chi$. In Fig.~\ref{FIG:fig8}, we show the theoretical $\left<\sigma v\right>$ and various experimental limits, where $\left<\sigma v\right>$ is the thermally averaged cross section for DM annihilation times velocity at present.

In subfigure (a) of Fig.~\ref{FIG:fig8}, the traditional WIMP $\left<\sigma v\right>$ namely $2\times 10^{-26}~\rm cm^3/s$  is mostly composed of channel $\chi\chi\to\rho\eta$. Assuming $\theta=0.05$, we find that $\rho\to \eta\eta$ is the dominant decay mode of $\rho$. The pseudo scalar $\eta$ could have keV scale mass due to quantum gravity effects~\cite{Lattanzi:2014mia}, which leads to the cascade decay $\eta\to \nu\nu$ in the global $U(1)_{B-L}$ scenario. In this way, the process $\chi\chi\to \rho\eta$ will generate neutrinos via cascade decays of $\rho$ and $\eta$. Currently, the experimental limit on the annihilation cross section of DM into neutrinos is at the order of $10^{-24}~{\rm cm}^3/s$, which has no limitation on theoretical parameter space. The sub-dominant decay mode $\rho\to W^+W^-$ has a branching ratio of 0.1 for $\theta=0.05$.  Therefore, $\left<\sigma v\right>$ of $\chi\chi\to \rho \eta\to W^+W^-\eta$ is roughly located near $10^{-27} \rm cm^3/s$, which is one order of magnitude smaller than the projected upper limit for $W^+W^-$ final state.

Within the resonance region, both $p$-wave $\chi\chi\to\eta\eta, VV$ and $s$-wave $\chi\chi\to NN$ channels contribute to the relic density of DM. At present, the cross sections of $p$-wave $\chi\chi\to\eta\eta,VV$ channels are velocity suppressed, resulting in the $\chi\chi\to NN$ the dominant channel. So the cross section of $\chi\chi\to NN$ is typically smaller than the WIMP value $\left<\sigma v\right>\sim2\times 10^{-26}~\rm cm^3/s$. With $\theta=0.05$, we also find that the samples with $\left<\sigma v\right> \gtrsim2\times10^{-28}\rm cm^3/s$ are excluded dominantly by LZ. Therefore, the $\chi\chi\to NN$ channel is also beyond the reach of indirect detection.

Different from scenario (a), smaller cross section $\left<\sigma v\right>$ in the resonance region for scenarios (b), (c) and (d) are dominated by the $s$-wave process $\chi\chi\to h\eta$, which is suppressed by the mixing angle $\theta=0.05$. So the the surviving samples with $\left<\sigma v\right>\lesssim\mathcal{O}(10^{-29})~\rm cm^3/s$ are difficult to detect by current or future experiments. We also find that there are few samples are dominant by the $\chi\chi\to W^+W^-$ process when $v_\phi$ is very large,  namely $\lambda_{H\phi}$ is much greater than $\lambda_\phi$. For the non-resonance region, the dominant annihilation mode is $\chi\chi\to\rho\eta$, which leads to similar results as scenario (a) for indirect detection. It should be noted that our indirect detection results are different from the previous study in Ref.~\cite{Escudero:2016tzx}, mainly because of the non-vanishing mixing $\theta$. In the simplified case with $\theta=0$, the processes into SM final states as $\chi\chi\to h\eta, W^+W^-$ mediated by $\rho$ are absent as in Ref.~\cite{Escudero:2016tzx}, then the $p$-wave $\chi\chi\to \eta\eta$ becomes the dominant contribution in the resonance region.

 \section{The Local $U(1)_{B-L}$ Scenario}\label{SEC:LUS}

\subsection{Relic Density and Leptogenesis}\label{SEC:LRD}

For the local scenario, the various dominant annihilation processes of $N$ and $\chi$ correspond to Feynman diagrams in Fig.\ref{FIG:FD} by replacing $\eta$ with $Z^\prime$ in panel (a)-(c). Whereupon the Boltzmann equations Eq.\ref{Eqn:GBE} can be rewritten as
\begin{eqnarray}
	\frac{dY_N}{dz} &= & -\frac{z}{sH(m_N)}\left(\frac{Y_N}{Y_N^\eq}-1\right)(\gamma_{N\to HL}+2\gamma_{NL\to qt}+4\gamma_{Nt\to qL})\nonumber \\
	&-&\frac{z}{sH(m_N)} \left(\left(\frac{Y_N}{Y_N^\eq}\right)^2-1\right)(2\gamma_{NN\to\rho\rho,Z^\prime Z^\prime}+2\gamma_{NN\to\rho Z^\prime,hZ'}+2\gamma_{NN\to \SM\SM})\nonumber \\
	&+&\frac{z}{sH(m_N)}\left(\left(\frac{Y_\chi}{Y_\chi^\eq}\right)^2-\left(\frac{Y_N}{Y_N^\eq}\right)^2\right)2\gamma_{\chi\chi\to NN},\\
	\frac{dY_\chi}{dz} &= & -\frac{z}{sH(m_N)} \left(\left(\frac{Y_\chi}{Y_\chi^\eq}\right)^2-1\right)(2\gamma_{\chi\chi\to\rho\rho,Z^\prime Z^\prime}+2\gamma_{\chi\chi\to\rho Z^\prime,hZ'}+2\gamma_{\chi\chi\to \SM\SM})\nonumber \\
	&-&\frac{z}{sH(m_N)}\left(\left(\frac{Y_\chi}{Y_\chi^\eq}\right)^2-\left(\frac{Y_N}{Y_N^\eq}\right)^2\right)2\gamma_{\chi\chi\to NN},\\
	\frac{dY_{B-L}}{dz} &= & \frac{z}{sH(m_N)} \left(\varepsilon_{\CP}\left(\frac{Y_N}{Y_N^\eq}-1\right)-\frac{Y_{B-L}}{2Y_L^{\eq}}\right)\gamma_{N\to HL}\nonumber \\
	&-&\frac{z}{sH(m_N)}\frac{Y_{B-L}}{Y_L^{\eq}}\left(\frac{Y_N}{Y_N^{\eq}}\gamma_{NL\to qt}+2\gamma_{Nt\to qL}\right),
\end{eqnarray}
where the notations are the same as those in Eq.\ref{Eqn:GBE}.  It should be emphasized that for TeV scale $N$ and $\chi$ in Fig.\ref{FIG:FD} (d),  the  SM fermion final states are dominantly mediated by $Z^\prime$, while the SM boson final states are dominantly mediated by $\rho$. For convenience to distinguish different contributions, we therefore use $NN(\chi\chi)\to VV (V=W,Z,h)$ and $NN(\chi\chi)\to f\bar{f}$ to represent the $\rho$ and $Z^\prime$ mediated channels respectively. The new reduced cross sections arising here, i.e., $NN\to\rho Z^\prime,hZ', Z^\prime Z^\prime, \SM\SM$ and $\chi\chi\to\rho Z^\prime,hZ', Z^\prime Z^\prime, \SM\SM, NN$  have specific expressions shown in Appendix~\ref{SEC:RCS}.  

Different from the global scenario, there are various tight constraints on the new gauge boson $Z'$ in the local scenario. Searches of $Z'$ in the dilepton final states have excluded $m_{Z'}\lesssim5.15$~TeV with 140 fb$^{-1}$ data at LHC \cite{ATLAS:2019erb,CMS:2021ctt}. Meanwhile, electroweak precision test limits from LEP require $m_{Z'}/g_{B-L}=2v_\phi\gtrsim7$ TeV \cite{Cacciapaglia:2006pk}. In the following study, we fix $m_{Z'}=7$ TeV for illustration.

\begin{figure} 
	\begin{center}
		\includegraphics[width=0.45\linewidth]{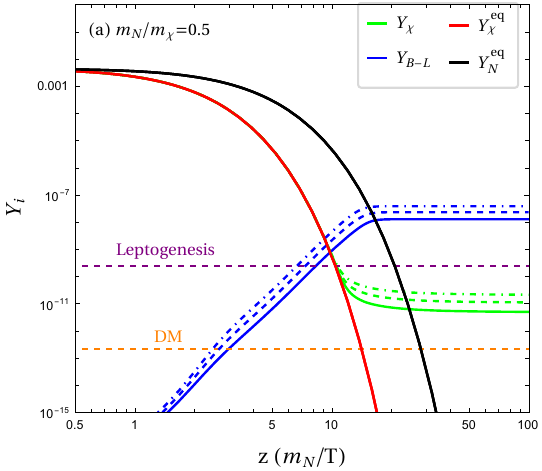}
		\includegraphics[width=0.45\linewidth]{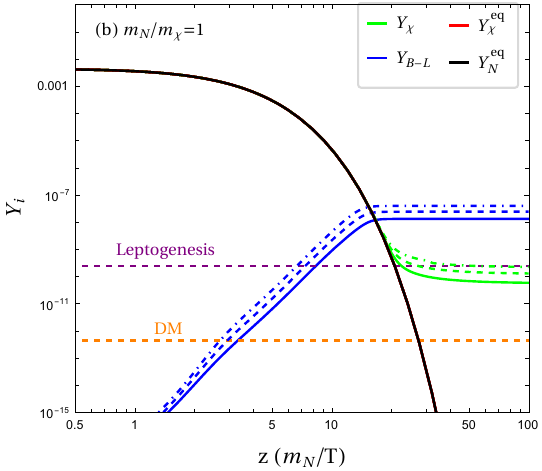}
		\includegraphics[width=0.45\linewidth]{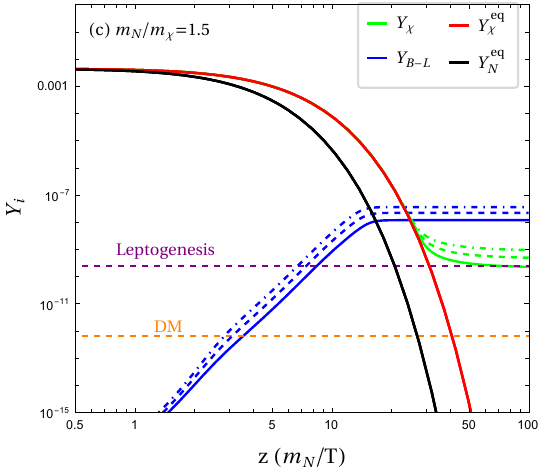}
		\includegraphics[width=0.45\linewidth]{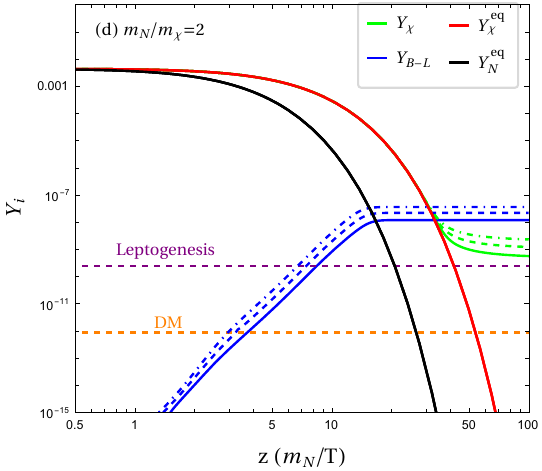}
	\end{center}
	\caption{The evolutions of abundances $Y_i$ with $m_N=1000 ~\GeV$, $m_\rho=500 ~\GeV$, $m_{Z^\prime}=7000~\GeV$, and $\varepsilon_{\CP}=0.1$ in the local scenario. Subfigures (a)-(d) correspond to the four cases with mass ratios $m_N/m_\chi$= 0.5,1,1.5 and 2. The markers of the curves are consistent with that in Fig.\ref{FIG:fig2} except for the blue and green lines, in which solid, dashed and dot-dashed correspond to $v_{\phi}=4000~\GeV$, $5000~\GeV$, and $6000~\GeV$,  respectively. 
	}
	\label{FIG:fig9}
\end{figure}

We first take some benchmarks to demonstrate the evolution of abundances of various particles in  Fig.~\ref{FIG:fig9}. In panel (a) for leptogenesis, the dominant $NN\to\rho\eta$ channel in the global scenario is replaced by the new process $NN\to\rho Z^\prime$. However, the contribution of $NN\to\rho Z^\prime$ to relic density is negligible, since $2m_N$ is much smaller than $m_\rho+m_Z'$ for the benchmark scenarios. We find that the process $NN\to f\bar{f}$  plays a major contribution,  while  $NN\to \rho\rho$  is subdominant. It is clear that decreasing $v_\phi$ results in a smaller $Y_{B-L}$ for a fixed value of $\varepsilon_{\CP}$. To meet the observation value of baryon asymmetry, $v_{\phi}$ typically needs to be smaller than $1000$ GeV for $\varepsilon_{\CP}=0.1$, which is already excluded by LEP-II ~\cite{Cacciapaglia:2006pk,ALEPH:2013dgf} in the local scenario. So the only viable method is reducing $\varepsilon_{\CP}$. For DM $\chi$, the dominant annihilation channel is $\chi\chi\to f\bar{f}$, as $\chi\chi\to \rho Z',Z'Z'$ are forbidden kinematically for the benchmark points.

The results for increasing  $m_N/m_\chi$ are shown in panels (b)-(d) of Fig.~\ref{FIG:fig9}. Although the reduced cross section of $NN\to\chi\chi$ should become larger, $Y_{B-L}$ has hardly changed since this channel is not the dominant contribution. Meanwhile, the decrease of $m_\chi$ leads to the increase of $Y_{\chi}$. Under the LEP-II constraint $v_{\phi}\gtrsim3500$ GeV, the DM abundances are much larger than the Planck observation in the non-resonance region of the benchmark points. Therefore, we can not find the common parameter space of leptogenesis and DM in the non-resonance region for the special scenario illustrated. We need a more efficient pathway to produce DM.

\begin{figure} 
	\begin{center}
		\includegraphics[width=0.45\linewidth]{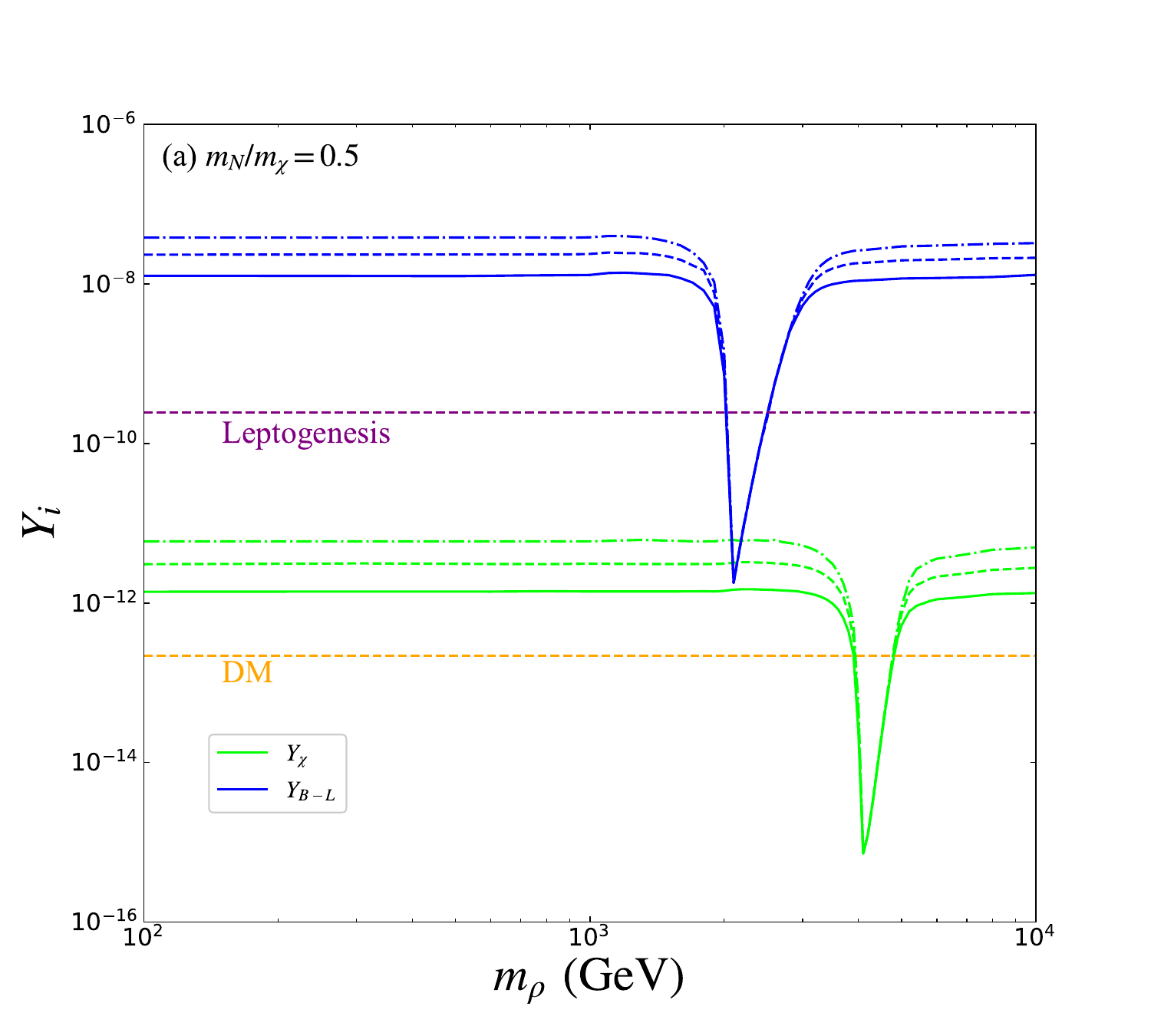}
		\includegraphics[width=0.45\linewidth]{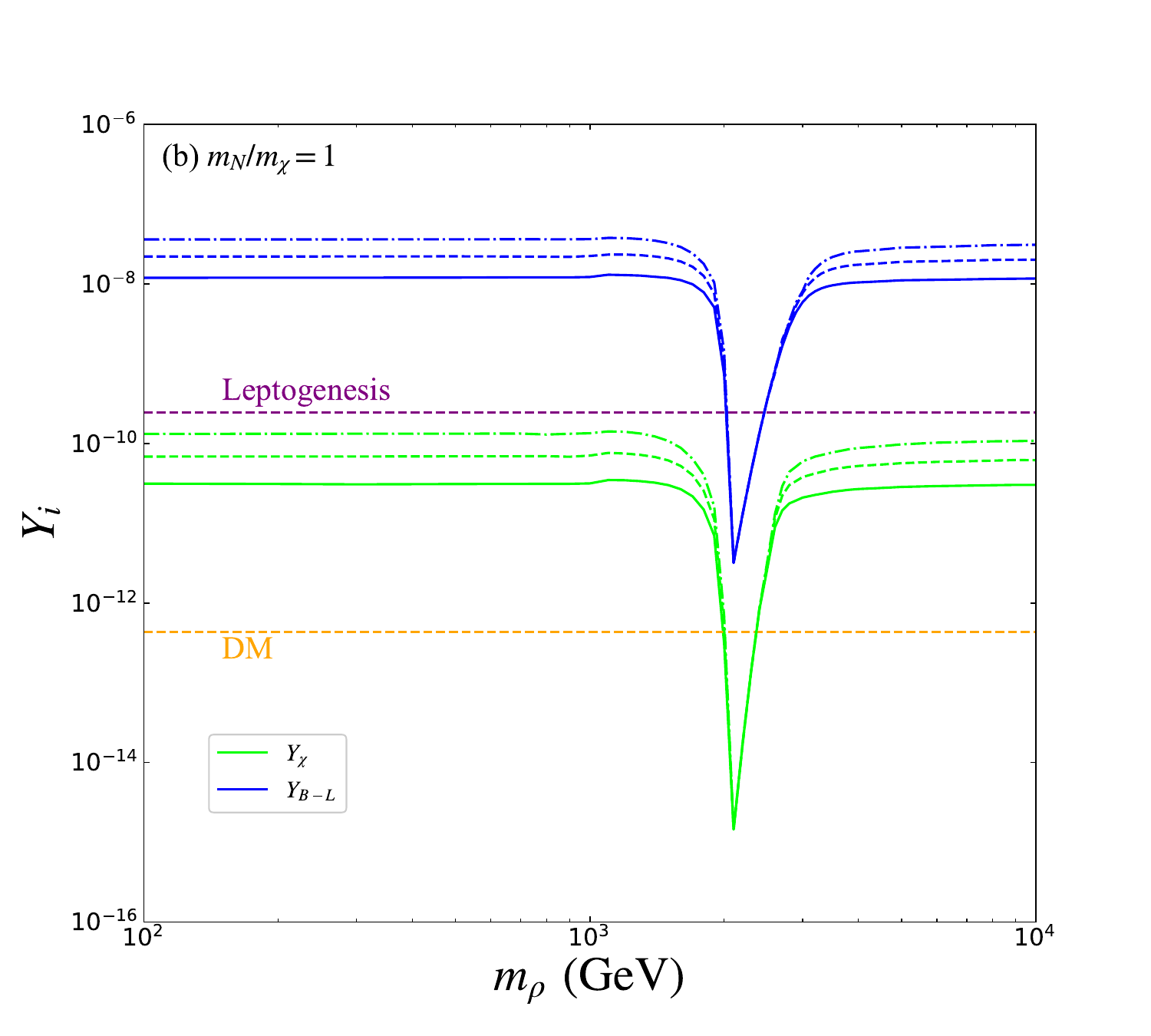}
		\includegraphics[width=0.45\linewidth]{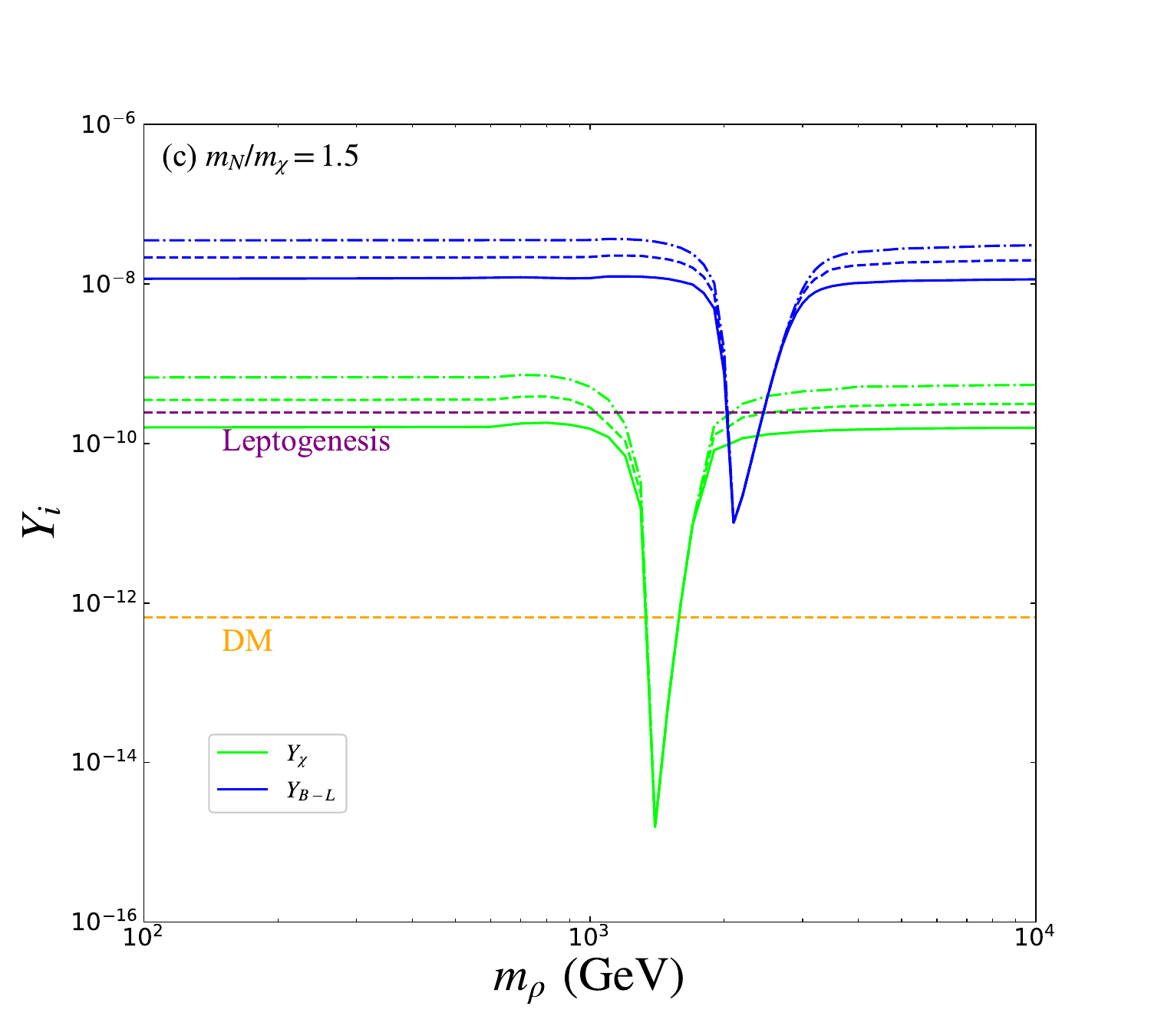}
		\includegraphics[width=0.45\linewidth]{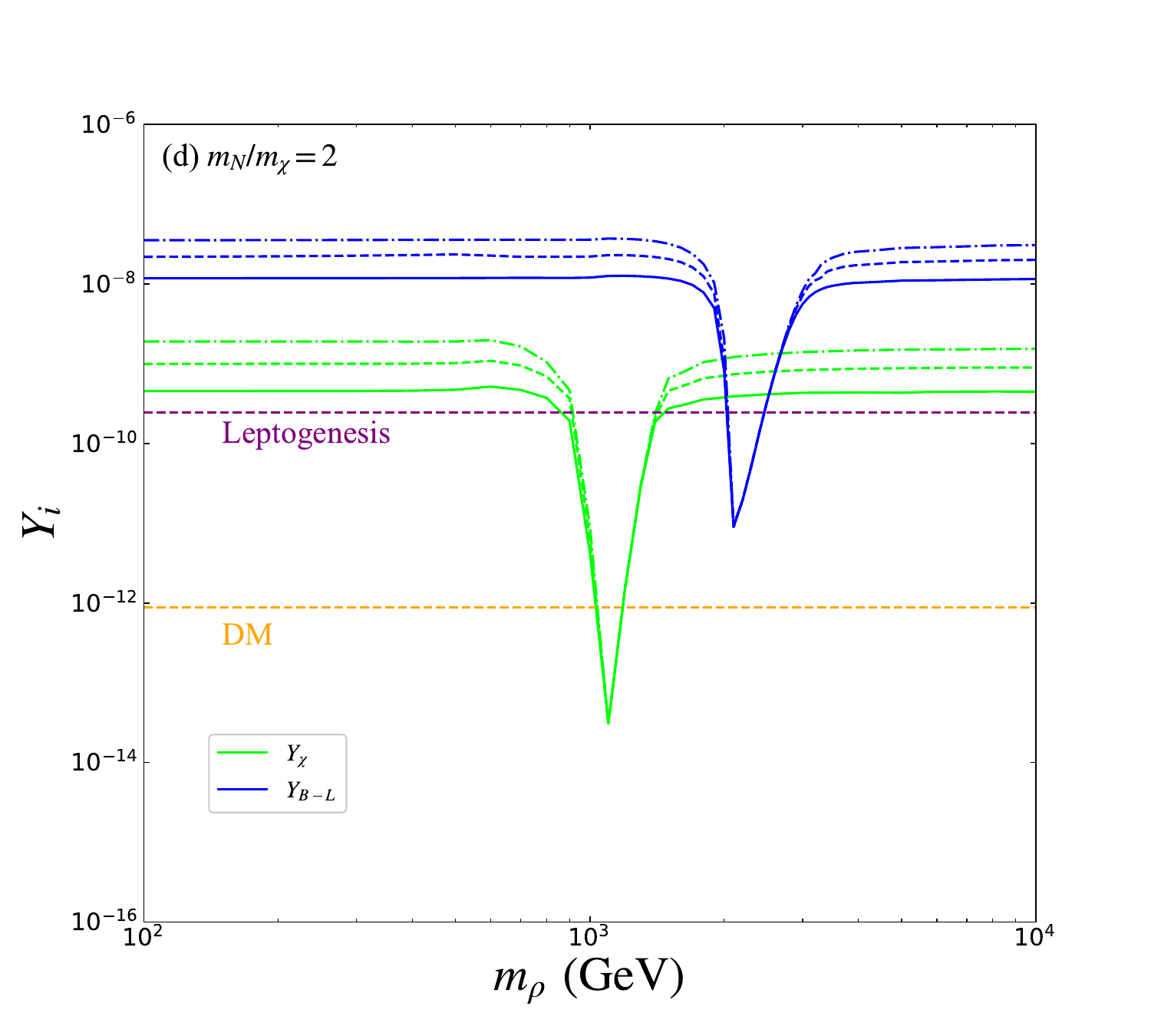}
	\end{center}
	\caption{$Y_i$ as a function of $m_\rho$ at present in the local scenario. $m_N$ and $m_{Z^\prime}$ are fixed as 1000 and 7000 GeV. The markers of the curves can be seen in Fig.~\ref{FIG:fig3} except for blue and green lines, in which the solid, dashed, and dot-dashed represent $v_{\phi}=4000~\GeV$, $5000~\GeV$, and $6000~\GeV$,  respectively.
	}
	\label{FIG:fig10}
\end{figure}

As shown in Fig.~\ref{FIG:fig3}, the abundances of DM can be greatly reduced in the resonance region for the global scenario, which portends that leptogenesis and DM may also coexist  in this region for the local scenario. Then, Fig.~\ref{FIG:fig10} shows the influences of $m_\rho$ on abundances $Y_i$, where the resonance effect is obvious. In the resonance region at $m_\rho\simeq2m_N$, the dominant channel is $NN\to \eta\eta/VV$ for the global scenario, while it is substituted for $NN\to VV$ in the local scenario. It should be noted that the $\rho$ portal channels depend on the mixing angle $\theta$, where we use $\theta=0.05$ in the calculations. In the decoupling limit with $\theta=0$, the resonance effect will be absent, because $\rho$ can not mediate $NN\to \SM\SM$ process in this special case for the benchmark points with $m_N/m_\chi=0.5$. Outside of the resonance region, the dominant channel is $NN\to \rho\rho$ for light $m_{\rho}$, and then becomes $NN\to f\bar{f}$ when $m_\rho>m_N$. 

Besides the resonance happens at $m_\rho\simeq 2 m_\chi$, the results of DM are similar to the sterile neutrino. In panel (a) of Fig.~\ref{FIG:fig10} with DM heavier than sterile neutrino, the $\chi\chi\to NN$ channel is always allowed, which also contributes to DM annihilation at the $\rho$ resonance. As for panel (b)-(d),  $Y_{B-L}$ shows an obvious resonance effect at $m_\rho\simeq2m_N$ under the influence of $NN\to VV$ and $NN\to \chi\chi$. Therefore, the symmetric $m_\rho$ in this region can successfully explain baryon asymmetry without modifying $\varepsilon_{\CP}$. In these three cases, DM is not heavier than sterile neutrino, which indicates that the impact of $\chi\chi\to NN$ can be ignored even in the resonance region. In addition, the overall increase of  $Y_{\chi}$ can not change the fact that the correct DM abundance can only be obtained at position $m_\rho\simeq2m_\chi$. These results confirm the condition that to simultaneously satisfy $Y_{B-L}$ and  $Y_{\chi}$, $m_\rho$ could appear in the DM resonance region as $m_\rho\simeq2m_\chi$, while changing $\varepsilon_{\CP}$.

As already shown in Fig.~\ref{FIG:FD}, DM and sterile neutrino $N$ also annihilate via the new gauge boson $Z'$. Therefore, correct DM abundance can be obtained at the $Z'$ resonance region as $m_{Z'}\simeq 2 m_\chi$ \cite{Blanco:2019hah}. Varying $m_{Z'}$ would lead to quite similar results as varying $m_{\rho}$ shown in Fig.~\ref{FIG:fig10}. In this paper, we have fixed $m_{Z'}=7000$ GeV, which means that the $Z'$ resonance appears at DM mass $m_\chi\simeq m_{Z'}/2=3500$ GeV. This result is clearly shown in Fig.~\ref{FIG:fig11}.

\subsection{Perturbation and CP-asymmetry} \label{SEC:LPC}

Similar to the global scenario, we also perform a scan in searching for the common origin of leptogenesis and DM in the parameter ranges as
\begin{eqnarray}\label{Eqn:LC}
	\begin{aligned}
		m_{\chi}\in[400,10000]~\GeV, m_\rho\in[100,10000]~\GeV, v_{\phi}\in[3500,20000]~\GeV, \varepsilon_{\CP}\in[10^{-6},1].
	\end{aligned}
\end{eqnarray}
During the scan, we have fixed $m_{Z'}=7000$ GeV and $\theta=0.05$ for illustration. The range of $v_\phi$ in the local scenario is different from the global scenario because $v_\phi\lesssim3500$ GeV is already excluded by LEP-II ~\cite{Cacciapaglia:2006pk,ALEPH:2013dgf}. The gauge coupling is determined by $g_{B-L}=m_{Z'}/(2v_\phi)$, which is smaller than one in the above range of $v_\phi$.

\begin{figure} 
	\begin{center}
		\includegraphics[width=0.45\linewidth]{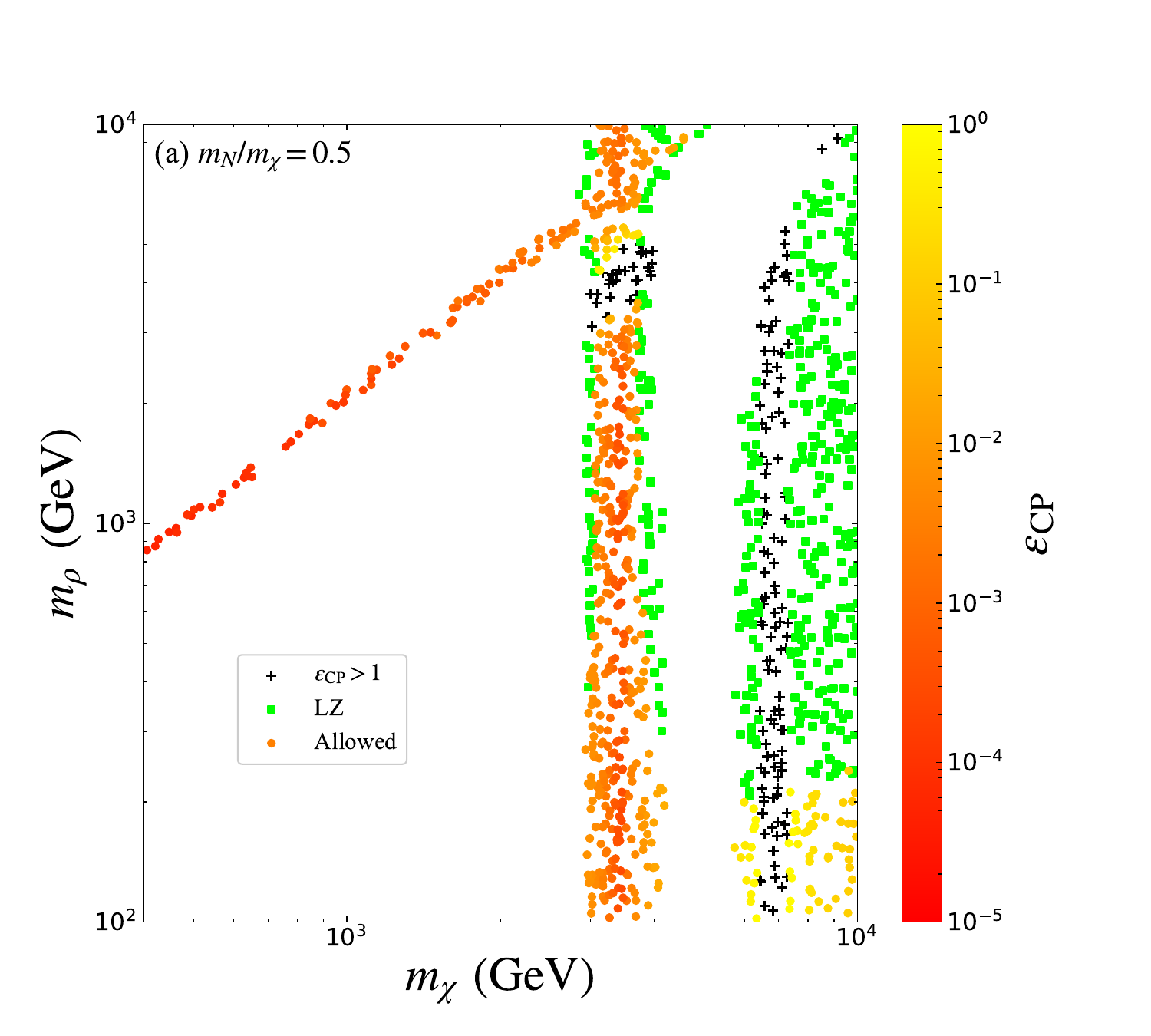}
		\includegraphics[width=0.45\linewidth]{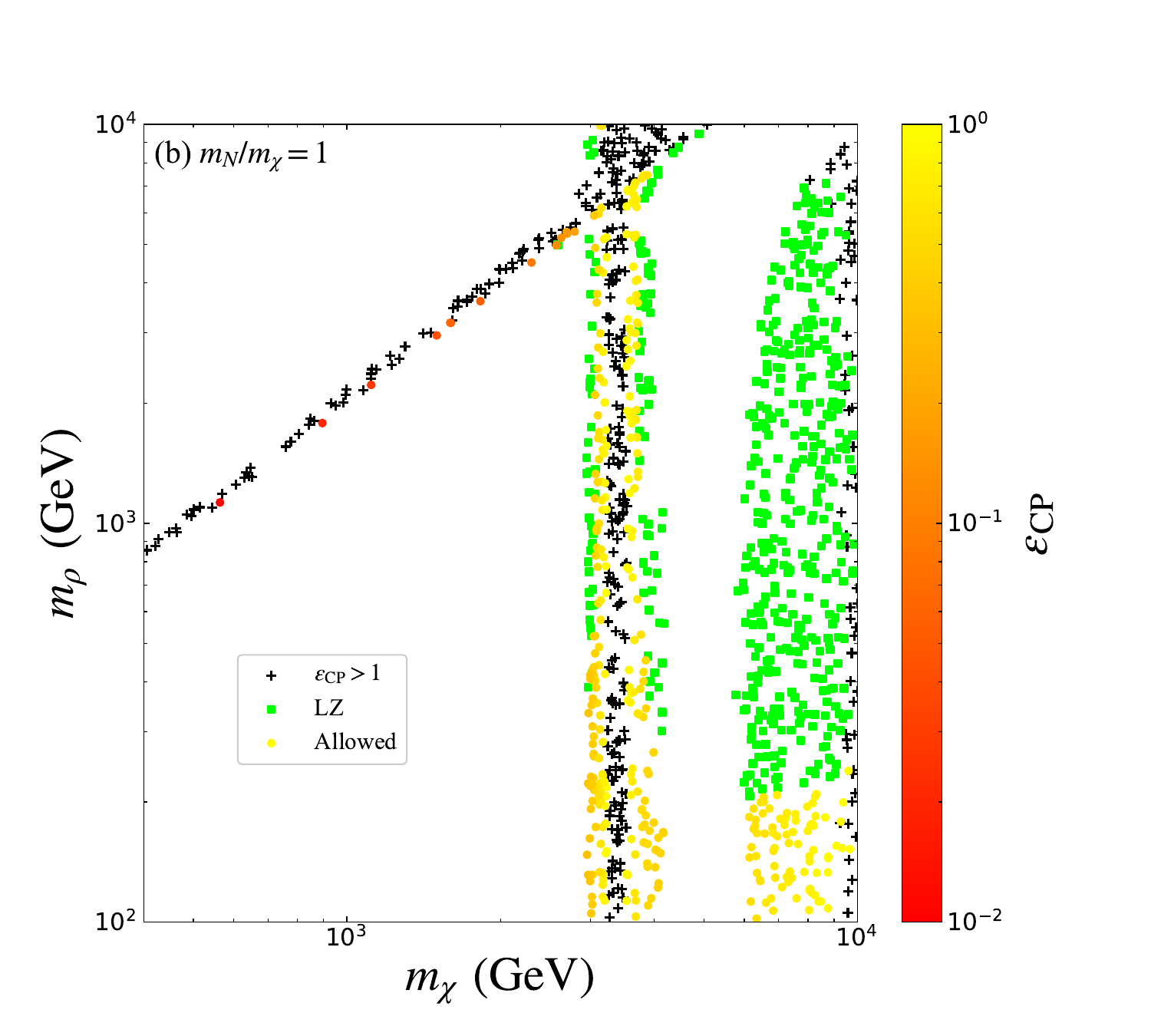}
		\includegraphics[width=0.45\linewidth]{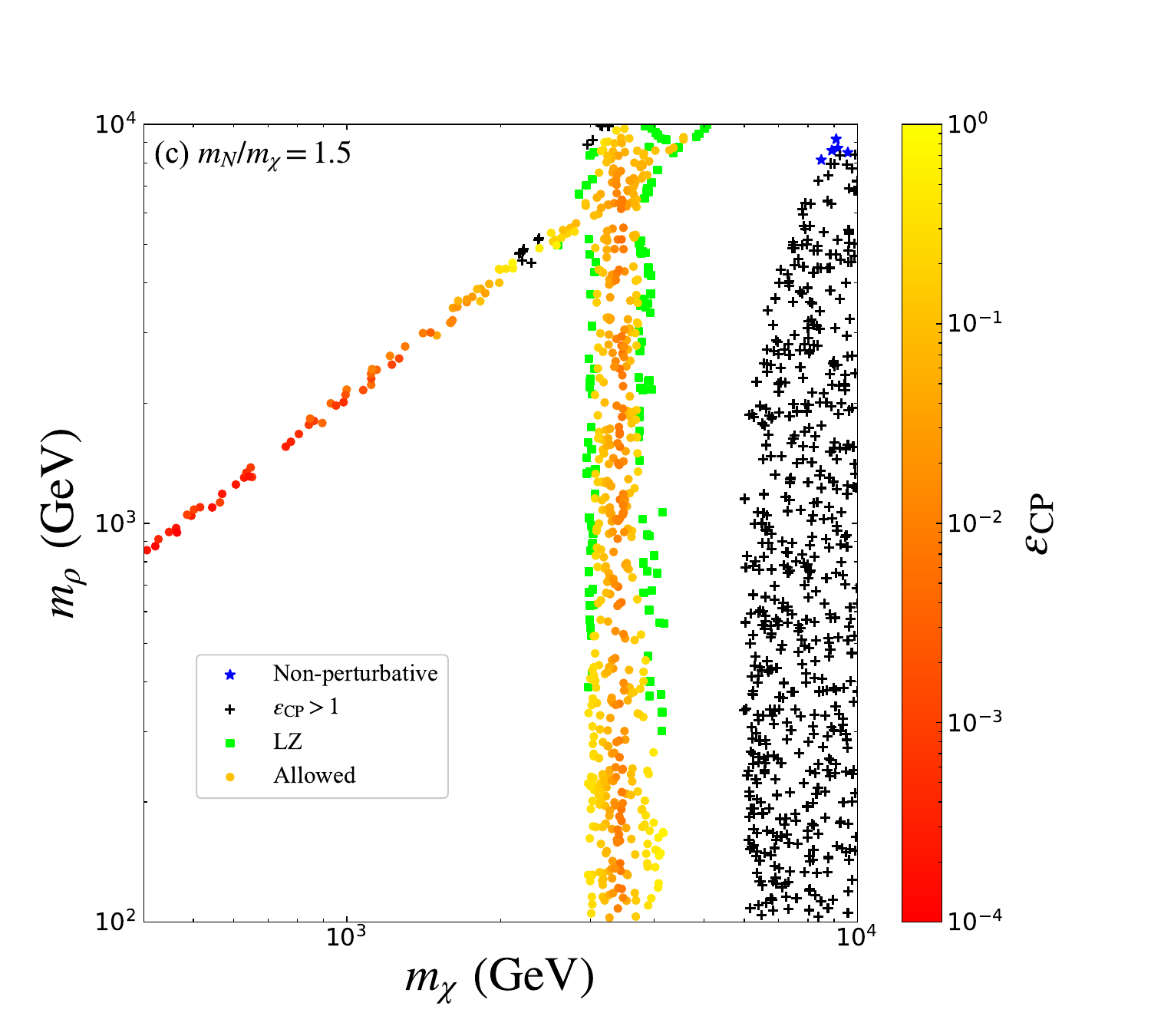}
		\includegraphics[width=0.45\linewidth]{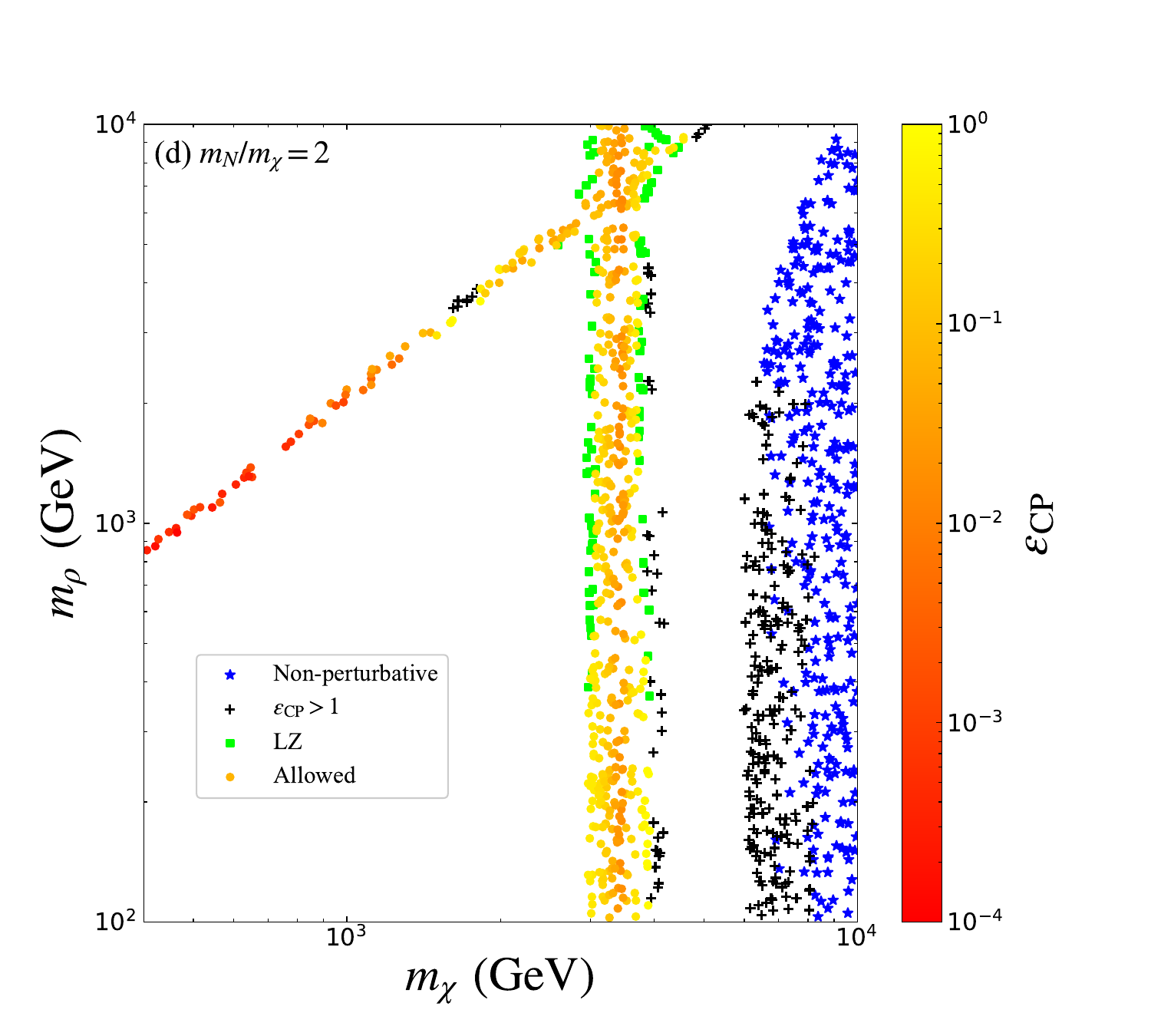}
	\end{center}
	\caption{Perturbation  and exceeding $\varepsilon_{\CP}$ constraints in the local scenario. (a)-(d) correspond to four cases with mass ratios $m_N/m_\chi = 0.5,1,1.5,2$. The markers of the samples are the same as those in Fig.~\ref{FIG:fig4}.
	}
	\label{FIG:fig11}
\end{figure}

In the local scenario, the perturbation constraints are $\lambda_{N,\chi}<\sqrt{4\pi}$ and $\lambda_{\phi}<2\pi$ \cite{Kahlhoefer:2015bea, Duerr:2016tmh}, meanwhile the CP-asymmetry is required to satisfy $\varepsilon_{\rm CP}<1$. The scanning results for the four specific cases with mass ratios $m_N/m_\chi=0.5, 1, 1.5, 2$  are displayed in Fig.~\ref{FIG:fig11}. Panel (a) is the case with light sterile neutrino, where all samples satisfy the perturbation constraints. A few samples with $\varepsilon_{\CP}>1$ occur at the $\rho$ or $Z^\prime$ resonance due to over dilution. The $\rho$ portal resonance is attributed to $NN\to VV$ with $m_\rho\simeq2m_N=m_\chi\sim3500$~GeV, while the $Z'$ portal resonance is dominant by $NN\to f\bar{f}$ with $m_{Z'}\simeq2m_N=m_\chi\sim7000$ GeV. For the DM, there are three viable regions for  the common origin in case (a), i.e., the $\rho$ resonance, the $Z'$ resonance, and the heavy DM $m_\chi\gtrsim6000$ GeV. Under the strict constraints of the present direct detection bound of LZ~\cite{LZ:2022lsv},  $m_\rho \lesssim 200$ GeV is required in the heavy DM region, where the $s$-wave $\chi\chi\to \rho Z^\prime$ is the dominant annihilation channel. Finally, these allowed samples are accompanied by relatively tiny $\varepsilon_{\CP}$ in the resonance region, approximately $\in[10^{-5},10^{-1}]$. But in the non-resonance region,  $\varepsilon_{\CP}$ has much larger values, namely $\varepsilon_{\CP}\in[10^{-1},1]$.

Increasing the mass ratio $m_N/m_\chi$ leads to different results, which are shown in panels (b)-(d) of Fig.~\ref{FIG:fig11}. The $\rho$ and $Z^\prime$ resonance regions for DM are still viable, where allowed samples have a larger $\varepsilon_{\CP}$ compared to panel (a). Meanwhile, the area excluded by perturbation constraints expands with $\lambda_N\geqslant\sqrt{4\pi}$ as $m_N$ increases. Due to excessive dilution of $N$, the positions of $\rho$ and $Z^\prime$ resonance related to $N$ are disfavored by $\varepsilon_{\CP}<1$. However, the  $\varepsilon_{\CP}>1$ excluded samples at resonance vary for different mass ratio $m_N/m_\chi$. For case (b) with $m_N=m_\chi$, the resonance regions for DM are the same as $N$, so a lot of samples in these resonance regions are also excluded by $\varepsilon_{\CP}>1$. The $Z'$ resonance related to $N$ excludes $m_\chi=2m_N/3\simeq m_{Z'}/3$ for case (c) and $m_\chi=m_N/2\simeq m_{Z'}/4$ for case (d) where $m_{Z'}=7000$ GeV.  The $\rho$ resonance related to $N$ in principle excludes  $m_\rho\simeq2m_N=3m_\chi\simeq3m_{Z'}/2$ for case (c) and $m_\rho\simeq2m_N=4m_\chi\simeq2m_{Z'}$ for case (d), however the latter case is out the scanned range of $m_\rho$. In panels (c) and (d) of Fig.~\ref{FIG:fig11}, we also find that the samples in the non-resonance region are fully excluded by the limitations of perturbation and $\varepsilon_{\CP}$. The sterile neutrino heavier than the DM is totally not allowed in the  non-resonance region  for the local scenario, which is different from the global scenario.

\subsection{Collider Signal and Direct detection} 

\begin{figure} 
	\begin{center}
		\includegraphics[width=0.45\linewidth]{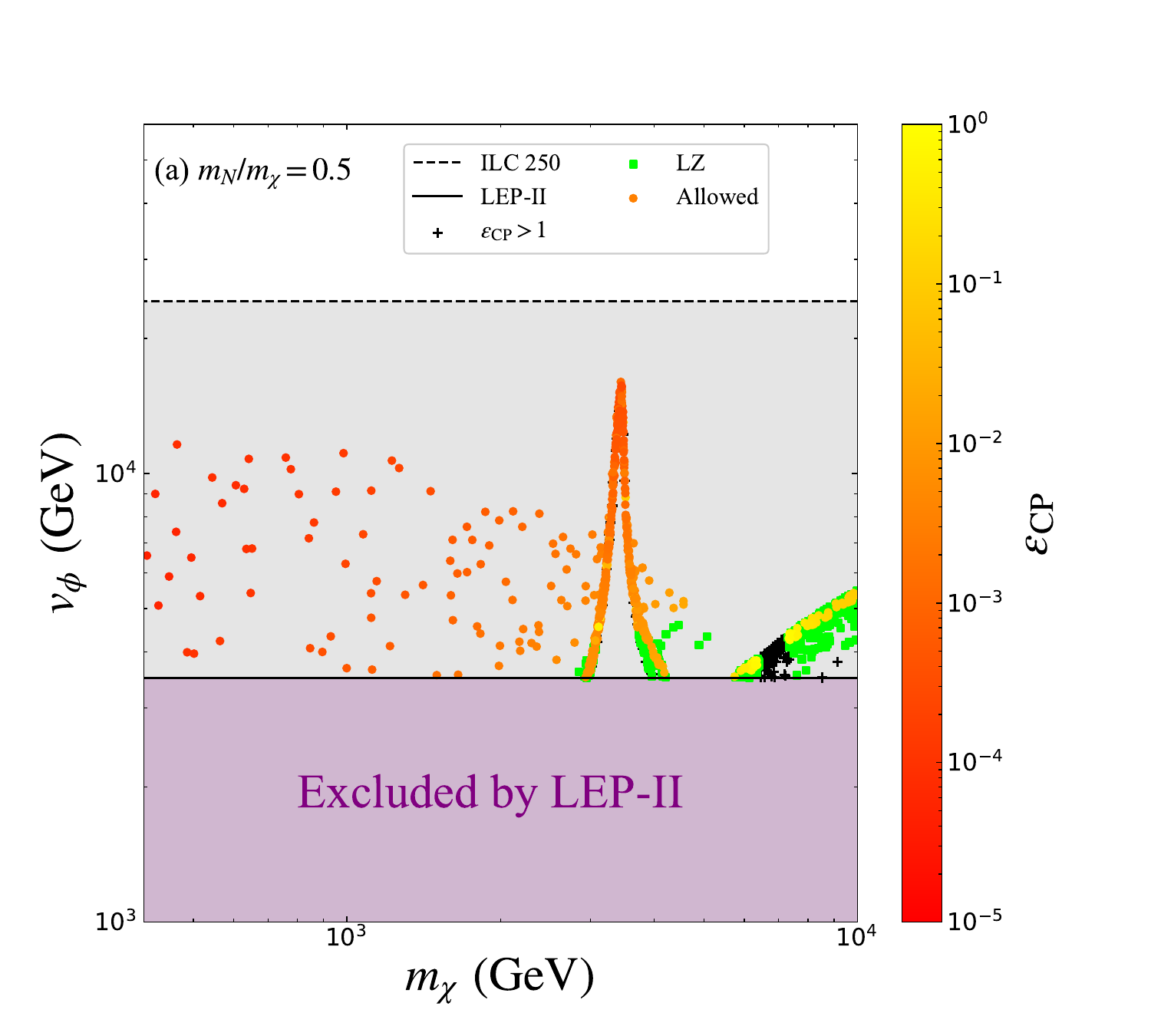}
		\includegraphics[width=0.45\linewidth]{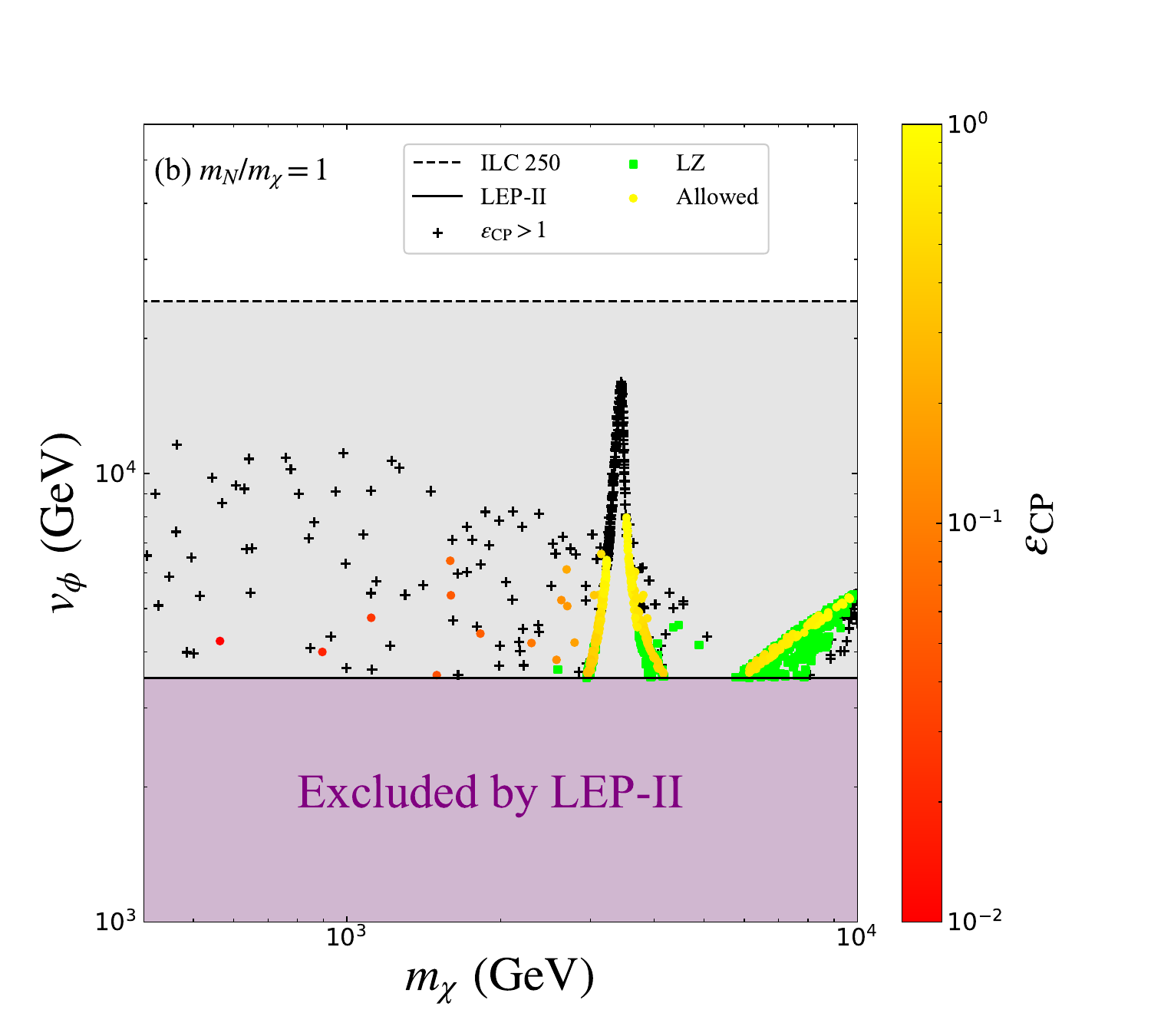}
		\includegraphics[width=0.45\linewidth]{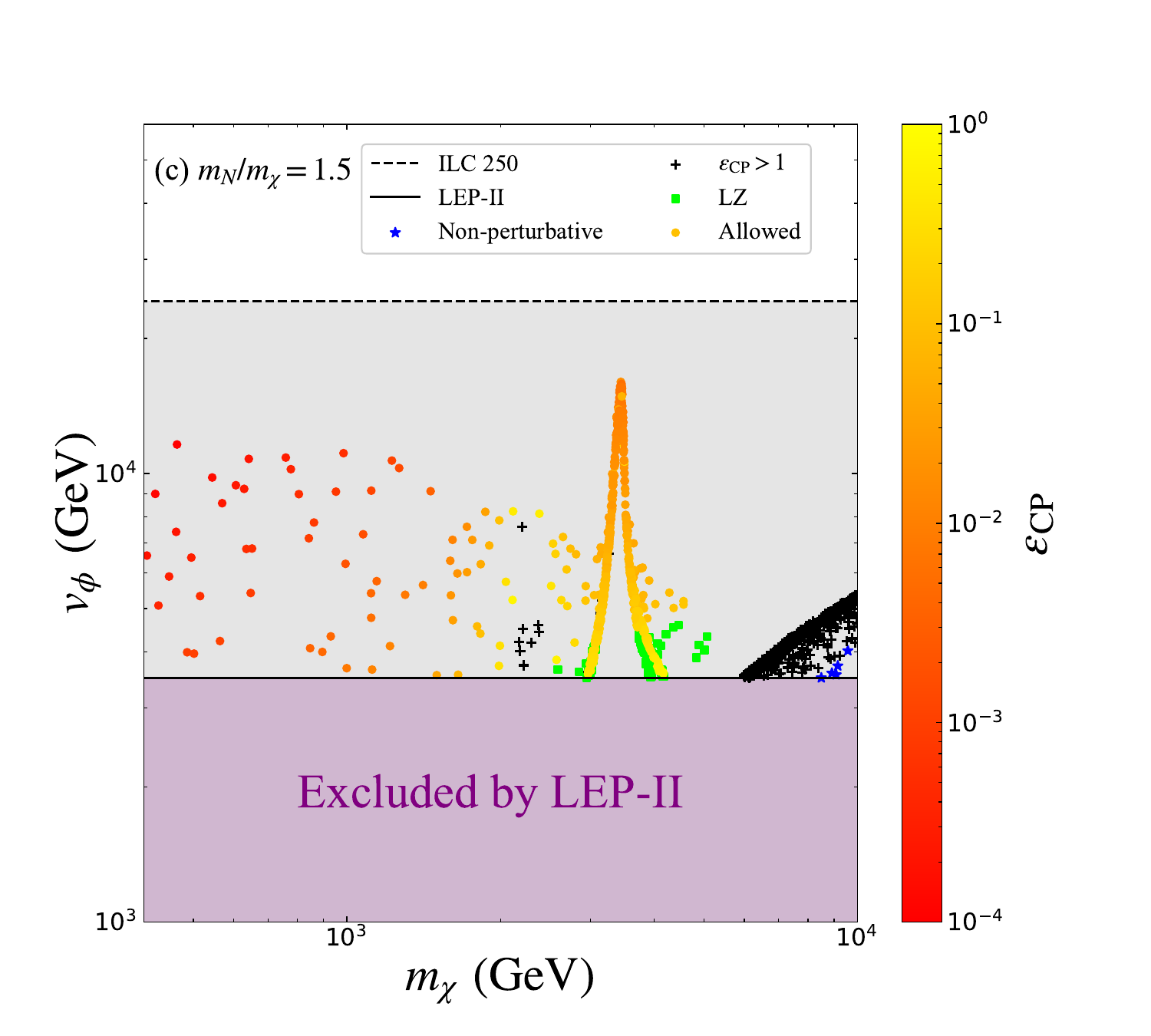}
		\includegraphics[width=0.45\linewidth]{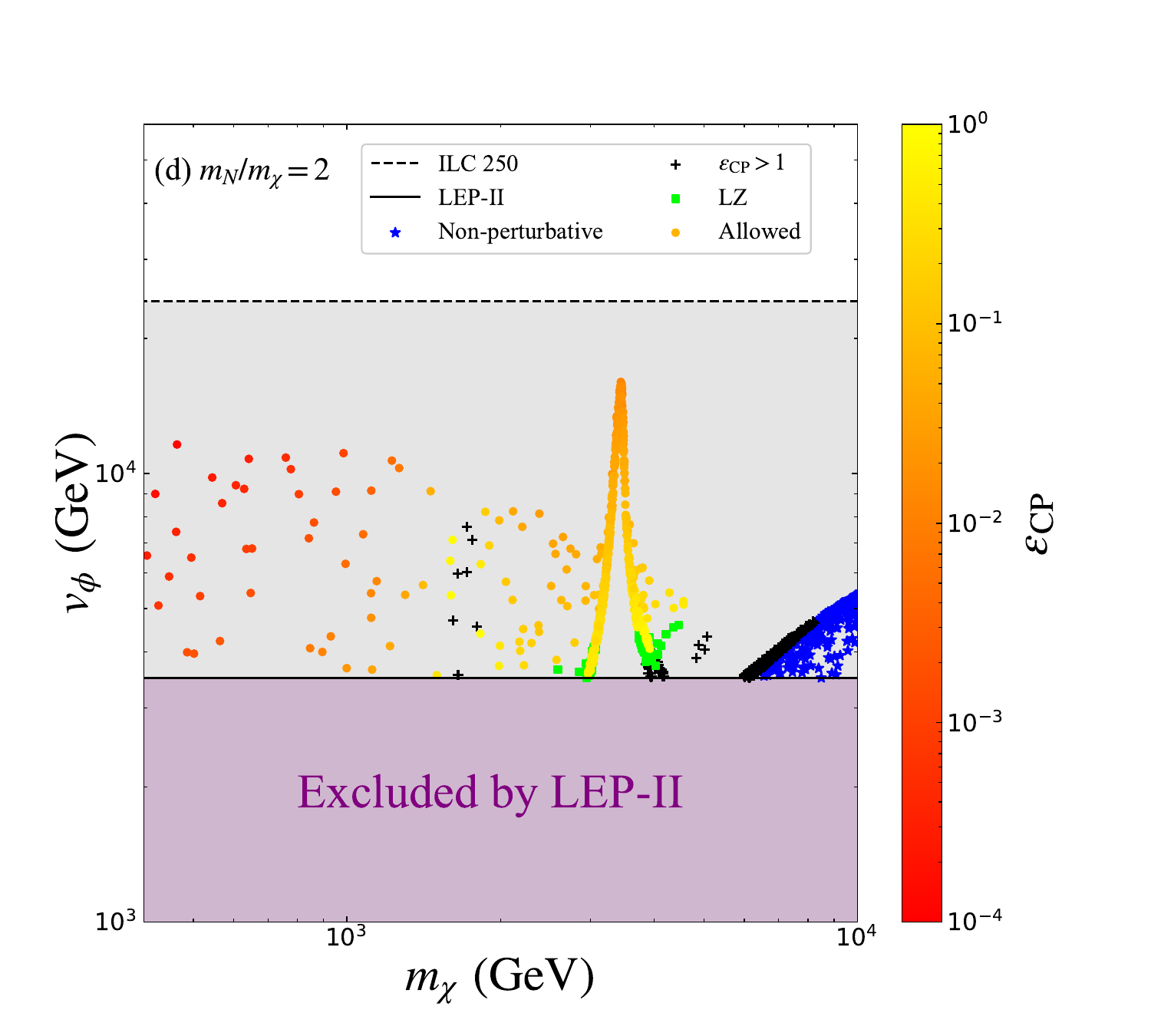}
	\end{center}
	\caption{The current LEP-II~\cite{Cacciapaglia:2006pk,ALEPH:2013dgf} and prospective ILC limits~\cite{LCCPhysicsWorkingGroup:2019fvj} in the local scenario. (a)-(d) correspond to four cases with mass ratios $m_N/m_\chi = 0.5,1,1.5,2$. The black dashed line represents the prospective ILC limit with $\sqrt{s}=250$ GeV. The markers of the samples are the same as that in Fig.~\ref{FIG:fig4}.
	}
	\label{FIG:fig12}
\end{figure}

The collider signature of gauged $U(1)_{B-L}$ is extensively studied \cite{Basso:2008iv,Iso:2009nw}. In the local scenario, the DM $\chi$ could contribute to Higgs invisible decay via $h\to \chi\chi$ when $m_\chi<m_h/2$, but it is forbidden kinematically with the parameter space in Eqn.~\eqref{Eqn:LC}.  The golden channel for discovering the gauge boson $Z'$ is the dilepton channel $pp\to Z'\to \ell^+\ell^-$,  which has excluded $m_{Z'}\lesssim5.15$~TeV at LHC \cite{ATLAS:2019erb,CMS:2021ctt}. The gauge boson $Z'$ can also mediate the pair production of sterile neutrino as $pp\to Z'\to NN$ \cite{FileviezPerez:2009hdc}, which has excluded the region with $m_N\lesssim1.4$ TeV and $m_{Z'}\lesssim4.4$ TeV \cite{CMS:2023ooo}. The choice of $m_{Z'}=7000$~GeV with TeV scale $m_N$ satisfies these current experimental limits. In the future, the 250 GeV ILC could exclude $v_\phi\lesssim22$ TeV via the process $e^+e^-\to f\bar{f}$ \cite{LCCPhysicsWorkingGroup:2019fvj,KA:2023dyz}.

The scanned results in the $m_\chi-v_\phi$ plane are shown in Fig.~\ref{FIG:fig12}. In panel (a), it is obvious that the allowed samples at the $\rho$ resonance satisfy $v_\phi\lesssim11$ TeV.  By comparison, $v_\phi$ in the $Z^\prime$ resonance can reach a larger value, but it will not exceed $20$ TeV. Meanwhile, the larger the $v_\phi$ is, the smaller the CP-asymmetry $\varepsilon_{\rm CP}$ is required, especially clear in the $Z^\prime$ resonance region. For 10 TeV scale heavy DM in the non-resonance region, survived samples require relatively larger coupling thus smaller $v_\phi$ to satisfy observed relic abundance, which is less than $5500$ GeV here.  Certainly, samples with smaller $v_\phi$ tend to have larger scattering cross section, hence are also more tightly constrained by LZ. Based on the above results, we can conclude that the future 250~GeV ILC is enough to completely exclude all the samples within the parameter space in Eqn.~\eqref{Eqn:LC}, even at the resonance region.

In panel (b) of Fig.~\ref{FIG:fig12}, we find that the excluded samples at the $Z'$ resonance by the constraint of exceeding $\varepsilon_{\CP}>1$ concentrate in the sharp peak, which leads to the allowed samples satisfying $v_\phi\lesssim 8000$~GeV. Meanwhile, the few survived samples via the $\rho$ resonance indicate $v_\phi\lesssim 6000$~GeV. In panels (c)-(d), as mentioned in Fig.~\ref{FIG:fig11}, only some samples at regions of $\rho$ and $Z^\prime$ resonance are left under various current limits. But they can not avoid the constraints of future 250 GeV ILC either. Except for the $Z'$ resonance related to $N$, it is clear that a lighter DM $\chi$, i.e., a lighter sterile neutrino $N$, requires smaller $\varepsilon_{\rm CP}$ for samples via the $\rho$ resonance.

\begin{figure} 
	\begin{center}
		\includegraphics[width=0.5\linewidth]{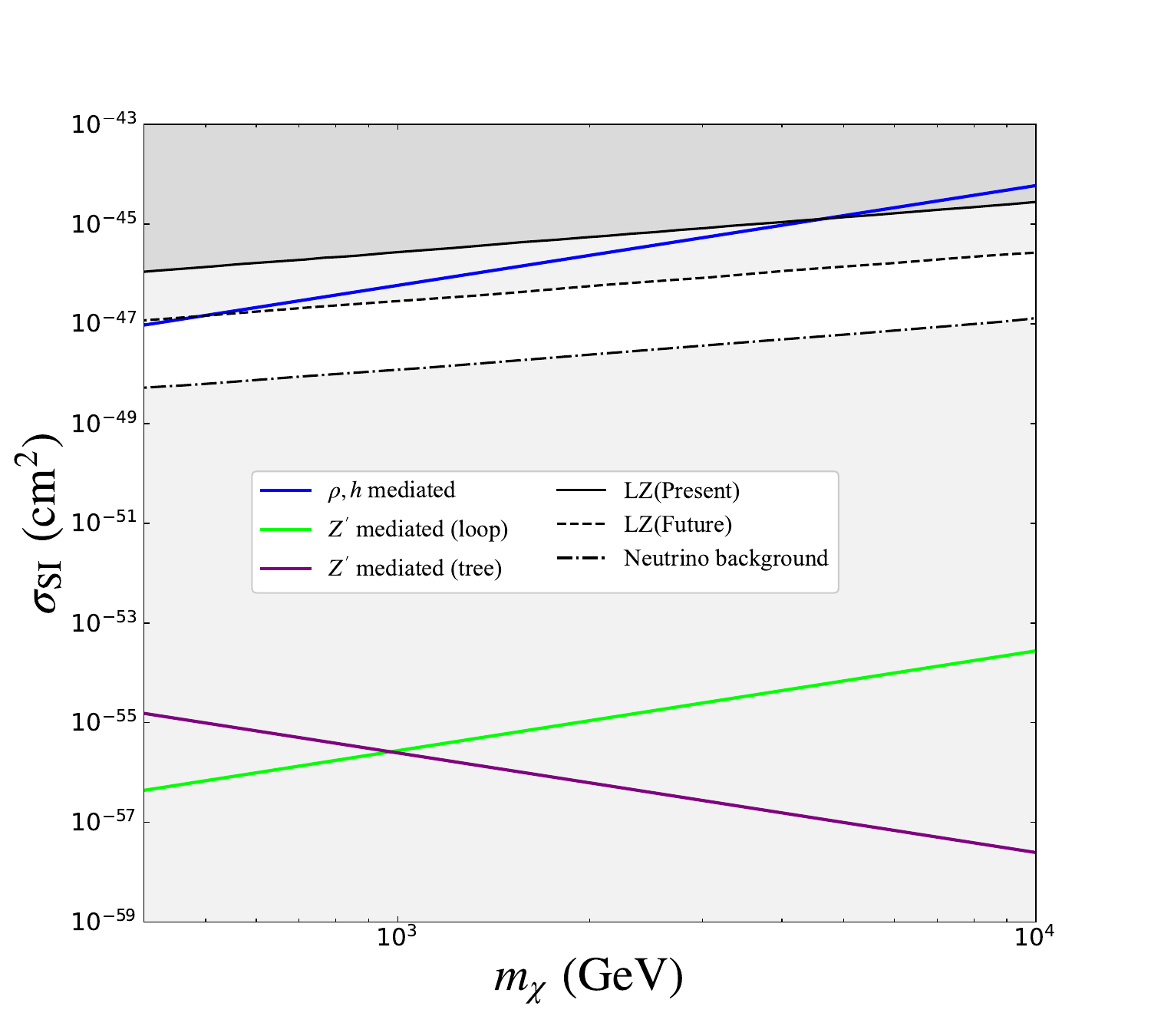}
	\end{center}
	\caption{Spin-independent scattering cross sections obtained through the Higgs portal in Eq.~\eqref{Eqn:dd}~(blue line), through tree-level $Z'$ in Eq.~\eqref{Eqn:ddt}~(purple line) and loop-level $Z'$ in Eq.~\eqref{Eqn:ddl}~(lime line). The parameters $\theta$, $v_{\phi}$, and $m_\rho$ are fixed as 0.05, 5000 GeV, and 1000 GeV, respectively.  The black solid, dashed and  dot-dashed lines represent the present LZ limit~\cite{LZ:2022lsv}, future~\cite{LZ:2015kxe} LZ limit, and the neutrino floor ~\cite{Billard:2013qya}. 
	}
	\label{FIG:fig13}
\end{figure}

We then consider the direct detection limits. The elastic scattering cross sections of $\chi$ can be divided into two kinds according to the mediators. The Higgs portal $\rho$ and $h$ mediated one are the same as the scenario discussed in the global scenario, which heavily depends on mixing angle $\theta$. The expression of the spin-independent scattering cross section can be seen in Eq.~\eqref{Eqn:dd}. The other one is mediated by $Z^\prime$. The spin-independent scattering cross section at the tree-level can be written as\cite{Hooper:2014fda}
\begin{eqnarray}\label{Eqn:ddt}
	\sigma_{\rm SI}^{\text {tree}}=\frac{2m_\chi^2 m_n^4 (g_\chi^A)^2 v^2}{\pi m_{Z^\prime}^4(m_\chi+m_n)^4 A^2}\left[Z(2g_u^V+g_d^V)+(A-Z)(g_u^V+2g_d^V)\right]^2,
\end{eqnarray}
where $v$ is the non-relativistic velocity of DM at present.  $Z$ and $A$ are the atomic number and atomic mass of the target, respectively. Since $v\sim10^{-3}$, the scattering cross section through tree-level $Z'$ is heavily suppressed. The velocity unsuppressed scattering via $Z'$ appears at the one-loop level. The corresponding spin-independent scattering cross section is calculated as ~\cite{Kahlhoefer:2015bea}
\begin{eqnarray}\label{Eqn:ddl}
	\sigma_{\rm SI}^{\text {loop}}=\frac{C^2 m_n^4 m_\chi^4 (g_\chi^A)^4 (g_q^V)^4 }{\pi^5(m_\chi+m_n)^2 m_\rho^4 m_{Z^\prime}^4},
\end{eqnarray}
where $q$ in $g_q^V$ represents quarks.

\begin{figure} 
	\begin{center}
		\includegraphics[width=0.45\linewidth]{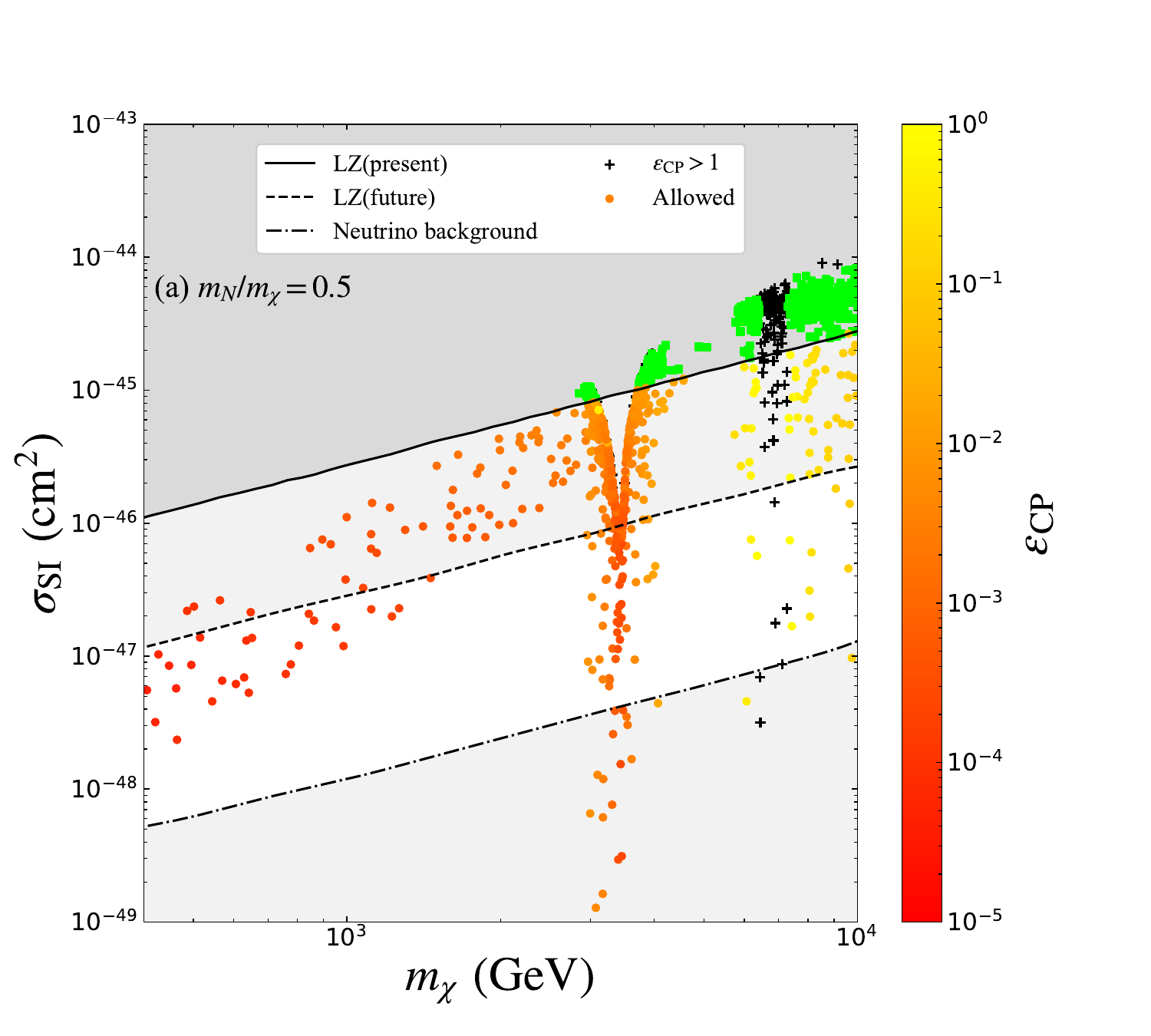}
		\includegraphics[width=0.45\linewidth]{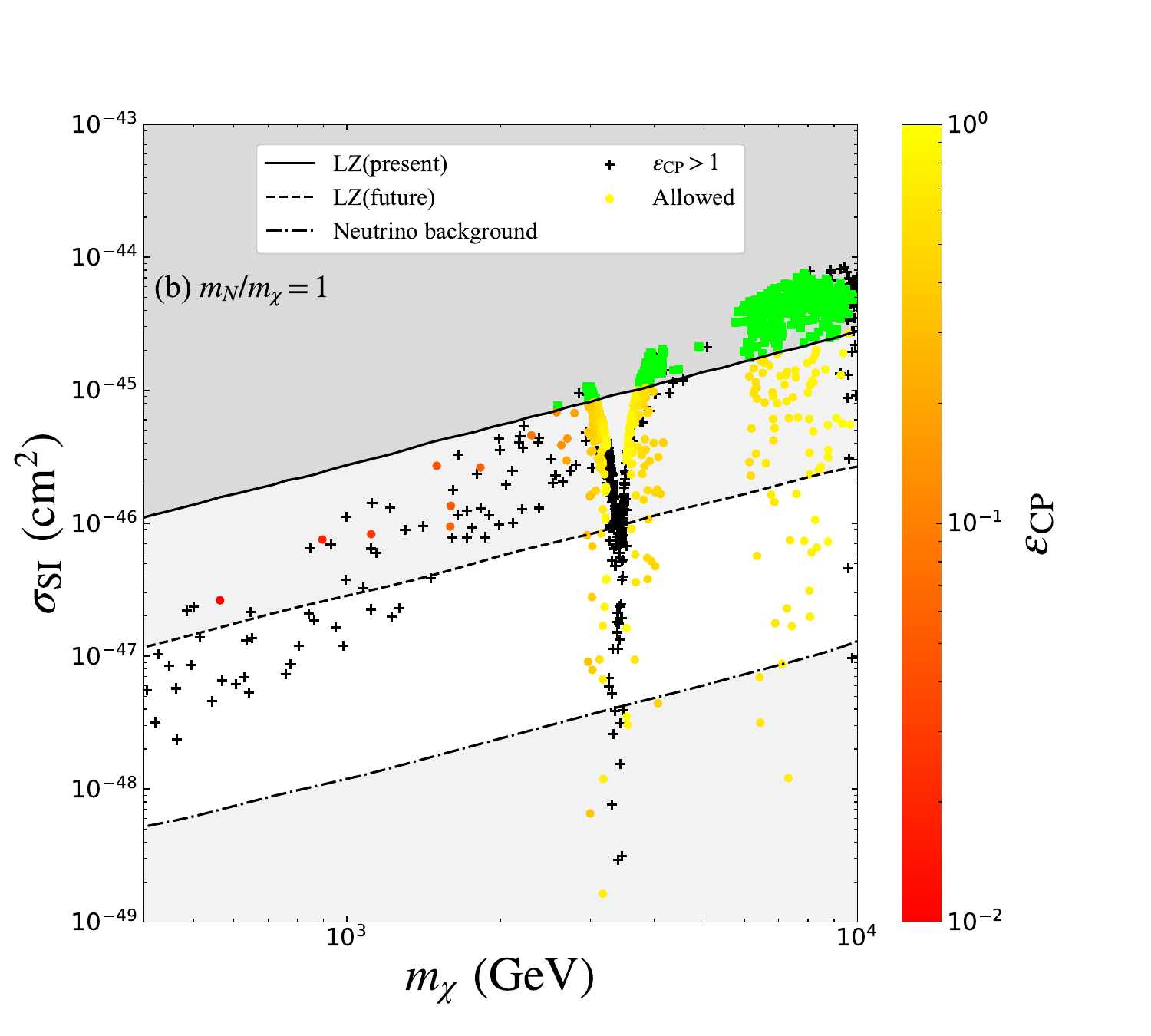}
		\includegraphics[width=0.45\linewidth]{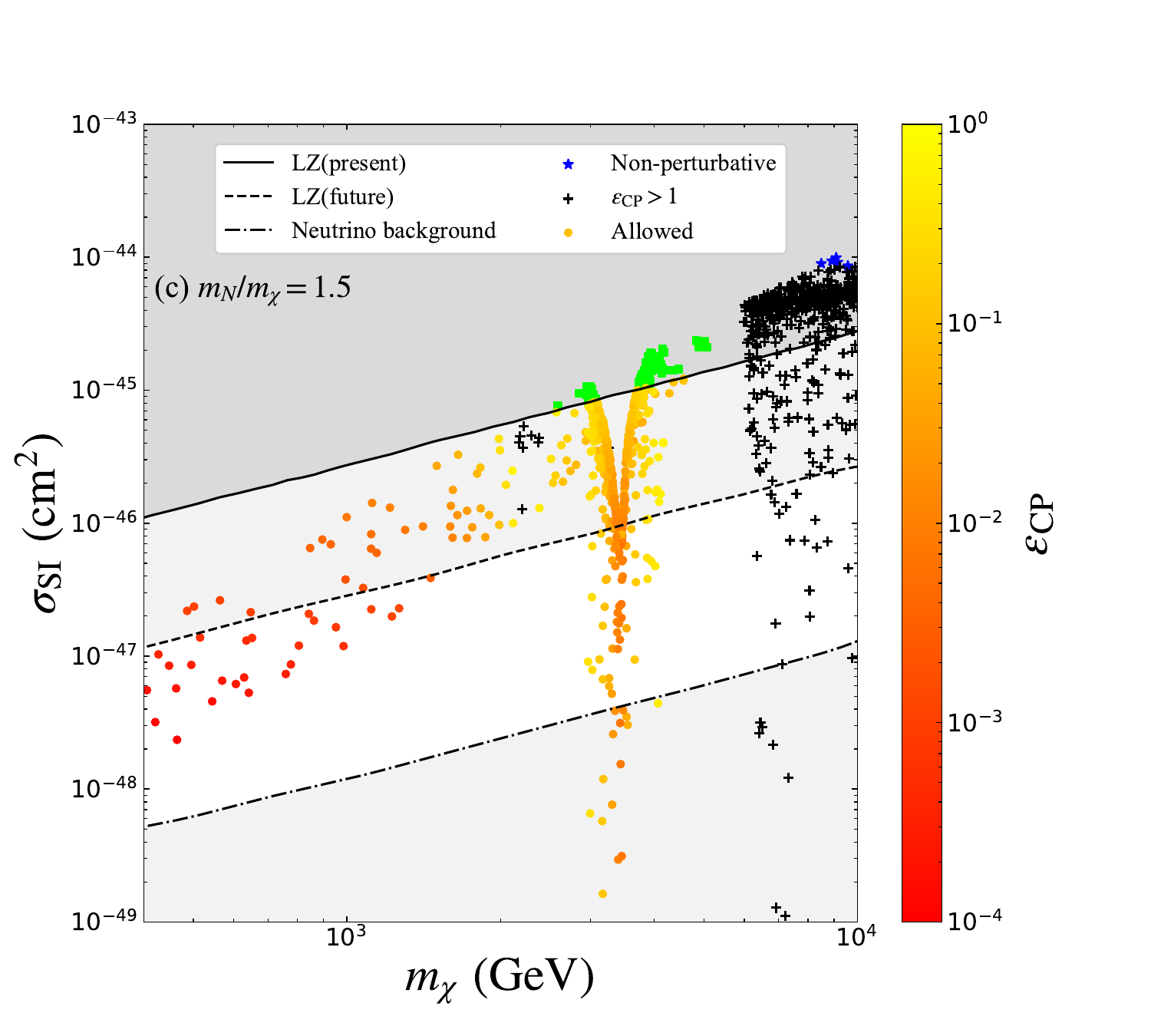}
		\includegraphics[width=0.45\linewidth]{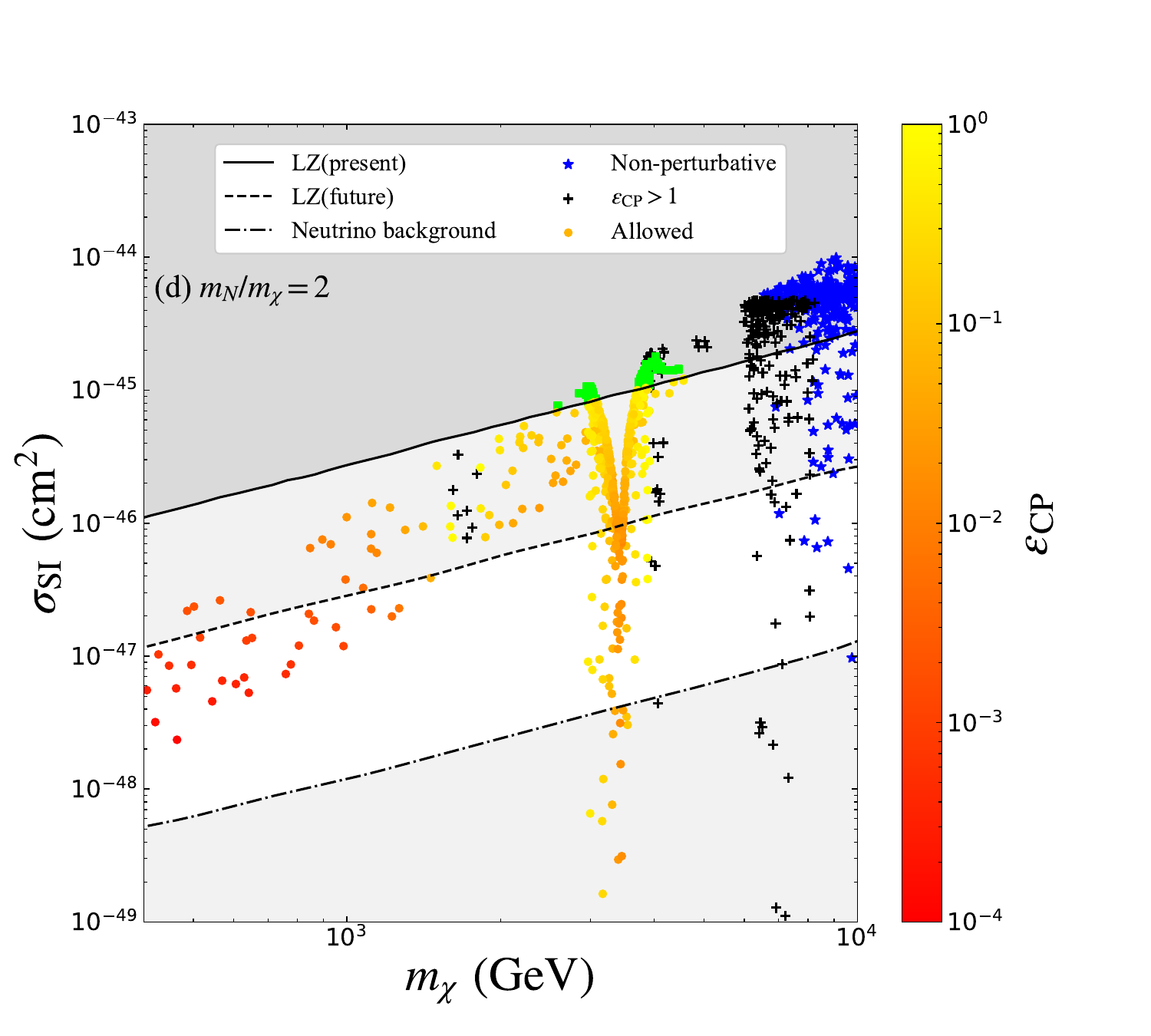}
	\end{center}
	\caption{Constraints from the direct detection experiments in the local scenario. $\theta$ is also fixed as $0.05$.
	}
	\label{FIG:fig14}
\end{figure}

The numerical results for a specific scenario are shown in Fig.\ref{FIG:fig13}. The blue curve for the Higgs portal scattering cross section is calculated with mixing $\theta=0.05$, which can be tested by future LZ experiment. For $m_\chi\lesssim1000$ GeV, the tree-level $Z'$ mediated result is larger than the loop-level one. And the loop induced scattering via $Z'$ becomes larger when DM $\chi$ is heavier than about 1000 GeV.  The scattering cross sections mediated by $Z^\prime$ at both tree and loop-level are smaller than $10^{-54}~\rm cm^2$, which are over eight orders of magnitudes smaller than the Higgs portal result. Therefore, we only consider the spin-independent scattering cross section through the Higgs portal for simplicity during the scan.

The scanned results of spin-independent scattering cross sections are shown in Fig.\ref{FIG:fig14}, where $\theta$ is fixed as 0.05. Except for different excluded samples under constraints from perturbation and $\varepsilon_{\rm CP}>1$, the theoretical values of $\sigma_{\rm SI}$ are similar for the four scenarios for different mass ratios of $m_N/m_\chi$. The current LZ limit tends to exclude DM heavier than 3000 GeV, while light DM is safe under the current limit. This is because $\sigma_{\rm SI}$ is proportional to $m_\chi^4$ according to Eqn.~\eqref{Eqn:dd}. For certain $v_\phi$ determined by DM relic abundance, the heavier DM thus leads to a larger scattering cross section. For DM via the $\rho$ portal, allowed samples predict $\sigma_{\rm SI} \gtrsim3\times10^{-48}~\rm cm^2$, which is over the neutrino background. A few $\rho$ portal samples with $m_\chi\lesssim1500$~GeV would escape the future LZ limit, but they are within the reach of future DARWIN experiment \cite{DARWIN:2016hyl}. However, in panel (b) with $m_N/m_\chi=1$, future LZ is likely to exclude all allowed samples at the $\rho$ resonance because of exceeding $\varepsilon_{\CP}$.  On the other hand, the $Z^\prime$ resonance and non-resonance samples could have relatively small $\sigma _{\rm SI}$, mainly due to the cancellation effect with $m_\rho\simeq m_h$. Once again, the above direction detection results heavily depend on the mixing angle $\theta=0.05$. A smaller mixing as $\theta=0.01$ would result in most of the samples under the LZ limit.

\subsection{Indirect Detection}

The theoretical DM annihilation channels in the local scenario are also shown in Fig.~\ref{FIG:FD}. Different from the global scenario, the dominant DM annihilation channels at present heavily depend on the mass spectrum of the new particles. Due to the Majorana nature of DM $\chi$, the DM annihilation channels into SM final states via $\rho$ and $Z'$ are $p$-wave suppressed. So most of the samples in the $\rho$ or $Z'$ resonance region are beyond the reach of indirect detection. For samples in the non-resonance region, the dominant annihilation channel is the $s$-wave $\chi\chi\to \rho Z'$, which is hopeful to be detected.

\begin{figure} 
	\begin{center}
		\includegraphics[width=0.45\linewidth]{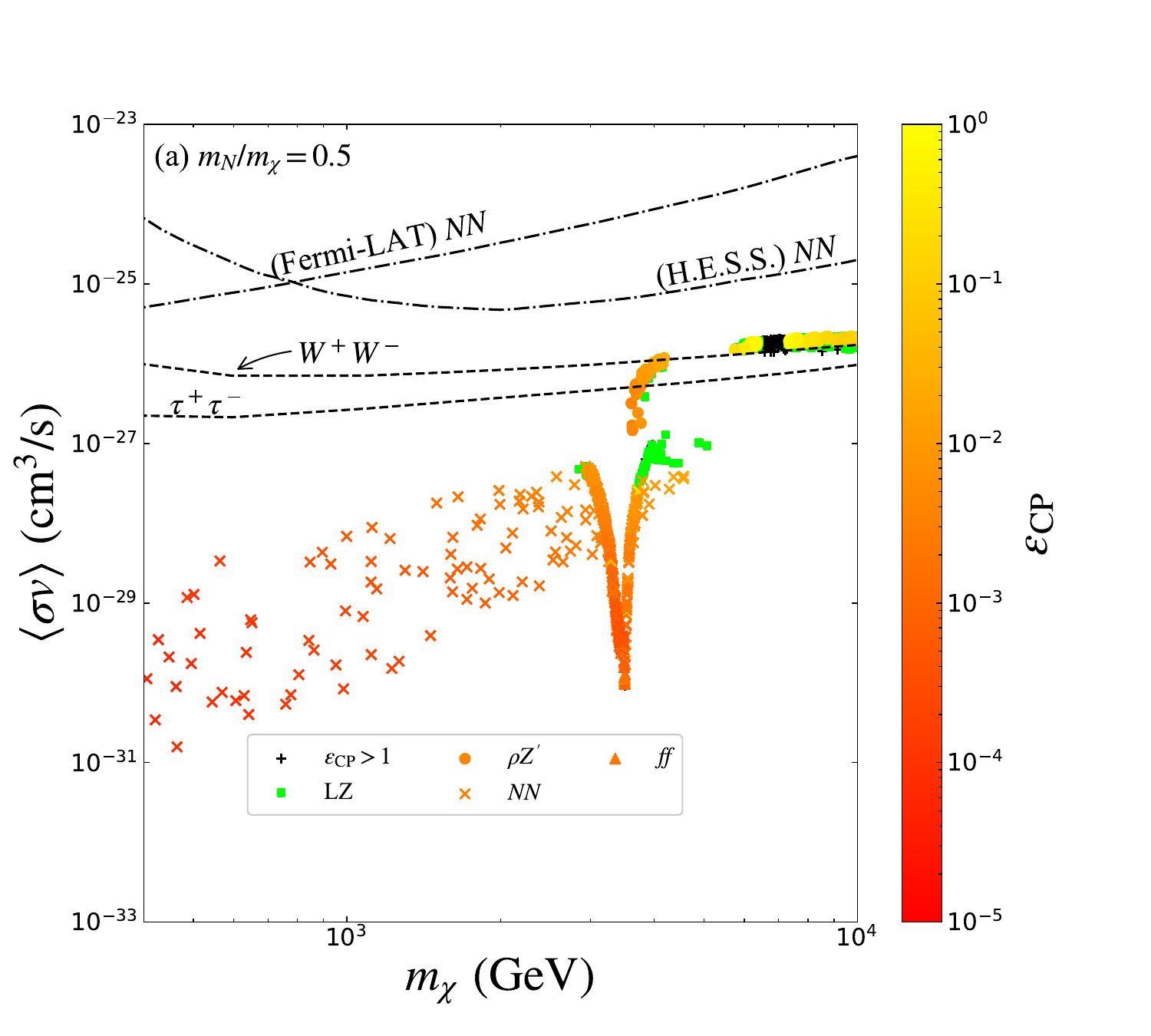}
		\includegraphics[width=0.45\linewidth]{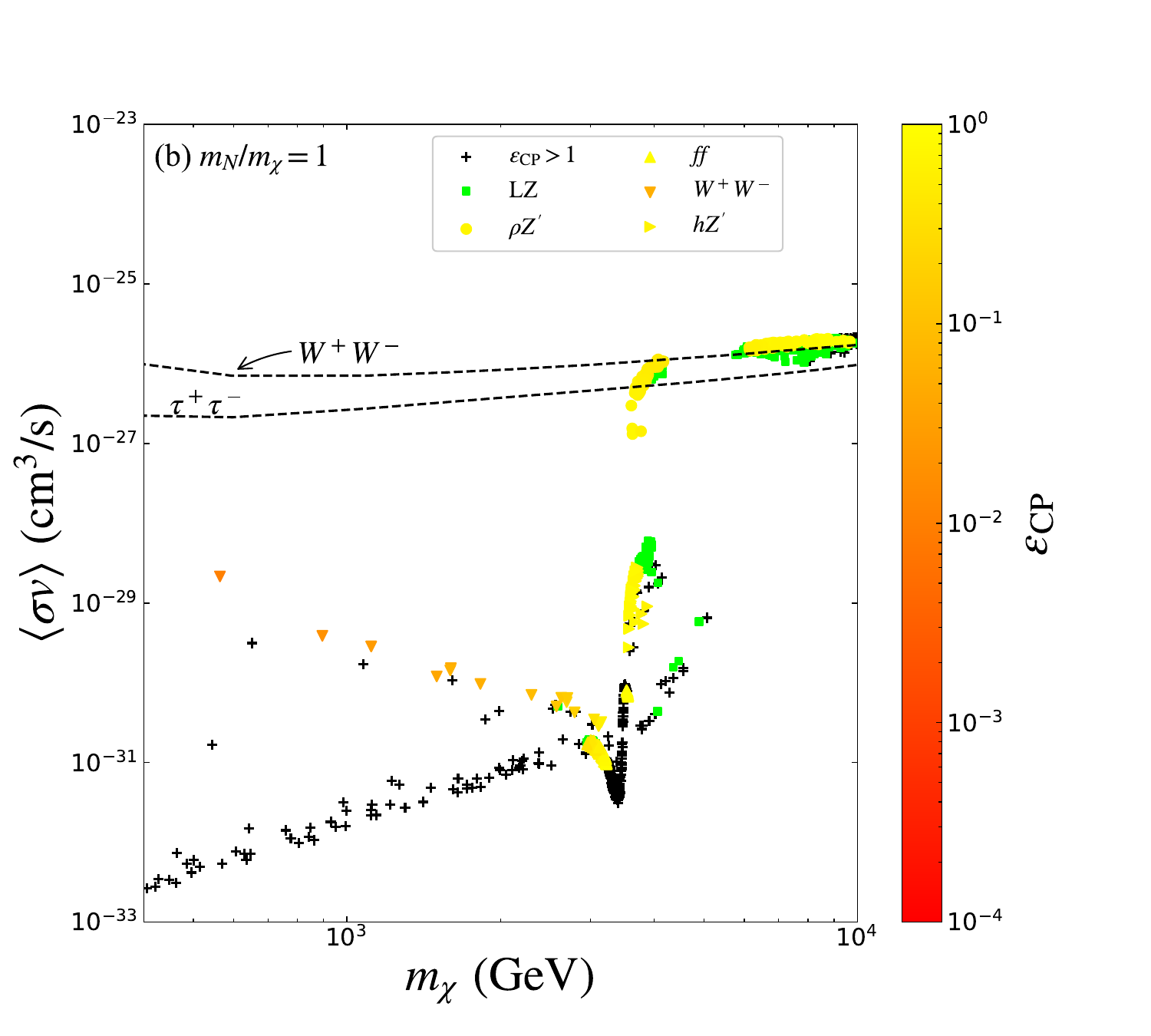}
		\includegraphics[width=0.45\linewidth]{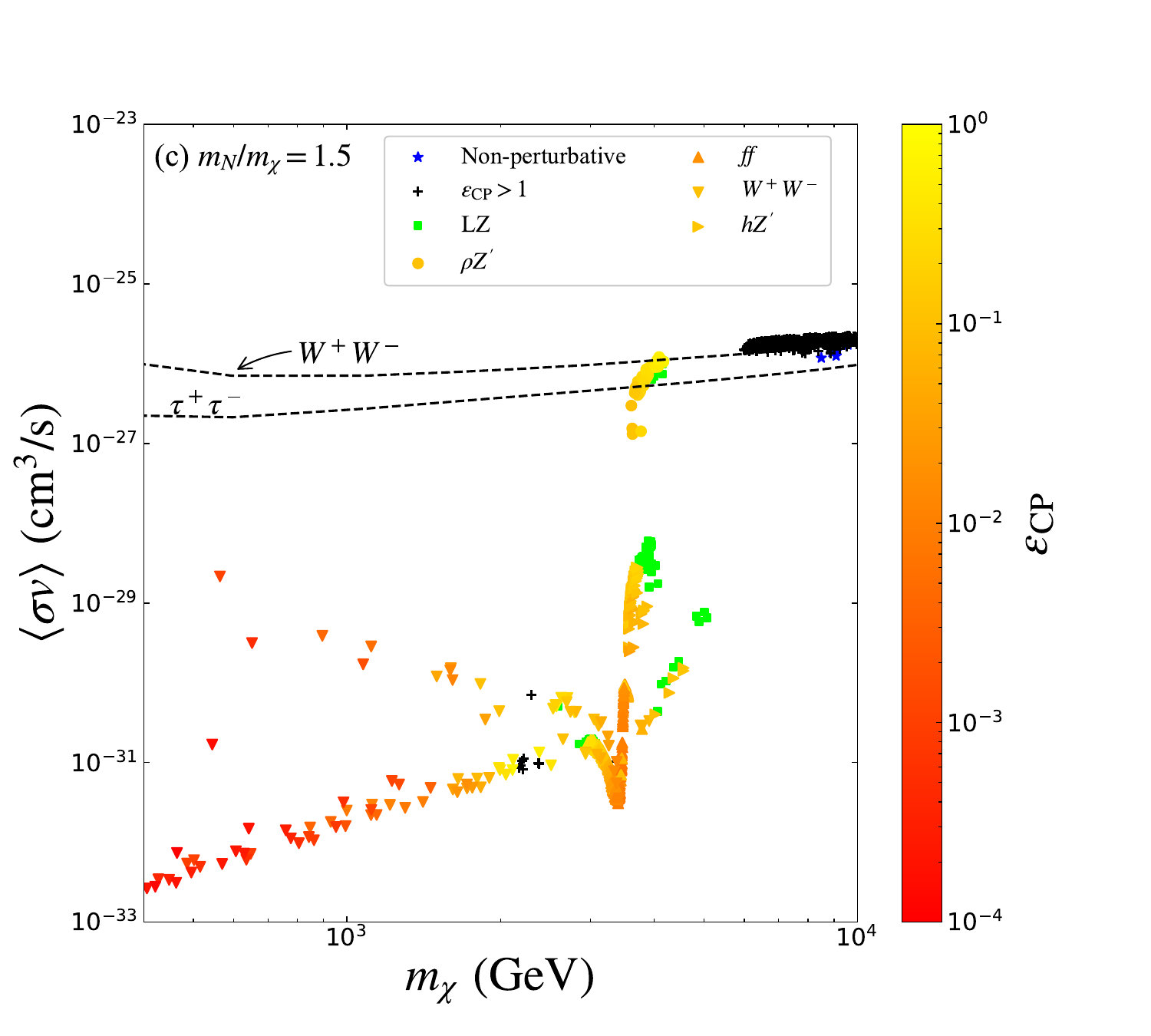}
		\includegraphics[width=0.45\linewidth]{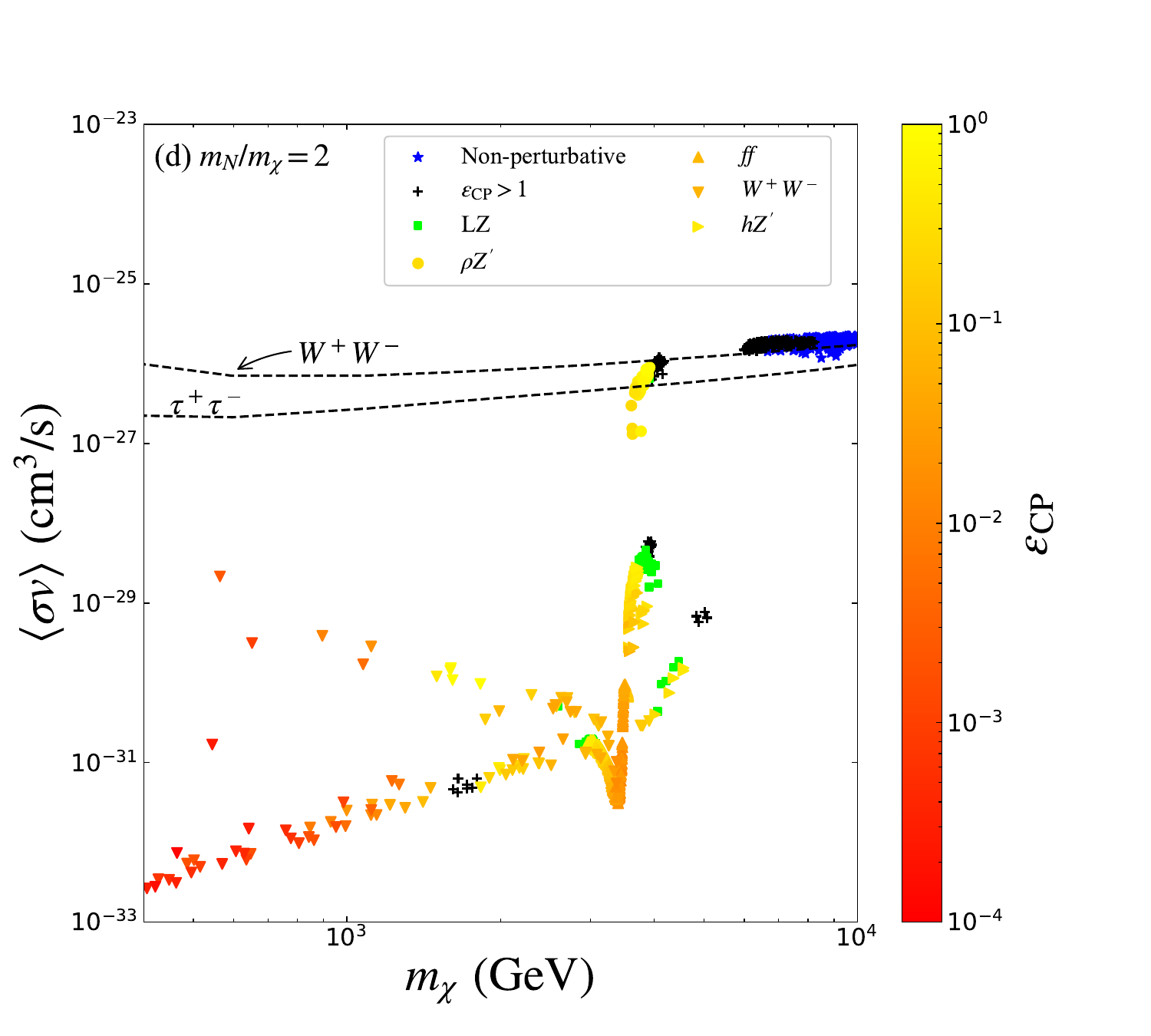}
	\end{center}
	\caption{Constraints from the indirect detection experiments in the local scenario. The newly added $\tau^+\tau^-$ dashed line represents the projected upper limits of $\tau^+\tau^-$ final states from the Cherenkov Telescope Array~\cite{CTA:2020qlo, Mangipudi:2021ivm}. Other labels are the same as that in Fig.~\ref{FIG:fig8}. 
	The allowed samples are marked with heterogeneous shapes to represent different dominant final states and colored by $\varepsilon_{\CP}$.
	}
	\label{FIG:fig15}
\end{figure}

The results of $\left<\sigma v\right>$ at present  with a variety of bounds are shown in Fig.~\ref{FIG:fig15}. In panel (a), we first illustrate the case with light sterile neutrino. At the time of DM freeze out, both $\chi\chi\to \text{SM SM}$ and $\chi\chi\to NN$ via $\rho$ or $Z'$ contribute to DM annihilation. As the former channel is $p$-wave suppressed at present, the dominant annihilation channel becomes $\chi\chi\to NN$. With a subdominant contribution to relic density, the $\chi\chi\to NN$ channel has an annihilation cross section $\left<\sigma v\right>$ less than $10^{-27}~\rm{cm^3/s}$, which is far below the constraints of Fermi-LAT and H.E.S.S. on the $NN$ final states~\cite{Campos:2017odj}. The small annihilation cross section $\left<\sigma v\right>$ of  $\chi\chi\to NN$ also makes it beyond the scope of future CTA sensitivity. Close to the sharp $Z'$ resonance, there are a few samples that dominantly annihilate into $f\bar{f}$ at present. However, their annihilation cross section is at the order of $\mathcal{O}(10^{-31})~\rm{cm^3/s}$, which is also difficult to be detected.

In panels (b), (c), and (d) of Fig.~\ref{FIG:fig15}, the $s$-wave $\chi\chi\to NN$ channel is kinematically disallowed at present due to $m_N\geq m_\chi$. Therefore, the permitted samples at the $\rho$ resonance dominantly annihilate into $W^+W^-$, while samples at the $Z'$ resonance annihilate into $f\bar{f}$. Both of these two channels are $p$-wave, so the corresponding current annihilation cross sections are less than $\mathcal{O}(10^{-29})~\rm{cm^3/s}$, which is far below the experimental sensitivity.

The promising annihilation channel is the $s$-wave process $\chi\chi\to\rho Z^\prime$ in the non-resonance region, which has a typical annihilation cross section of $2\times10^{-26}~\rm{cm^3/s}$.  Although part of such samples may suffer tight constraints from direct detection, they can always escape this limit by lowing $\theta$ while keeping $\left<\sigma v\right>$ of $\chi\chi\to\rho Z'$ unchanged. Different from the global scenario, the additional scalar singlet $\rho$ does not decay into the invisible Majoron. From Fig.~\ref{FIG:fig11}, we know that $\rho\lesssim200$ GeV when $\theta=0.05$. In this way, $\rho\to W^+W^-$ becomes the dominant decay mode, since $\rho\to Z^\prime Z^\prime$ and $\rho\to NN$ are kinematically disallowed. Meanwhile, the new gauge boson mainly decays into fermion pair as $Z'\to f\bar{f}$. The detailed calculation of the gamma-ray spectrum in the full dark matter annihilation channel of $\chi\chi\to \rho Z'$ is beyond the scope of this paper. For illustration, we consider the future CTA limit of the individual $W^+W^-$ and $\tau^+\tau^-$ final states. It is clear that the samples with $m_\chi \gtrsim 5800$ GeV are promising in the future CTA experiment for cases (a) and (b) of Fig.~\ref{FIG:fig15}. Although the $\chi\chi\to \rho Z'$ channel could also lead to an observable signature at CTA for case (c) and (d) with sterile neutrino heavier than DM, such samples are already excluded by perturbation and exceeding $\varepsilon_{\CP}$.

Furthermore, close to the $Z'$ resonance with $2m_\chi\gtrsim m_{Z^\prime}+m_\rho$ for light $m_\rho$, although some samples are generated through $\chi\chi\to f\bar{f}$ via $Z'$ at freeze out, the current annihilation of them are dominated by the $s$-wave process $\chi\chi\to \rho Z^\prime$. Due to the subdominant contribution to the DM abundance, the annihilation cross section of these samples is typically less than $10^{-26}~\rm{cm^3/s}$, which is below the CTA $W^+W^-$ sensitivity. Meanwhile, a few samples are above the $\tau^+\tau^-$ sensitivity, so these samples might still be promising when considering the full decays of $\chi\chi\to \rho Z^\prime$. On the other hand, when $2m_\chi<m_{Z'}+m_\rho$ at the $Z'$ resonance, the current dominant channels would be the $s$-wave process $\chi\chi\to h Z^\prime$. However, the corresponding annihilation cross section of certain samples are suppressed by the mixing $\theta=0.05$, which results in $\left<\sigma v\right>\lesssim 10^{-28}~\rm{cm^3/s}$. So such samples are clearly beyond the experimental reach.

\section{Conclusion}\label{SEC:CL}

This $U(1)_{B-L}$ model further extends the SM with sterile neutrinos $N$, Majorana dark matter $\chi$, and complex scalar $\phi$, which are charged under the $U(1)_{B-L}$ symmetry. An additional $Z_2$ symmetry is also employed to keep the stability of DM $\chi$. After the spontaneous breaking of $U(1)_{B-L}$ symmetry, both $N$ and $\chi$ obtain masses through the Yukawa interactions related to $\phi$. Besides generating the tiny neutrino masses via type-I seesaw, the out-of-equilibrium decays of TeV scale $N$ create the baryon asymmetry through the  resonance leptogenesis. Furthermore, the dilution of $N$ has a significant impact on leptogenesis with new particles in the $U(1)_{B-L}$.  Since the evolution of $N$ and $\chi$ involve quite similar processes in the calculation of baryon asymmetry and DM relic density, the focus of this paper is to explore the common parameter space of these two aspects under various constraints. Both the global $U(1)_{B-L}$ and local $U(1)_{B-L}$ scenario are studied in this work.

In the global $U(1)_{B-L}$ scenario, we have a scalar singlet $\rho$ and a (nearly) massless Majoron $\eta$ after the symmetry breaking. The evolution of $N$ and $\chi$ are dominantly determined by the pair annihilation processes $NN,\chi\chi\to\rho\rho,\rho\eta, \eta\eta$. Meanwhile, the conversion process $\chi\chi \leftrightarrow  NN$ mediated by $\rho$ and $\eta$ could also impact. We have chosen four scenarios, namely $m_N/m_\chi=0.5,1,1.5,2$, to illustrate the phenomenology. In addition, the mixing angle $\theta$ between $\rho$ and Higgs $h$ is important for DM direct detection, which is assigned to an appropriate value of $0.05$. Under various constraints, correct DM abundance and baryon asymmetry can be realized at the resonance region of $m_\rho\simeq 2 m_\chi$ or the non-resonance region of $m_\rho\lesssim200$ GeV. In the resonance region, the required CP-asymmetry $\varepsilon_{\CP}$ can be as small as $\mathcal{O}(10^{-5})$, while the non-resonance samples usually need $\varepsilon_{\CP}\gtrsim10^{-2}$. For the non-resonance region, the perturbation and exceeding $\varepsilon_{\CP}$ constraints become tighter when $m_N/m_\chi$ increases. Although the Higgs invisible decay induced by $h\to \eta\eta$ can not be detected by LHC, the future CEPC may have a positive signature. With $\theta=0.05$, most of the samples could lead to detectable DM-nucleon scattering cross section. For DM annihilation at present, the $s$-wave process $\chi\chi\to \rho\eta$ mainly decays into neutrinos, which is far below experimental limits. Due to additional contribution from $p$-wave $\chi\chi\to\eta\eta$ at freeze out, the $\chi\chi\to NN$ channel typically has an annihilation cross section smaller than $2\times 10^{-26}~\rm cm^3/s$, which leads to most samples beyond the reach of CTA. So the global $U(1)_{B-L}$ scenario is usually unpromising in indirect detection experiments.

In the local $U(1)_{B-L}$ scenario, the Majoron $\eta$  is replaced by the gauge boson $Z^\prime$. We have fixed $m_{Z'}=7000$ GeV in this study to respect current experimental  limits. The processes as $NN,\chi\chi\to \SM \SM$ via $\rho$ and $Z'$ become important. Four scenarios with $m_N/m_\chi=0.5,1,1.5,2$ are also chosen for illustration. Under various experimental limits, the common origin of DM and leptogenesis is viable at the $\rho$ resonance $m_\rho\simeq 2 m_\chi$ with $\varepsilon_{\CP}\gtrsim\mathcal{O}(10^{-5})$ or the $Z^\prime$ resonance $m_{Z'}\simeq 2 m_\chi$ with $\varepsilon_{\CP}\gtrsim\mathcal{O}(10^{-3})$. For the non-resonance region, we find that it is only possible for $m_\chi\gtrsim6000$ GeV and $m_N/m_\chi\leq1$ with $\varepsilon_{\CP}\gtrsim\mathcal{O}(10^{-1})$. When the sterile neutrino $N$ is heavier than the DM, the non-resonance samples are excluded by exceeding $\varepsilon_{\CP}$. The existence of the new gauge boson $Z'$ makes this scenario more promising at colliders. In the future, the 250 GeV ILC could probe all the allowed parameter space by precise measurement of the $e^+e^-\to f\bar{f}$ process. The DM-nucleon scattering is dominant by the Higgs portal processes, so we fix $\theta=0.05$ for illustration as in the global scenario. In this way, the future direct detection experiment would test a large portion of the allowed parameter space. At both resonance regions, the $\chi\chi\to \SM\SM$ processes are $p$-wave suppressed at present due to the  Majorana nature of $\chi$, thus are hard to detect. The promising case for the future CTA experiment is the non-resonance region with $\chi\chi\to \rho Z'$, where $\rho\to W^+W^-$ and $Z'\to f\bar{f}$ are the dominant decay modes.

To summarize, both the global and local $U(1)_{B-L}$ scenario favor the resonance region for common origin of DM and leptogenesis. However, the resonance region is hard to test at indirect detection experiments for both scenarios. On the other hand, correct relic abundance and baryon asymmetry can also be simultaneously satisfied in the non-resonance region. For the global $U(1)_{B-L}$, electroweak scale DM is viable, but is still hard to be  detected at indirect detection. For the local $U(1)_{B-L}$, DM must be over TeV scale, but is within the future CTA limit. The Majoron $\eta$  could induce testable Higgs invisible decay in the global scenario, while the gauge boson $Z'$ could be detected at colliders in the local scenario. These differences can be used to distinguish the global and local scenarios.

\section*{Acknowledgments}
This work is supported by the National Natural Science Foundation of China under Grant  No. 12375074, No. 12175115, and No. 11805081, Natural Science Foundation of Shandong Province under Grant No. ZR2019QA021 and No. ZR2022MA056, the Open Project of Guangxi Key Laboratory of Nuclear Physics and Nuclear Technology under Grant No. NLK2021-07.

\allowdisplaybreaks

\appendix
\section{The reduced cross sections}  \label{SEC:RCS} 
The reduced cross sections of annihilation processes involved in the Boltzmann equations are shown in this appendix. We use CalcHEP \cite{Belyaev:2012qa} throughout to calculate them.
In the global $U(1)_{B-L}$ scenario, our results match with the Ref. \cite{Dev:2017xry}. The reduced cross sections of sterile neutrino pair  annihilation processes $NN\to\rho\rho,\rho\eta,\eta\eta$ are 
\begin{eqnarray}
	\hat{\sigma}(NN\to\rho\rho)	&=&\frac{\lambda_N^4}{32\pi} \bigg(\frac{9 \beta_{\rho\rho}(x_N-4)r_{\rho N}^2}{(x_N-r_{\rho N})^2}-\frac{24\beta_{\rho\rho}x_N r_{\rho N}}{x_N|x_N-r_{\rho N}|} + \frac{12r_{\rho N}(x_N +2r_{\rho N}-8)}{x_N|x_N-r_{\rho N}|}	\nonumber \\
	&\times& \log\left(\frac{(1-\beta_{\rho\rho})x_N-2r_{\rho N}}{(1+\beta_{\rho\rho})x_N-2r_{\rho N}}\right)-4\beta_{\rho\rho} - \frac{8\beta_{\rho\rho}(r_{\rho N}-4)^2}{(x_N-2r_{\rho N})^2-\beta_{\rho\rho}^2x_N^2}-\frac{2}{x_N(x_N-2r_{\rho N})} \nonumber \\
	&\times&\Big(x_N^2-2(r_{\rho N}-4)(2x_N-3r_{\rho N}-4)\Big)\log\left(\frac{(1-\beta_{\rho\rho})x_N-2r_{\rho N}}{(1+\beta_{\rho\rho})x_N-2r_{\rho N}}\right)\bigg),\\
	\hat{\sigma}(NN\to\rho\eta)	&=&\frac{\lambda_N^4}{16\pi x_N^2}\left(\beta_{\rho\eta}r_{\rho N}^2 x_N-2(x_N-r_{\rho N})(x_N-2r_{\rho N})\log\left(\frac{(1-\beta_{\rho\eta})x_N-r_{\rho N}}{(1+\beta_{\rho\eta})x_N-r_{\rho N}} \right)\right),\\
	\hat{\sigma}(NN\to\eta\eta)	&=&\frac{\lambda_N^4}{32\pi(x_N-r_{\rho N})^2}\left(\beta_{\eta\eta}x_N(r_{\rho N}^2-4x_N)-2(x_N^2-r_{\rho N}^2)\log\left(\frac{1-\beta_{\eta\eta}}{1+\beta_{\eta\eta}} \right)\right),
\end{eqnarray}
where $x_i=s^{\prime}/m_i^2$, $r_{ij}=m_i^2/m_j^2$. In the above equations, we have considered a vanishing mixing $\theta$ as Ref. \cite{Dev:2017xry} to obtain the simplified expressions. The Yukawa coupling $\lambda_N$ is evaluated as $\lambda_N=m_N/v_\phi$ in the calculation. For $N$ pair annihilation, the $\beta_{ij}$ function is defined as
\begin{eqnarray}
	\beta_{ij}=x_N^{-1}\sqrt{(1-4x_N^{-1})(x_N^2+r_{iN}^2+r_{jN}^2-2x_N r_{iN}-2x_N r_{jN}-2r_{iN}r_{jN})}
\end{eqnarray}	

On the other hand, the annihilation channels into SM final states depend on the mixing $\theta$. The $NN\to VV$ ($V=W, Z, h$) process is dominant by the $\rho$ portal for TeV scale $N$, which is calculated as
\begin{eqnarray}
	\hat{\sigma}(N N\to VV) &=&\frac{1}{32\pi v_\phi^2(x_N-r_{\rho N})^2}\bigg(\beta_{WW}(x_N-4)\cos^2\theta\Big( \cos2\theta(\lambda_{H\phi}^2v_\phi^2-4\lambda_H^2v_H^2)
	\nonumber \\
	&+&\frac{g_{e}^2\sin^2\theta}{\sin^4\theta_w}(2g_{e}^2v_H^2-m_W^2\sin^2\theta_w-2m_N^2\sin^2\theta_wx_N)+\lambda_{H\phi}^2v_\phi^2
	\nonumber \\
	&+&4\lambda_H^2v_H^2+4\lambda_{H\phi}\lambda_Hv_\phi v_H\sin2\theta
	\Big)
	\nonumber \\
	&+&\frac{\beta_{ZZ}(x_N-4)\cos^2\theta}{2}\Big(\cos2\theta
	(\lambda_{H\phi}^2v_\phi^2-4\lambda_H^2v_H^2)+\frac{g_{e}^2\sin^2\theta}{\sin^4\theta_w\cos^4\theta_w}
	\nonumber \\
	&\times&(2g_{e}^2v_H^2-m_Z^2\sin^2\theta_w\cos^2\theta_w-2m_N^2\sin^2\theta_w\cos^2\theta_w x_N)+\lambda_{H\phi}^2v_\phi^2
	\nonumber \\
	&+&4\lambda_H^2v_H^2+4\lambda_{H\phi}\lambda_Hv_\phi v_H\sin2\theta
	\Big)	
	\nonumber \\
	&+&\beta_{hh}(x_N-4)\cos^2\theta\Big(\lambda_{H\phi}v_\phi\cos^3\theta+\lambda_{H\phi}v_H\sin^3\theta+2(3\lambda_{H}-\lambda_{H\phi})v_H
	\nonumber \\
	&\times&\cos^2\theta\sin\theta+2(3\lambda_{H}-\lambda_{H\phi})v_{\phi}\sin^2\theta\cos\theta\Big)^2
	\bigg),
\end{eqnarray}
where $g_{e}$ and $\theta_w$ are the electric coupling constant and Weinberg angle, respectively. At present, the $s$-wave $NN\to h\eta$ channel is also important, which is calculated as
	\begin{eqnarray}	
		\hat{\sigma}(NN\to h\eta)&=&\frac{\lambda_N^2}{16\pi x_N^2 m_N^2}\bigg(4\lambda_{H\phi}m_Nv_H\lambda_{N}\sin \theta\cos\theta(r_{hN}-x_N)\log\left(\frac{(1-\beta_{h\eta})x_N-r_{h N}}{(1+\beta_{h\eta})x_N-r_{h N}} \right) \nonumber \\
		&+& 2m_N^2\sin^2\theta\bigg(2\beta_{h\eta}\lambda_{\phi}^2 x_N+\lambda_{N}^2(r_{hN}-x_N)(4\lambda_{\phi}+x_N\lambda_{N}^2)\nonumber \\
		&\times&\log\left(\frac{(1-\beta_{h\eta})x_N-r_{h N}}{(1+\beta_{h\eta})x_N-r_{h N}} \right)\bigg)+\beta_{h\eta}\lambda_{H\phi}^2v_H^2 x_N\lambda_{N}^2 \cos^2\theta \nonumber \\
		&+&2\beta_{h\eta}\lambda_{H\phi}\lambda_{\phi}\lambda_{N} m_N v_Hx_N\sin2\theta
		\bigg)
\end{eqnarray}

In the local $U(1)_{(B-L)}$ scenario, the reduced cross section of $NN\to \rho Z',Z'Z'$ are
\begin{eqnarray}
	\hat{\sigma}(NN\to\rho Z^{\prime})	&=&\frac{1}{48\pi x_N}\bigg(\frac{\beta_{\rho Z^\prime}x_N}{\lambda_N^2 (x_N-r_{Z^\prime N})^2}\Big(384 g_{B-L}^6(x_N-6)+3\lambda_N^6 r_{\rho N}^2 x_N-12\lambda_N^4 g_{B-L}^2 \nonumber \\
	&\times&(2r_{\rho N}^2-2r_{\rho N}r_{Z^\prime N}+2r_{\rho N}x_N-r_{Z^\prime N}x_N+2x_N^2)+ 4\lambda_N^2 g_{B-L}^4\big(3r_{\rho N}^2 + 3r_{Z^\prime N}^2 \nonumber \\
	&-&6r_{Z^\prime N}x_N- 6r_{\rho N}(x_N+r_{Z^\prime N}-8)+x_N(3x_N-\beta_{\rho Z^\prime}^2 x_N+60)\big)\Big) \nonumber \\
	&-&\frac{64\lambda_N^2\beta_{\rho Z^\prime}x_N}{(r_{\rho N}+r_{Z^\prime N}+x_N(\beta_{\rho Z^\prime}-1))(r_{\rho N}+r_{Z^\prime N}-x_N(\beta_{\rho Z^\prime}+1))}\Big(\lambda_N^2\big(r_{\rho N}^2+ r_{Z^\prime N}^2 \nonumber \\
	&+&r_{\rho N}(6r_{Z^\prime N}-2x_N)-x_N^2(\beta_{\rho Z^\prime}^2-1)-2r_{Z^\prime N}(x_N+8)\big)+2g_{B-L}^2\big(96+r_{\rho N}^2 \nonumber \\
	&+&6r_{\rho N}r_{Z^\prime N}+r_{Z^\prime N}^2+x_N^2-\beta_{\rho Z^\prime}^2x_N^2-2r_{\rho N}(12+x_N)-2r_{Z^\prime N}(8+x_N)\big)\Big) \nonumber \\
	&+&\big(12\lambda_N^2 g_{B-L}^2(r_{\rho N}+r_{Z^\prime N}-x_N-12)+6\lambda_N^4(r_{\rho N}+r_{Z^\prime N}-x_N)\big) \nonumber \\
	&\times&\log\left(\frac{(1-\beta_{\rho Z^\prime})x_N-r_{\rho N}-r_{Z^\prime N}}{(1+\beta_{\rho Z^\prime})x_N-r_{\rho N}-r_{Z^\prime N}}\right)+6\lambda_N^2\beta_{\rho Z^\prime}x_N(\lambda_N^2-2g_{B-L}^2)
	\nonumber \\
	&+&\frac{12\lambda_N^2}{r_{\rho N}+r_{Z^\prime N}-x_N}\Big(-\lambda_N^2 r_{Z^\prime N}(r_{\rho N}-4)+2g_{B-L}^2\big(r_{\rho N}(r_{Z^\prime N}-x_N+4) \nonumber \\
	&-&2(r_{Z^\prime N}-3x_N+12)\big)\Big)\log\left(\frac{(1-\beta_{\rho Z^\prime})x_N-r_{\rho N}-r_{Z^\prime N}}{(1+\beta_{\rho Z^\prime})x_N-r_{\rho N}-r_{Z^\prime N}}\right)-\frac{12}{r_{Z^\prime N}-x_N} \nonumber \\
	&\times&\bigg(-48\beta_{\rho Z^\prime}g_{B-L}^4 x_N + 2\beta_{\rho Z^\prime} \lambda_{N}^2 g_{B-L}^2(r_{\rho N}+r_{Z^\prime N})x_N+\Big(\lambda_N^4 r_{\rho N}(r_{Z^\prime N}-r_{\rho N} \nonumber \\
	&+&x_N)+16g_{B-L}^4(2r_{\rho N}+r_{Z^\prime N}+x_N-12)-4\lambda_N^2 g_{B-L}^2\big(r_{\rho N}(r_{Z^\prime N}+x_N-2) \nonumber \\
	&-&2(r_{Z^\prime N}+x_N)\big)\Big)\log\left(\frac{(1-\beta_{\rho Z^\prime})x_N-r_{\rho N}-r_{Z^\prime N}}{(1+\beta_{\rho Z^\prime})x_N-r_{\rho N}-r_{Z^\prime N}}\right)\bigg)
	\bigg),
\end{eqnarray}	

\begin{eqnarray}
	\hat{\sigma}(NN\to Z^{\prime} Z^{\prime})&=&\frac{1}{32\pi x_N}\bigg(-2\beta_{Z^\prime Z^\prime}\lambda_N^2(\lambda_N^2+4g_{B-L}^2)x_N + \frac{\beta_{Z^\prime Z^\prime}(x_N-4)x_N}{(x_N-r_{\rho N})^2}\Big(256g_{B-L}^4   \nonumber \\
	&+& \lambda_N^4r_{\rho N}^2-8\lambda_N^2g_{B-L}^2(r_{Z^\prime N}+2x_N)\Big)-\frac{2\beta_{Z^\prime Z^\prime}x_N}{4r_{Z^\prime N}x_N-4r_{Z^\prime N}^2+(\beta_{Z^\prime Z^\prime}^2-1)x_N^2}
	\nonumber \\
	&\times&\Big(\lambda_N^4\big(4r_{Z^\prime N}x_N-8r_{Z^\prime N}^2+(\beta_{Z^\prime Z^\prime}^2-1)x_N^2\big)-4\lambda_N^2g_{B-L}^2\big(8r_{Z^\prime N}^2-(\beta_{Z^\prime Z^\prime}^2-1)x_N^2
	\nonumber \\
	&-&4r_{Z^\prime N}(6+x_N)\big)-4g_{B-L}^4\big(144+8r_{Z^\prime N}^2+x_N^2-\beta_{Z^\prime Z^\prime}^2x_N^2-4r_{Z^\prime N}(12+x_N)\big)\Big)
	\nonumber \\
	&+&\frac{4}{2r_{Z^\prime N}-x_N}\Big(\lambda_N^4r_{Z^\prime N}^2+4g_{B-L}^4\big(2r_{Z^\prime N}(x_N-1)-3(x_N-4)\big)+4\lambda_N^2g_{B-L}^2r_{Z^\prime N}
	\nonumber \\
	&\times&\!(r_{Z^\prime N}\!-x_N\!-2)\Big)\!\log\left(\frac{(1-\beta_{Z^\prime Z^\prime})x_N\!-r_{Z^\prime N}}{(1+\beta_{Z^\prime Z^\prime})x_N\!-r_{Z^\prime N}} \right)\!+2\Big(\lambda_N^4(2r_{Z^\prime N}\!-x_N)\!+4\lambda_N^2g_{B-L}^2
	\nonumber \\
	&\times&(2r_{Z^\prime N}-x_N)+g_{B-L}^4\big(8r_{Z^\prime N}-4(x_N+18)\big)\Big)\log\left(\frac{(1-\beta_{Z^\prime Z^\prime})x_N\!-r_{Z^\prime N}}{(1+\beta_{Z^\prime Z^\prime})x_N\!-r_{Z^\prime N}} \right)
	\nonumber \\
	&+&\frac{4}{r_{\rho N}-x_N}\bigg(2\beta_{Z^\prime Z^\prime}x_N(16g_{B-L}^4+\lambda_N^4r_{\rho N}-2\lambda_N^2g_{B-L}^2x_N)+\Big(8\lambda_N^2g_{B-L}^2
	\nonumber \\
	&\times&\big(r_{Z^\prime N}(x_N-2)-x_N\big)+\lambda_N^4r_{\rho N}(x_N-2r_{Z^\prime N})-32g_{B-L}^4(r_{Z^\prime N}+x_N-6)\Big)
	\nonumber \\
	&\times&\log\left(\frac{(1-\beta_{Z^\prime Z^\prime})x_N-r_{Z^\prime N}}{(1+\beta_{Z^\prime Z^\prime})x_N-r_{Z^\prime N}} \right)
	\bigg)\bigg).
\end{eqnarray}	

As couplings of $\rho$ to SM fermions are suppressed by the small mixing $\theta$, we only consider the $Z'$ contribution in the process of $NN\to f\bar{f}$. The result is
\begin{eqnarray}
	\hat{\sigma}(N N\to f\bar{f}) &=&\frac{N_c(f)Q_f^2\beta_{ff}g_{B-L}^4\Big(12r_{fN}(x_N-4)+x_N(3x_N+\beta_{ff}^2x_N-12)\Big)}{24\pi(x_N-r_{Z^\prime N})^2},
\end{eqnarray}
where $N_c(f)$ and $Q_{f}$ represent
the color multiplicity and the $B-L$ charge of the fermion $f$, respectively. Meanwhile, the s-wave $NN\to h Z^\prime$ induced by the small mixing $\theta$ is calculated as
\begin{eqnarray}
	\hat{\sigma}(NN\to h Z^{\prime})&=&\frac{1}{8\pi x_N m_N^2}\bigg(
	\frac{\beta_{h Z^\prime}x_N}{6\lambda_N^2 (x_N-r_{Z^\prime N})^2}
	\bigg(4m_N^2 \sin^2\theta\Big(96 g_{B-L}^6(x_N-6)+3\lambda_\phi^2 \lambda_{N}^2 x_N\nonumber \\
	&+&\lambda_N^2 g_{B-L}^4\big(3r_{h N}^2 + 3r_{Z^\prime N}^2 - 6r_{Z^\prime N}x_N- 6r_{h N}(x_N+r_{Z^\prime N}-8)\nonumber \\
	&+&x_N(3x_N-\beta_{h Z^\prime}^2 x_N+60)\big)+3\lambda_N^2 g_{B-L}^2\big(4\lambda_{\phi}(r_{h N}-r_{Z^\prime N})  \nonumber \\
	&+&x_N \lambda_{N}^2(r_{Z^\prime N}-r_{h N}-2x_N)\big)\Big) +3\lambda_{H\phi}v_H \lambda_N^3\Big(\lambda_{H\phi}v_H \lambda_N x_N\cos^2\theta \nonumber \\
	&+& 2m_N\sin 2\theta(2g_{B-L}^2(r_{h N}-r_{Z^\prime N})+\lambda_\phi x_N)\Big)\bigg)	\nonumber \\
	&-&	\frac{2m_N\sin\theta}{r_{Z^\prime N}-x_N}\bigg(-2\beta_{h Z^\prime}g_{B-L}^2 m_N x_N \sin\theta\big(24g_{B-L}^2-\lambda_N^2(r_{h N}+r_{Z^\prime N})\big)\nonumber \\
	&+&\!\!\Big(\lambda_{H\phi}v_H \lambda_N^3\!\cos\theta(r_{h N}\!-r_{Z^\prime N}\!-x_N)\!+\!2m_N\!\sin\theta\big(8g_{B-L}^4\!(2r_{h N}\!+r_{Z^\prime N}\!+x_N\!-12) \nonumber \\
	&+&\!\lambda_{\phi}\lambda_N^2(r_{h N}\!-r_{Z^\prime N}\!-x_N)-2g_{B-L}^2 y_N^2(r_{h N}(r_{Z^\prime N}+x_N-2)-2r_{Z^\prime N}-2x_N)\big)\Big)\nonumber \\
	&\times&\log\left(\frac{(1-\beta_{h Z^\prime})x_N-r_{h N}-r_{Z^\prime N}}{(1+\beta_{h Z^\prime})x_N-r_{h N}-r_{Z^\prime N}}\right)\bigg)\nonumber \\
	&-&\bigg(\frac{\lambda_N^2\beta_{h Z^\prime}x_Nm_N^2 \sin^2\theta}{(r_{h N}+r_{Z^\prime N}+x_N(\beta_{h Z^\prime}-1))(r_{h N}+r_{Z^\prime N}-x_N(\beta_{h Z^\prime}+1))}\Big(\lambda_N^2\big(r_{h N}^2+ r_{Z^\prime N}^2 \nonumber \\
	&+&r_{h N}(6r_{Z^\prime N}-2x_N)-x_N^2(\beta_{h Z^\prime}^2-1)-2r_{Z^\prime N}(x_N+8)\big)+2g_{B-L}^2\big(96+r_{h N}^2 \nonumber \\
	&+&6r_{h N}r_{Z^\prime N}+r_{Z^\prime N}^2+x_N^2-\beta_{h Z^\prime}^2x_N^2-2r_{h N}(12+x_N)-2r_{Z^\prime N}(8+x_N)\big)\Big)\nonumber \\
	&-&m_N^2\lambda_{N}^2 \sin^2\theta\big(2 g_{B-L}^2(r_{h N}+r_{Z^\prime N}-x_N-12)+\lambda_N^2(r_{h N}+r_{Z^\prime N}-x_N)\big)\nonumber \\
	&\times&\log\left(\frac{(1-\beta_{h Z^\prime})x_N-r_{h N}-r_{Z^\prime N}}{(1+\beta_{h Z^\prime})x_N-r_{h N}-r_{Z^\prime N}}\right)\bigg)\nonumber \\
	&+&\bigg(\frac{2 m_N^2\lambda_N^2\sin\theta^2}{r_{h N}+r_{Z^\prime N}-x_N}\Big(-\lambda_N^2 r_{Z^\prime N}(r_{h N}-4)+2g_{B-L}^2\big(r_{h N}(r_{Z^\prime N}-x_N+4) \nonumber \\
	&-&2(r_{Z^\prime N}-3x_N+12)\big)\Big)\log\left(\frac{(1-\beta_{h Z^\prime})x_N-r_{h N}-r_{Z^\prime N}}{(1+\beta_{h Z^\prime})x_N-r_{h N}-r_{Z^\prime N}}\right)\nonumber \\
	&+&m_N^2\lambda_N^2\sin\theta^2 \beta_{h Z^\prime}x_N(\lambda_{N}^2-2g_{B-L}^2)\bigg)
	\bigg)
\end{eqnarray}
 In addition, for the $s$-channel processes with $\rho$ and $Z^\prime$ resonance, we have added the decay width $\Gamma_\rho$ and $\Gamma_Z^\prime$ at the denominator of Breit-Wigner propagator in the calculations.

As for the annihilation processes of $\chi\chi$,  the reduced cross section can refer to the corresponding one of $N$ pair annihilation, i.e., replacing $\lambda_N$ and $m_N$ with $\lambda_\chi$ and $m_\chi$, respectively. Meanwhile, the $\beta_{ij}$ function for $\chi$ pair annihilation is defined as
\begin{eqnarray}
	\beta_{ij}=x_\chi^{-1}\sqrt{(1-4x_\chi^{-1})(x_\chi^2+r_{i\chi}^2+r_{j\chi}^2-2x_\chi r_{i\chi}-2x_\chi r_{j\chi}-2r_{i\chi}r_{j\chi})}
\end{eqnarray}

The reduced cross section of the conversion process $\chi\chi\to NN$ is
\begin{eqnarray}
	\hat{\sigma}(\chi\chi\to N N)_{\rm global} &=&\frac{\lambda_N^2\lambda_\chi^2}{16\pi (x_\chi-r_{\rho \chi})^2}\beta_{NN}\Big( r_{\rho\chi}(r_{\rho\chi}-2x_\chi)-4r_{N\chi}(x_\chi-4)+2x_\chi(x_\chi-2)\Big),\\
\hat{\sigma}(\chi\chi\to N N)_{\rm local} &=&\frac{\beta_{NN}}{48\pi}\bigg(\frac{3\lambda_\chi^2\lambda_N^2(x_\chi-4)(x_\chi-4r_{N\chi})}{(x_\chi-r_{\rho\chi})^2}+\frac{1}{(x_\chi-r_{Z^\prime \chi})^2}\Big(\!-24\lambda_\chi\lambda_N g_{B-L}^2 r_{N\chi}^{1/2}x_\chi
\nonumber \\
&+&3\lambda_\chi^2\lambda_N^2 x_\chi^2+g_{B-L}^4\big(-12r_{N\chi}(x_\chi-8)+x_\chi(-12+3x_\chi+\beta_{NN}^2x_\chi)\big)\Big)\bigg).
\end{eqnarray}


\end{document}